\documentclass[12pt]{article}
\pdfoutput=1
\usepackage{jheppub}
\usepackage{subfigure}

\usepackage{setspace}
\usepackage{array}

\usepackage{color}

\newcommand{\N}{\mathcal{N}}
\newcommand{\ti}{\widetilde}

\newcommand{\be}{\begin{equation}}
\newcommand{\ee}{\end{equation}}
\newcommand{\bea}{\begin{eqnarray}}
\newcommand{\eea}{\end{eqnarray}}
\newcommand{\ba}{\begin{aligned}}
\newcommand{\ea}{\end{aligned}}

\newcommand{\tr}{\textrm{Tr} \,}

\newcommand{\ol}{\overline}

\newcommand{\bC}{\ensuremath{\mathbb{C}}}

\newcommand{\bH}{\ensuremath{\mathbb{H}}}

\newcommand{\bR}{\ensuremath{\mathbb{R}}}

\newcommand{\bZ}{\ensuremath{\mathbb{Z}}}

\newcommand{\scB}{\ensuremath{\mathcal{B}}}
\newcommand{\scC}{\ensuremath{\mathcal{C}}}

\newcommand{\scM}{\ensuremath{\mathcal{M}}}

\newcommand{\scO}{\ensuremath{\mathcal{O}}}

\newcommand{\scR}{\ensuremath{\mathcal{R}}}

\numberwithin{equation}{section}       

\title{The Space of Vacua of 3d $\mathcal{N}=3$ Abelian Theories}

\author{Benjamin Assel}
\affiliation{CERN, Theoretical Physics Department, CH-1211 Geneva 23, Switzerland}

\abstract{We use brane techniques to study the space of vacua of abelian 3d $\mathcal{N}=3$ gauge theories. The coordinates on these spaces are the vevs of chiral monopole and meson operators, which are realized in the type IIB brane configuration of the theory by adding semi-infinite $(1,k)$ strings or F1 strings. The study of various brane setups allows us to determine a basis of chiral operators and chiral ring relations relevant to each branch of vacua, leading to the algebraic description of these branches. The method is mostly graphical and does not require actual computations. We apply it and provide explicit results in various examples. For linear quivers we find that the space of vacua has in general a collection of Coulomb-like branches, a Higgs branch and mixed branches. For circular quivers we find an extra branch, the geometric branch, parametrized by monopoles with equal magnetic charges in all $U(1)$ nodes and meson operators. We explain how to include FI and mass deformations. We also study $\mathcal{N}=3$ theories realized with $(p,q)$ 5-branes.}

\emailAdd{benjamin.assel@gmail.com}

\begin{document}

\vspace*{-2cm} 
\begin{flushright}
{\tt  CERN-TH-2017-120} 
\end{flushright}

\maketitle

\section{Introduction}
\label{sec:Introduction}

It is a central question in quantum field theories to determine the vacuum configurations and the low-energy physics as one flows above each vacuum. In supersymmetric gauge theories, it often happens that the vacua are not discrete isolated points in the space of field configurations but instead are parametrized by continuous moduli. This moduli space of vacua -- or vacuum space for short -- 
may have singularities indicating the presence of additional massless states and possibly signaling infrared interacting fixed points. 
Studying the vacuum space of a supersymmetric theory is a difficult task, since vacua can be lifted by quantum effects, in particular by non-perturbatively generated superpotentials. The task can become more tractable when the amount of supersymmetry is large enough to fix uniquely the superpotential of the theory. The low energy effective action is then determined from the metric on the vacuum space, which can still receive perturbative and non-perturbative quantum corrections.

In this work we are interested in three-dimensional gauge theories with $\N=3$ supersymmetry.  They are closely related to $\N=4$ gauge theories which have received some attention recently and it is useful to first review a few properties of those theories and the recent progress in understanding their vacuum space.
The space of vacua of 3d $\N=4$ gauge theories can be conveniently described as a complex algebraic variety (see \cite{Seiberg:1996nz} for an early example). This description is closely connected to the notion of chiral ring in the theory: the vevs of gauge invariant chiral operators are the holomorphic functions on the vacuum space and the chiral ring relations become algebraic relations on the coordinate ring (the set of holomorphic functions).  The vacuum space is then simply described in terms of a finite number of complex coordinates, which are the vevs of a basis of chiral operators, and a set of polynomial relations, inherited from the chiral ring relations. In this context the choice of a chiral subalgebra is correlated to the choice of a complex structure on the vacuum space. 
One major advantage of the algebraic description, compared to the description in terms of a metric for instance, is that the ring relations are independent of the gauge couplings and therefore are not subject to quantum corrections.

The space of vacua in $\N=4$ theories has a Higgs branch and a Coulomb branch, as well as mixed branches, all hyperk\"ahler manifolds. The Higgs branch is the same as in 4d $\N=2$ theories. It can be defined as a hyperk\"ahler quotient by the gauge group action of the space parametrized by the vevs of the chiral scalars in the hypermultiplets. Its metric is protected from quantum corrections \cite{Intriligator:1996ex}. 
The Coulomb branch is parametrized by the vevs of chiral scalars in the vector multiplets and chiral monopole operators, which are specific to three-dimensional theories and capture the moduli associated to the dual photons.\footnote{In three dimensions, abelian gauge fields can be dualized to periodic scalars, usually called dual photons, which can take vevs \cite{Aharony:1997bx}.} These monopole operators are disorder operators. Their insertion is defined by prescribing a supersymmetric Dirac monopole singularity for the vector multiplet fields at a point in 3d space in the path integral formulation of the theory. To obtain an algebraic description of the Coulomb branch, one must find the chiral ring relations obeyed by the monopole operators. They are purely quantum relations, which arise from the dynamics of the theory rather than from a superpotential, and are in general difficult to determine. The first monopole chiral ring relations were found in \cite{Borokhov:2002cg} for the $\N=4$ SQED theory using CFT methods.  Recently several methods have been proposed to find these monopole relations. One is based on Coulomb branch Hilbert series \cite{Cremonesi:2013lqa,Cremonesi:2014uva,Hanany:2016ezz}, which are protected indices counting monopole operators. After writing the Hilbert series as a plethystic exponential, one is able to extract a basis of monopole generators and their charges under global symmetries. When the global charges allow to distinguish between all operators in the basis, one can also extract the chiral ring relations up to coefficients. Another new approach to determining the exact Coulomb branch monopole relations has been presented in \cite{Bullimore:2015lsa}, using the so-called {\it abelianization map} and {\it abelianized relations} of the theory.
Finally we proposed a third approach in \cite{Assel:2017hck} based on the brane realization of the 3d theory in IIB string theory, where a basis of chiral operators and the chiral ring relations between them are extracted from reading a selection of brane setups with additional D1 strings for the Coulomb branch or additional F1 strings for the Higgs branch. This last method was developed for $\N=4$ abelian theories with a brane realization in type IIB, and confirmed results that could be obtained by the other methods.

Three-dimensional theories with $\N=3$ supersymmetry have comparatively received much less attention, although the amount of supersymmetry is large enough to study many aspects of these theories. 
In the present work we propose an analysis of the space of vacua of $\N=3$ theories, focusing on abelian quiver theories. 
The $\N=3$ gauge theories have the same field content as $\N=4$ gauge theories and the same Lagrangian, except for the presence of Chern-Simons terms which breaks the supersymmetry to $\N=3$. The space vacua is still hyperk\"ahler and can be parametrized by vevs of chiral operators in the theory, subject to ring relations. The presence of the Chern-Simons terms induce gauge charges for the chiral monopole operators. To define gauge invariant operators, they must be dressed with matter scalars. Such monopole operators appeared  already in \cite{Gaiotto:2009tk,Benini:2009qs,Jafferis:2009th,Benini:2011cma}, where a class of 3d $\N=3$ and $\N=2$ circular quiver theories and their holographic duals were studied. The chiral ring relations involving monopole operators were guessed in order to map the space of vacua with the transverse geometry probed by M2 branes in the M-theory dual backgrounds. 
To our knowledge, the only technique available to find  chiral ring relations between monopole operators in $\N=3$ theories comes from the Hilbert series which were recently studied in \cite{Cremonesi:2016nbo}. However this applies only to theories where global symmetries allow to guess the relations, which is not the generic case. 

In this paper we extend the brane techniques of \cite{Assel:2017hck} to $\N=3$ abelian quiver theories with linear and circular shapes and  provide an algorithm to solve for their space of vacua. The $\N=3$ theories have a brane realization with D3 branes suspended between $(1,k_i)$ 5-branes -- we call them 5$_{(1,k_i)}$ branes --  and intersecting D5 branes. The Chern-Simons levels $\kappa_i$ in the gauge theory are given by the differences $\kappa_i = k_i - k_{i-1}$. The orientations of the 5$_{(1,k_i)}$ branes in directions transverse to the D3 branes depend on the value of $k_i$ \cite{Gauntlett:1997pk,Kitao:1998mf}. We find a dictionary between chiral operator insertions in the 3d gauge theory and brane setups in type IIB. In particular the insertion of chiral monopole operators in the theory are realized by adding semi-inifinite $(1,k_i)$ strings -- we call them 1$_{(1,k_i)}$ strings --  stretched between 5$_{(1,k_i)}$ branes and ending on the D3 segments. For circular quivers the D3 branes wrap a compact direction in space and additional monopole operators can be inserted be letting semi-infinite 1$_{(r,s)}$ strings wrap the compact direction and end on the D3s. Chiral operators of mesonic type are inserted by extra D3 branes intersecting the initial D3 segments at points. To find chiral ring relations one has to consider a set of brane setups with 1$_{(1,k_i)}$ strings stretched between 5$_{(1,k_i)}$ branes, and 1$_{(r,s)}$ strings intersecting the D3 segments for circular quivers. Each of these brane setups has two interpretations in terms of chiral operator insertions, involving products of chiral operators. Identifying the two insertions yield the chiral ring relations. The essential ideas can be rapidly understood from the simple examples of Sections \ref{ssec:CSTSU2}, \ref{ssec:U1cube}, \ref{ssec:T3}.

The algorithm to solve for the space of vacua can be summarized as follows for linear quiver theories.
\begin{itemize}
\item From the brane configuration realizing the 3d theory, one can identify the branches of vacua, using and extending the techniques of \cite{Hanany:1996ie,Gaiotto:2008sa,Gaiotto:2008ak}. These are associated to the possible motions of D3 segments along 5$_{(1,k_i)}$ branes. When all the 5 branes are of different type ($k_i \neq k_j$ for $i\neq j$) the D3 segments cannot move and the space of vacua is trivial (a single point). Branches of vacua appear when several 5 branes are of the same type, allowing D3 segments stretched between them to move along their worldvolume directions. This is tied to the existence of sequences of CS levels in the quiver satisfying $\sum_{n=i}^j \kappa_n =0$. For each collection of 5 branes of the same type there is a {\it maximal branch} associated to D3 motions along these 5 branes. In $\N=4$ theories, these are the Coulomb branch (NS5 branes) and the Higgs branch (D5 branes). In $\N=3$ theories there is an arbitrary number of these maximal branches. Generically D3 segments can split into several groups moving along 5branes of different types, corresponding to mixed branches of vacua, which are direct products of submanifolds of the maximal branches. For each branch of vacua, with associated D3 motions, one is able to read from the brane configuration the set of operators which can take non-zero vev on the branch. Except for the Higgs branch, all the maximal branches are parametrized by monopole operators and some meson operators. This provides the global branch structure of the vacuum space.
\item For each maximal branch one must identify the ring relations.  The operators parametrizing the branch associated to a type of 5 branes, say 5$_{(1,k)}$ branes, are a set of monopole operators $V^\pm_{[i,j]}$, with $i<j$, realized by 1$_{(1,k)}$ strings stretched between the $i$th and $j$th 5$_{(1,k)}$ branes and ending on the D3, and a set of mesons $X_n$.\footnote{For the branch associated to NS5 branes, the $X_n$ can also be vector multiplet scalars. On other branches the vector multiplet scalars are identified with mesons due to F-term constraints.} The $\pm$ label distinguishes monopoles realized with the string ending on the D3 branes ``from above" and ``from below" (see later sections) and the two corresponding monopoles have opposite magnetic charges.
To find the ring relations there are two sets of brane setups to consider. One group of brane setups have a 1$_{(1,k)}$ string stretched between two consecutive 5$_{(1,k)}$ branes and crossing the D3 segments. They yield relations of the type
\be
V^+_{[i,i+1]}V^-_{[i,i+1]} = P_i(X_n) \,,
\ee
where $P_i(X_n)$ are polynomials of the mesons $X_n$. The insertion of $P_i(X_n)$ comes from integrating out fermionic modes of mass $X_n$ arising in the brane configurations.
The second group of brane setups have a semi-infinite 1$_{(1,k)}$ string stretched between two distant 5$_{(1,k)}$ branes and ending on the D3 branes. They yield relations of the type
\be
V^\pm_{[i,j]}V^\pm_{[j,k]} = V^\pm_{[i,k]} P_{ijk}(X_n) \,, \quad i<j<k \,,
\ee
with $P_{ijk}(X_n)$ are polynomials in the $X_n$.
These relations are not the full set of ring relations on the maximal branch -- we call them {\it pre-relations} -- however the missing relations are found by manipulating the pre-relations and allowing for division by $X_n$ operators appearing on both sides of a relations. In the simple cases studied in Section \ref{sec:SimpleCases} the pre-relations are directly the full set of ring relations. In all cases the ring relations can be determined completely.
\item Once the algebraic description of the maximal branches is found, the mixed branches are easily obtained as direct products of (hyperk\"ahler) submanifolds of the maximal branches. The relevant submanifolds of a given maximal branch are obtained by setting to zero some of the coordinates/operators.
\end{itemize}
For circular quivers with the sum of the CS levels vanishing, $\sum_i \kappa_i = 0$, there is an extra maximal branch which is parametrized by the vevs of monopole operators $V^\pm_{(r,s)}$, realized with 1$_{(r,s)}$ strings wrapping the compact direction and ending on the D3 branes, and some meson operators $X_n$. This branch of vacua has been identified with the transverse geometry probed by M2 branes in M-theory dual backgrounds and we call it the {\it geometric branch} following \cite{Benini:2009qs}. The chiral ring relations for the operators on the geometric branch are read from the brane setups with an infinite 1$_{(r,s)}$ string crossing the D3 branes and are of the form
\be
V^+_{(r,s)}V^-_{(r,s)} = P_{r,s}(X_n) \,,
\ee
with polynomials $P_{r,s}$. To determine the geometric branch one still has to eliminate redundant operators. The $V^\pm_{(r,s)}$ can all be expressed as products of operators of in a finite basis. To find this basis one must study the possible deformations of the 1$_{(r,s)}$ strings into string junctions with other 1 strings and relate the monopole operators insertions. A simple example of this analysis is presented in Section \ref{ssec:ABJM} to find the geometric branch of the abelian ABJM theory \cite{Aharony:2008ug}. 

The Higgs branch is found in the same way as any maximal branch, with monopole operators replaced by meson operators made out of fundamental matter scalars, inserted by adding F1 strings stretched between D5 branes, as already explained in \cite{Assel:2017hck}. The ring relations that we find for the Higgs branch are compatible with the classical analysis (F-term constraints) and with the definitions of the meson operators in terms of elementary matter fields.

The $\N=3$ theories admit massive deformations by mass and FI parameters which lift some branches and deform the surviving ones. In the brane setup these deformations correspond to 5 brane displacements along transverse directions \cite{Hanany:1996ie}. We explain how to incorporate these effect in the analysis and derive the deformed chiral ring relations.

\medskip

This procedure allows to determine, for the first time, the branch structure of the vacuum space and the algebraic description of each branch for any $\N=3$ abelian linear or circular quiver. In particular it provides a basis of generators for each branch and all the quantum ring relations involving monopole operators. The method is essentially graphical and does not necessitate computations. For instance it does not necessitate to work out the spectrum of gauge invariant monopole operators and to compute their dimension (or R-charge). These informations are directly encoded in the brane setups. The only drawback of the method is that it only applies to theories with a brane realization in type IIB string theory, and so far only to abelian theories.

Throughout the paper we study several theories of increasing complexity, including Chern-Simons quiver theories with enhanced $\N=4$ supersymmetry \cite{Gaiotto:2008sa,Hosomichi:2008jd}. This allows us to test our results using dualities with standard $\N=4$ gauge theories for which the space of vacua is known (and to confirm the dualities). 
We also consider theories realized with 5$_{(p,q)}$ branes. For abelian theories a description of the low-energy quiver theory on the D3 branes was proposed in \cite{Gaiotto:2008ak}, involving $\N=4$ BF couplings between $U(1)$ gauge nodes.\footnote{For non-abelian theories the description is non-Langrangian.} Our method extends naturally to such theories and we determine the vacuum space in two simple examples.
\medskip

The rest of the paper is organized as follows. In Section \ref{sec:CSthReview} we review basics of 3d $\N=3$ gauge theories, chiral monopole operators and the algebraic description of the space of vacua. In Section \ref{sec:Branes} we review the brane realization of these theories and we find the brane and strings inserting local chiral operators in the theories. We then start our analysis with simple theories in Section \ref{sec:SimpleCases}, developing the techniques to extract the branch structure and the ring relations. We also explain how to include FI and mass deformations in the analysis. In Section \ref{sec:LinQuiv} we explain the general structure of the moduli space of linear quiver theories and study  a seven-nodes quiver theory with four maximal branches, to illustrate our algorithm in a generic theory. In Section \ref{sec:CircQuiv} we focus on circular quivers and study the geometric branch of vacua. In Section \ref{sec:5pq} we generalize the technique to studying quiver theories realized with $5_{(p,q)}$ branes, providing two simple examples. We conclude in Section \ref{sec:Future} with proposals for future work.


\section{Chern-Simons $\N=3$ gauge theories and their spaces of vacua}
\label{sec:CSthReview}

In this section we review the construction of 3d $\N=3$ gauge theories and we explain that the space of vacua is parametrized by the vevs of chiral monopole and meson operators, subject to classical and quantum constraints.
\medskip

In three dimensions one can construct gauge theories with $\N=3$ supersymmetry (i.e.~six Poincar\'e supercharges) by taking a standard $\N=4$ Yang-Mills gauge theory and adding a supersymmetric Chern-Simons deformation which breaks the supersymmetry down to $\N=3$, as we review below. Since the Yang-Mills coupling in 3d is dimensionful -- it has the dimension of the square-root of a mass -- the theory is not conformal, however in the low-energy limit the Yang-Mills kinetic term can be dropped and massive vector multiplet fields can be integrated out, leaving a low-energy Chern-Simons theory with supersymmetric matter which is classically (and quantum mechanically \cite{Gaiotto:2007qi}) conformal. In some cases it happens that the low-energy SCFT has enhanced $\N \ge 4$ supersymmetry. We will encounter quite a few occurrences of this phenomenon in this paper, in particular enhancements to $\N=4$ supersymmetry.

We start by reviewing the gauge theory construction of $\N=3$ theories, following \cite{Kapustin:1999ha, Gaiotto:2007qi}. To begin with we consider an $\N=4$ theory, with gauge group $G$ and hypermultiplet in a representation $\scR$ of $G$. The theory has an $\N=4$ vector multiplet which is composed in $\N=2$ language of a vector multiplet $V \simeq  (A_\mu, \lambda, \sigma)$ and an adjoint chiral multiplet $\Phi \simeq (\Phi, \chi)$, where $\lambda, \chi$ are two-component Dirac spinors, $\sigma$ is a real scalar and $\Phi$ is a complex scalar\footnote{We use the common abuse of notation where the chiral superfield and the complex scalar bottom component have the same name.}, all transforming in the adjoint representation. 
The matter fields come in a hypermultiplet, which is composed in $\N=2$ language of two chiral multiplets $Q \simeq (Q, \psi)$, $\ti Q \simeq (\ti Q, \ti\psi)$, with $Q, \ti Q$ two complex scalars and $\psi, \ti\psi$ two Dirac fermions, transforming in complex conjugate representations $\scR, \scR^\ast$. 
In term of this data -- $G$ and $\scR$ -- the Langrangian is fixed by the requirement of $\N=4$ supersymmetry. In particular the superpotential is given by 
\be 
W_{\N=4} =  \ti Q \Phi Q \,.
\ee
The $\N=4$ theory has $SU(2)_C \times SU(2)_H$ R-symmetry, with $(Re(\Phi), \sigma, Im(\Phi))$ forming a triplet of $SU(2)_C$ and $(Q,\ti Q)$ forming a doublet of $SU(2)_H$.\footnote{The fermions also transform under the R-symmetry. The real components of $(\lambda,\chi)$ recombine and transform in the $({\bf 2}, {\bf 2})$ of  $SU(2)_C \times SU(2)_H$ and $(\psi, \ti\psi)$ transform as a doublet of $SU(2)_C$.}

Supersymmetry can be broken to $\N=3$ by adding for each $U(N)$ (or $SU(N)$) gauge factor a supersymmetric Chern-Simons term
\be\ba
S_{CS, \, \N=3} &= S_{CS, \, \N=2} - \frac{\kappa}{4\pi} \int d^2\theta d^3x \, \tr \Phi^2  \, + \text{c.c.}       \,, \cr
S_{CS, \, \N=2}  &= \frac{\kappa}{4\pi} \int d^3 x \, \tr \big[ A \wedge dA + \frac{2}{3} A\wedge A \wedge A - \bar\lambda\lambda + 2 D\sigma \big]  \,,\label{CSterm}
\ea\ee
where $D$ is a real auxiliary field in the $\N=2$ vector multiplet, $\tr$ denotes the trace in the fundamental representation of $U(N)$ and $\kappa \in \bZ$ is the quantized Chern-Simons level. The deformation $S_{CS, \, \N=2}$ preserves $\N=2$ supersymmetry. Adding the superpotential term $- \frac{\kappa}{4\pi} \tr \Phi^2$ leads to $\N=3$ supersymmetry.
The R-symmetry group is now $SU(2)_R = $diag$(SU(2)_C \times SU(2)_H)$ \footnote{See \cite{Kao:1995gf} for an expression of the full $\N=3$ Chern-Simons term in component fields with manifest $SU(2)_R$ symmetry.}.
It is also possible to add a mixed Chern-Simons coupling, or BF term, for two abelian vector multiplets \cite{Kapustin:1999ha}, compatible with $\N=4$ supersymmetry.

The CS deformation \eqref{CSterm} gives a mass to the vector multiplet fields. At low energies the fermions $\lambda,\chi$ and the scalar fields $\sigma,\Phi$ can be integrated out as auxiliary fields, leading to a quartic potential for the hypermultiplet scalars. The resulting theory has $\N=3$ superconformal symmetry. However, for our purposes it will be more convenient to work with the UV description in terms of the Yang-Mills-Chern-Simons theory.

 In this paper we will only consider gauge groups which are products of abelian factors $\prod_{i}U(1)_{\kappa_i}$ where the index $\kappa_i$ indicates the Chern-Simons level ($\kappa_i=0$ indicates a pure Yang-Mills gauge factor).
 \medskip
 
 Due to the supersymmetry the moduli space of vacua of $\N=3$ Chern-Simons theories is hyperk\"ahler with $SU(2) \simeq SU(2)_R$ action. More precisely, each branch of vacua is a hyperk\"ahler cone with $SU(2)$ isometry. The vacuum configurations are parametrized by the vevs of scalars in the theory obeying constraints from minimizing the action potential. 
For a theory with $\prod_i U(1)_{\kappa_i}$ gauge group, vector multiplet scalars $(\varphi_i,\sigma_i)$ and hypermultiplets $(Q_\alpha,\ti Q_\alpha)$ of charge $q_{\alpha i}$ under $U(1)_{\kappa_i}$, the vacuum equations are
\be\ba
&  \sum_i q_{\alpha i}\varphi_i  Q_\alpha  =   0 \,,  \quad   \sum_i q_{\alpha i}  \varphi_i \ti Q_\alpha  = 0   \,, \quad  \sum_\alpha q_{\alpha i} Q_\alpha \ti Q_\alpha   = \kappa_i \varphi_i \,,   \cr
& \sum_i q_{\alpha i}\sigma_i  Q_\alpha  =   0 \,,  \quad   \sum_i q_{\alpha i}  \sigma_i \ti Q_\alpha  = 0   \,, \quad  
\sum_\alpha q_{\alpha i} (Q_\alpha  Q^{\dagger}_\alpha - \ti Q_\alpha^\dagger \ti Q_\alpha )  = \kappa_i \sigma_i \,,  
\label{PotentialConstraints}
\ea\ee
These constraints combine into triplets of $SU(2)_R$. Due to the supersymmetry they do not receive quantum corrections \footnote{The superpotential of the theory (in $\N=2$ language) is fixed by supersymmetry in $\N \ge 3$ theories. The possible shifts of Chern-Simons levels at one-loop in $\N=2$ theories do not arise with $\N\ge 3$ because the matter content is non-chiral.} . 
In addition there are moduli related to the vector fields, which can be understood as follows: 
the abelian gauge fields can be dualized to periodic scalars $\gamma_i$ called dual photon, $\star dA^{(i)} = d\gamma_i$, which can take non-zero vevs in vacuum configurations. Due to one-loop quantum corrections the circles parametrized by these scalars can shrink and the moduli space is not easily described. Moreover this picture does not extend straightforwardly to non-abelian theories.

Instead the space of vacua can be more conveniently parametrized by the vevs of gauge invariant chiral operators in the theory,\footnote{Although this seems to be widely believed, we are not aware of a proof of this statement.} subject to constraints. This contains the chiral scalars $\varphi_i$ and the gauge invariant polynomials of the chiral scalars $ Q_\alpha, \ti Q_\alpha$, the mesons, satisfying the vacuum constraints \eqref{PotentialConstraints}. 
The moduli carried by $\sigma_i$ and $\gamma_i$ are exchanged for the vevs of chiral monopole operators $u_i \sim e^{\frac{2\pi}{g^2}(\sigma_i + i \gamma_i)}$ dressed with scalars. The insertion at a point $x_0$ in 3d space of a monopole operator is equivalent to imposing in the path integral formulation of the theory, a Dirac monopole singulary at $x_0$
\be
\frac{1}{2\pi} \oint_{x_0} dA^{(i)} = n_i \in \bZ \,,
\label{MonopSing}
\ee
with $n=(n_i)$ a set of quantized magnetic charges.
A corresponding singularity is prescribed for the real scalars $\sigma_i \sim \frac{n_i}{|x-x_0|}$ in order to preserve at least two supercharges. This defines a {\it bare} chiral monopole operator $\ol V_n$, with magnetic charge $n=(n_i)$. The choice of $n$ can be rephrased as the choice of an embedding  $\rho_n: U(1) \to \prod_i U(1)_{\kappa_i}$. 
The Chern-Simons terms  give gauge charges $q_i$ to the bare monopole with
\be
q_i(\ol V_n) = - \kappa_i n_i \,.
\label{MonopElCharge}
\ee
In order to define gauge invariant operators 
the bare monopole must  be ``dressed" with the insertion of hypermultiplet scalars $Q_\alpha,\ti Q_\alpha$ transforming with the total opposite charges $\kappa_i  n_{i}$ for each gauge node. 
In order to solve the BPS monopole equations, it is important that the dressing scalars are not charged under $\rho_n(U(1))\subset \prod_{i}U(1)_{\kappa_i}$. This implies that the monopole gauge charge under $\rho_n(U(1))$ must be trivial (otherwise there would be no way to make it gauge invariant):
\be
\sum_i \ \kappa_i n_{i} = 0 \,.
\label{MonopConstraint1}
\ee
This procedure defines gauge invariant chiral monopole operators  $V_{n,P}$ labeled by a vector of integer magnetic charges $n=(n_i$ and a dressing polynomial $P$ in the hypermultiplet complex scalars. 
The monopole operators with zero magnetic charge $V_{0,P}$ are simply the chiral mesons mentioned before. 
This description of the monopole operators as chiral pointlike defects generalizes to the non-abelian theories and therefore is useful to describe the space of vacua of general $\N\ge 3$ gauge theories.

It will be useful to know the $U(1) \subset SU(2)_R$ R-charge of the chiral operators, which is the same as the (protected) conformal dimension at the infrared fixed point\footnote{The theories that we study in this paper are ``good" according to the classification of \cite{Gaiotto:2008ak}, implying that the UV and IR R-symmetries coincide.}. The hypermultiplet scalars $Q_\alpha,\ti Q_\alpha$ have canonical dimension $\frac 12$. The vector multiplet scalar $\varphi_i$ has dimension $1$. For monopole operators the dimension is computed as the sum of the dimensions of matter scalars and of the bare monopole entering in its definition. The dimension of a bare monopole $\ol V$ in $\N=4$ theories was computed in \cite{Borokhov:2002cg, Gaiotto:2008ak} and a more general formula for $\N=2$ theories with Chern-Simons couplings can be found in \cite{Cremonesi:2016nbo}. For abelian theories the formula reduces to
\be 
\Delta(\ol V_{n}) = \frac 12 \sum_{\alpha} \big| \sum_{i} n_i q_{\alpha i} \big| \,.
\label{MonopDim}
\ee

The moduli space of vacua is then parametrized by the vev of the $V_{n,P}, \varphi_i$ subject to the vacuum equations \eqref{PotentialConstraints}. The constraints on $\sigma_i$ gets replaced by constriaints on the monopole operators. They follow from the fact that the BPS monopole solution with magnetic charge $n = (n_{i})$ is not compatible with giving  a vev to a chiral field charged under $\rho_n(U(1)) \subset \prod_i U(1)_{\kappa_i}$. 
Therefore we have the constraints
\be\ba
& V_{n,P} Q_\alpha = V_{n,P} \ti Q_{\alpha} = 0 \,, \quad \text{if} \quad \sum_i n_i q_{\alpha i} \neq 0 \,.
\label{MonopConstraint2}
\ea\ee

 In addition the monopole operators can be subject to quantum relations, namely relations which do not follow from a superpotential but from the quantum dynamics of the theory, as happens in the more familiar context of Yang-Mills theories \cite{Borokhov:2002cg,Bullimore:2015lsa}.
 This was recently described in \cite{Cremonesi:2016nbo}, where the Hilbert series counting chiral operators in $\N\ge 2$ Chern-Simons theories were computed.
The emerging picture is then a moduli space of vacua parametrized by the vevs of gauge invariant dressed monopoles $V_{n,P}$, subject to ``classical" ring relations \eqref{PotentialConstraints}, \eqref{MonopConstraint2} and from purely quantum relations. As we will see, the classical relations are responsible for the splitting of the space of vacua into several branches\footnote{This typically happens in $\N=4$ Yang-Mills theories with Coulomb, Higgs and mixed branches.}

It is important to note that the description of the space of vacua, or of its various branches, as a complex algebraic variety is not subject to quantum corrections. This follows from the fact that the Yang-Mills coupling constant $g_{YM}$ cannot appear in the complex ring relations, since in 3d it cannot be promoted to a chiral background field \cite{Bullimore:2015lsa}. Moreover, with $\N \ge 3$, there can be no dynamically generated superpotential which could affect the ring relations.
Being independent of the coupling constant, the algebraic description is also independent of the RG scale and thus describes equally well the space of vacua of the infrared theory, which is sometimes subject to dualities.  This is to be contrasted with the metric on the moduli space of vacua (or rather on a given branch), which can receive quantum corrections in $\N=3$ theories, although these are severely constrained by hyperk\"ahlerity (see for instance \cite{Jafferis:2008em}). 

The quantum relations, those involving monopole operators of non-zero charge, are not given by any standard method and it is the object of this paper to derive them from the brane approach of \cite{Assel:2017hck} in a selected class of theories. We will work our way through examples of increasing complexity, focusing on abelian theories.


\section{Brane setups and chiral operator insertions}
\label{sec:Branes}

The approach of \cite{Assel:2017hck} is based on the type IIB brane constructions realizing 3d $\N=4$ quiver theories and the insertion of chiral operators in the theory by the addition of extended branes and strings to the configurations. In this section we introduce the brane configurations realizing $\N=3$ theories, following in particular \cite{Kitao:1998mf} \footnote{See also \cite{1997NuPhB.500..133G} and \cite{Gauntlett:1997cv} for a review on intersecting brane configurations.}, and we find the additional branes that can be responsible for the insertion of chiral operators. The results that will be used in later sections are summarized in Table \ref{tab:BraneOrientationsFinal}. The reader already familiar with brane constructions and interested mostly in applications to gauge theory can safely go to the next section, trusting the results in Table \ref{tab:BraneOrientationsFinal}.

\begin{figure}[h!]
\centering
\includegraphics[scale=0.75]{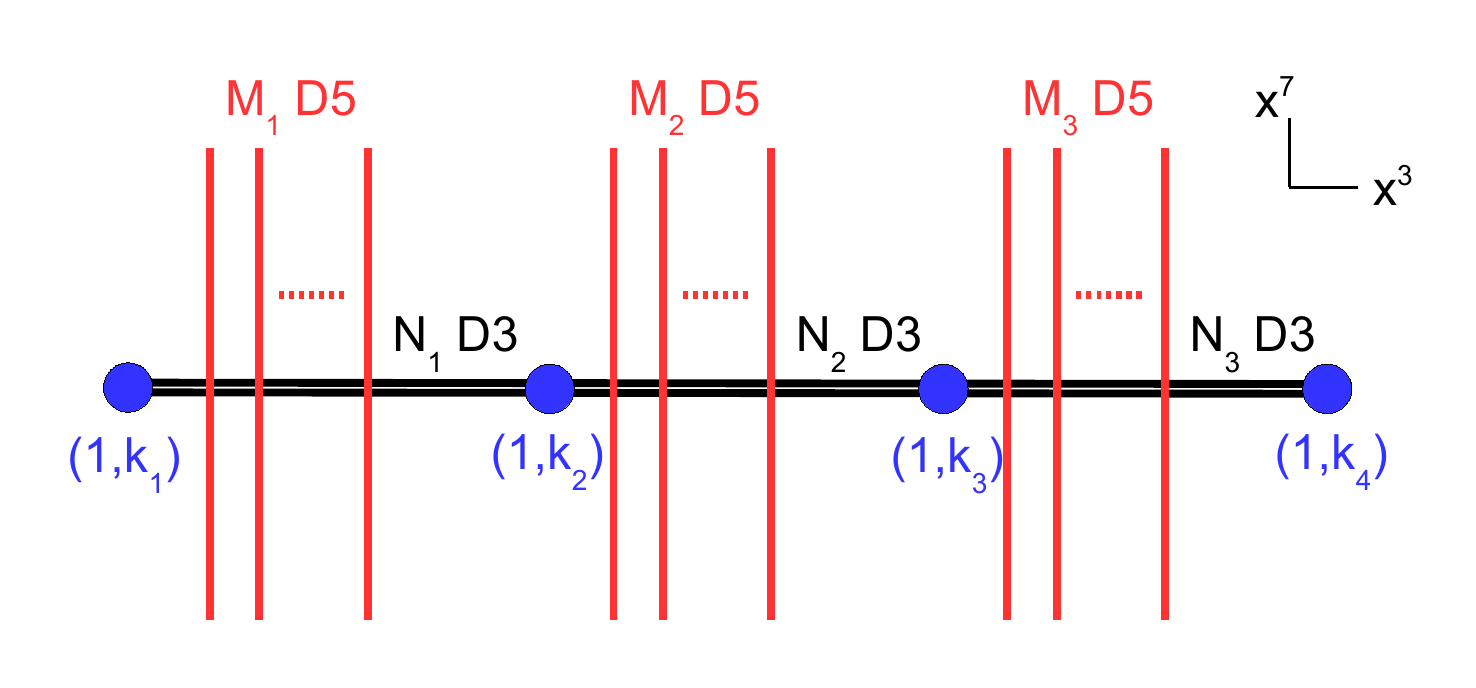} 
\vskip -0.5cm
\caption{\footnotesize Brane configuration realizing a general $\N=3$ linear quiver with three nodes, in the $(x^3,x^7)$ plane.}
\label{linquivbrane}
\end{figure}

Three-dimensional $\N=3$ theories can be realized from type II brane configurations involving D3 branes, NS5 branes and $(p,q)$5-branes, which are bound states of $p$ NS5 branes and $q$ D5 branes with $p,q$ coprime integers and which we denote $5_{(p,q)}$ branes. In order to preserve six supercharges the branes must be arranged with the orientations displayed in Table \ref{tab:BraneOrientations}, with the $5_{(p,q)}$ brane lying at an angle $\theta$ 
in the three planes $(x^4, x^7), (x^5,x^8)$ and $(x^6,x^9)$, namely spanning the lines
\be\ba 
& \underline{(p,q) \text{5-brane}:}  \quad
\sin \theta \, x^4 = \cos \theta \, x^7  \,, \quad \sin \theta \, x^5 = \cos \theta \, x^8  \,, \quad \sin \theta \, x^6 = \cos \theta \, x^9 = 0  \,. 
\ea\ee
When the type IIB axion background in vanishing $\chi=0$, the value of the angle $\theta$ must be tuned to $\tan\theta = \frac{q}{p}$ in order to preserve six supercharges. 
\begin{table}[h]
\begin{center}
\begin{tabular}{|c||c|c|c|c|c|c|c|c|c|c|}
  \hline
      & 0 & 1 & 2 & 3 & 4 & 5 & 6 & 7 & 8 & 9 \\  \hline
  D3  & X & X & X & X &   &   &   &   &   &   \\
  NS5 & X & X & X &   & X  & X  & X  &  &  &  \\
  $5_{(p,q)}$ & X & X & X &   & $\theta_{[4,7]}$  & $\theta_{[5,8]}$  & $\theta_{[6,9]}$  & $\theta_{[4,7]}$  & $\theta_{[5,8]}$  & $\theta_{[6,9]}$  \\ 
   \hline
\end{tabular}
\caption{\footnotesize Brane array realizing 3d $\N=3$ theories.}
\label{tab:BraneOrientations}
\end{center}
\end{table}
 \begin{figure}[h!]
\centering
\includegraphics[scale=0.8]{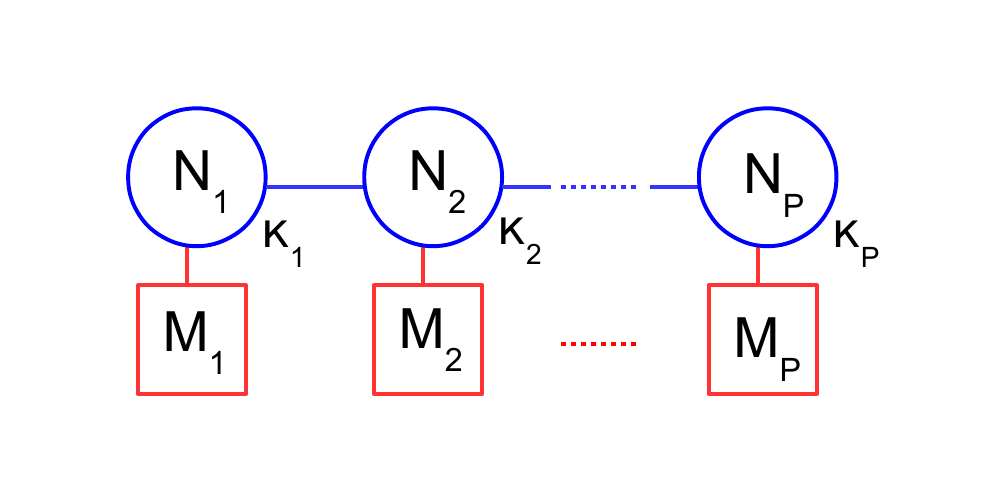} 
\vskip -0.5cm
\caption{\footnotesize $\N=3$  linear quiver diagram. The labels $\kappa_i$ indicate the CS levels.}
\label{linquiv}
\end{figure}
One can engineer $\N=3$ Chern-Simons linear quiver theories, described by the general quiver diagram of Figure \ref{linquiv}, by suspending D3 branes between NS5 branes and $5_{(1,k_i)}$ branes, and adding D5 branes (i.e. 5$_{(0,1)}$ branes) crossing the D3s, and circular quivers by further compactifying the $x^3$ direction and allowing D3 branes to wrap the $x^3$ circle. The gauge theory corresponds to the low-energy worldvolume theory on the D3 branes.
Figure \ref{linquivbrane} shows the brane configuration realizing a linear quiver theory with gauge group $\prod_{i=1}^P U(N_i)_{\kappa_i}$, Chern-Simons terms at level $\kappa_i$, and $M_i$ fundamental hypermulitplets in the $U(N_i)$ node, in the three-nodes case $P=3$. It involves $P+1$ 5$_{(1,k_i)}$ branes.
The $U(N_i)_{\kappa_i}$ gauge node lives on a stack of $N_i$ D3s stretched between the 5$_{(1,k_i)}$ brane and the 5$_{(1,k_{i+1})}$ brane, with 
\be
\kappa_i = k_{i+1}  - k_i \,, \quad i=1, \cdots, P \,.
\ee
There is a freedom to shift all $k_i$ by the same integer without affecting the low-energy theory on the D3 worldvolumes and we will always chose to fix $k_1=0$, namely that the first 5 brane is an NS5 brane.
The $M_i$ fundamental hypermultiplets are sourced by the $M_i$ D5 branes crossing the stack of $N_i$ D3s.

We are interested in adding elements to the brane construction which can insert local chiral operators in the 3d low-energy theory on the D3 branes. In \cite{Assel:2017hck} a similar problem was addressed where the objects inserting half-BPS local operators in 3d $\N=4$ theories associated to D3-D5-NS5 configurations were found to be F1, D1 and D3 branes with certain orientations. We expect in the D3-NS5-$5_{(p,q)}$ systems under consideration that the objects inserting local chiral operators are $(r,s)$1-strings and D3 branes with specific orientations preserving two supercharges. Here the $(r,s)$-strings,  that we will call $1_{(r,s)}$ strings, are bound states of $r$ D1 branes and $s$ F1 strings, with $r,s$ coprime integers.

In order to perform the supersymmetry analysis it is useful to revisit the derivation of \cite{Kitao:1998mf} where the brane system is uplifted to M-theory. The M-theory setup  involves M2 branes, M5 branes and M5' branes oriented as in Table \ref{tab:BraneOrientationsMth0}, where the M5' brane is oriented at angles $\theta, \varphi, \psi, \rho$ in the planes $(x^2,x^{10})$, $(x^4,x^7)$, $(x^5,x^8)$, $(x^6,x^9)$ respectively.
\begin{table}[h]
\begin{center}
\begin{tabular}{|c||c|c|c|c|c|c|c|c|c|c|c|}
  \hline
      & 0 & 1 & 2 & 3 & 4 & 5 & 6 & 7 & 8 & 9 & 10 \\  \hline
 M2  & X & X &  & X &   &   &   &   &   &  &  \\
M5 & X & X & X &   & X  & X  & X  &  &  &  & \\
  M5' & X & X & $\theta_{[2,10]}$  &   & $\varphi_{[4,7]}$  & $\psi_{[5,8]}$  & $\rho_{[6,9]}$  & $\varphi_{[4,7]}$  & $\psi_{[5,8]}$  & $\rho_{[6,9]}$ & $\theta_{[2,10]}$ \\ 
   \hline
\end{tabular}
\caption{\footnotesize M-theory brane array}
\label{tab:BraneOrientationsMth0}
\end{center}
\end{table}
The 11d supersymmetries are parametrized by a 32 components spinor $\varepsilon$. The M2, M5 and M5' branes impose the projections
\be\ba
&\underline{M2}: \quad  \epsilon = \Gamma^{013}\epsilon \,, \cr
&\underline{M5}: \quad  \epsilon = \Gamma^{012456}\epsilon \,,   \cr 
&\underline{M5'}: \quad  \epsilon = R\Gamma^{012456}R^{-1}\epsilon \,, 
\label{BasicSetupProj}
\ea\ee
where $\Gamma^{A_1A_2\cdots A_n} \equiv \Gamma^{[A_1}\Gamma^{A_2}\cdots\Gamma^{A_n]}$, with $\Gamma^A$ the 11d gamma matrices, and $R = \exp\big[ \frac{\theta}{2}\Gamma^{2,10} + \frac{\varphi}{2}\Gamma^{47} + \frac{\psi}{2}\Gamma^{58} + \frac{\rho}{2}\Gamma^{69}  \big]$. 
The analysis of this system of equations reveals that six supercharges can be preserved if the angles are tuned to satisfy 
\footnote{There are other (equivalent) choices preserving the same number of supercharges, e.g. $\varphi=\psi=-\rho = -\theta$.}
\be
\varphi=\psi=\rho = \theta \,,
\label{AnglesConstr}
\ee
for any angle $\theta$. We can call M5$_{\theta}$ the M5 brane at the angle $\theta$. Importantly the preserved supersymmetries $\varepsilon$ are independent of the actual value of $\theta$ so that a configuration with several M5$_{\theta_i}$ branes at different angles $\theta_i$ still preserves those six supersymmetries.

It should also be noted that eight supercharges can be preserved by further setting $\theta=\frac{\pi}{2}$, in which case the addition of the M5$_{\frac{\pi}{2}}$ brane to the M2-M5 system does not break additional supersymmetries.

\begin{figure}[h!]
\centering
\includegraphics[scale=0.8]{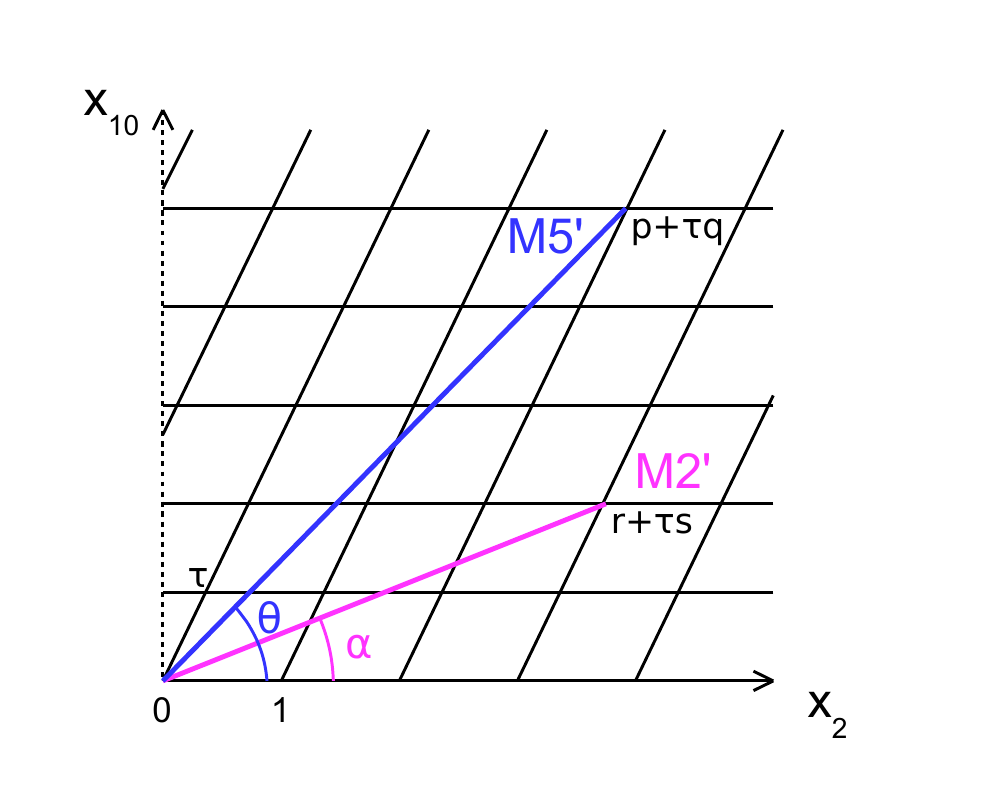} 
\vskip -0.5cm
\caption{\footnotesize Schematic view of the M5'$\cong$ M5$_\theta$ brane (blue) wrapping the $p \, \scC_x + q \, \scC_y$ cycle and M2'$\cong$ M2$_\alpha$ brane (purple) wrapping the $r \, \scC_x + s \, \scC_y$ cycle, in the $\tau$-torus. The branes intersect $|ps-qr|$ times.}
\label{TorusSpan}
\end{figure}

The M-theory setup is related to the type IIB setup by compactifying the $(x^2,x^{10})$ plane to a torus of modular complex parameter $\tau$, shrinking the torus to zero size, and T-dualizing along $x^2$. Under this duality the M-theory branes of Table \ref{tab:BraneOrientationsMth0} are mapped to the IIB branes of Table \ref{tab:BraneOrientations}: M2 branes become D3 branes, M5 branes become NS5 branes and M5$_\theta$ branes become $5_{(p,q)}$ branes, with the coprime integers $(p,q)$ being related to the angle $\theta$ and the modular parameter $\tau$. The relation comes from the fact that the M5$_{\theta}$ brane must wrap a circle in the $(x^2,x^{10})$ torus, implying that the angle $\theta$ is quantized, as depicted in Figure \ref{TorusSpan}. The complex coordinate on the $\tau$-torus is $z=x + \tau y$, with $x\sim x+1$, $y \sim y+1$. If we denote by $\scC_x$ and $\scC_y$ the cycles parametrized by $x$ and $\tau y$ respectively, we can wrap the M5$_\theta$ around a cycle $p \, \scC_x + q \, \scC_y$, with $(p,q)$ coprime integers. Under the string dualities it will become a $5_{(p,q)}$ brane. The angle $\theta$ is then given by
\be
\tan \theta = \frac{q \tau_2}{p + q\tau_1} \,,
\ee
with $\tau = \tau_1 + i \tau_2$. 
Under the duality, the modular parameter $\tau$ gets identified with the axio-dilaton background $\tau = \chi + i e^{-2\phi}$. Usually one considers the background $\tau = i$, which has vanishing axion $\chi=0$ (the choice is irrelevant to the low-energy 3d theory on the D3 branes). In this case the angle $\theta$ is given by
\be
\tan \theta = \frac{q}{p} \qquad (\tau=i)\,,
\label{ThetaQuantize}
\ee
and the special orthogonal orientation $\theta = \frac{\pi}{2}$ preserving eight supercharges, namely $\N=4$ in the 3d theory, corresponds to D5 branes ($(p,q)=(0,1)$).

As already noticed in \cite{Aharony:2008ug}, when the theory is engineered with only two types of 5 branes,  there is the possibility to choose the relative orthogonal orientation $\theta=\frac{\pi}{2}$ which preserves eight supercharges, by tuning $\tau$ to appropriate values. This choice of $\tau$ does not affect the deep infrared limit of the 3d theory on the D3 branes, therefore the 3d infrared fixed point must have enhanced $\N=4$ supersymmetry, even if we consider a brane configuration which preserves only $\N=3$. This is specific to theories engineered with only two types of 5 branes. When there are three or more types it is not possible to reach an configuration preserving eight supercharges and the infrared fixed point has a priori only $\N=3$ supersymmetry. 
 \medskip

To this setup we wish to add M2' branes preserving two supercharges, which become $1_{(r,s)}$ strings after dualizing to the IIB setup. The M2' branes must then be oriented as in Table \ref{tab:M2pM5ppOrientations} with the angles $\alpha,\beta$ tuned to preserve two supercharges. We chose the M2' to span a line in the $(x^4,x^7)$ plane but we could have chosen equivalently the M2' to span a line in the $(x^5,x^8)$ or $(x^6,x^9)$ plane.
\begin{table}[h]
\begin{center}
\begin{tabular}{|c||c|c|c|c|c|c|c|c|c|c|c|}
  \hline
      & 0 & 1 & 2 & 3 & 4 & 5 & 6 & 7 & 8 & 9 & 10 \\  \hline
  M2' &  &  & $\alpha_{[2,10]}$  &  X & $\beta_{[4,7]}$  &   &   & $\beta_{[4,7]}$  &  &  & $\alpha_{[2,10]}$ \\ 
  \hline
   M5'' &  &  & X  &   & X  & $\gamma_{[5,8]}$  & $\delta_{[6,9]}$  & X  &  $\gamma_{[5,8]}$  & $\delta_{[6,9]}$  & X  \\ 
   \hline
\end{tabular}
\caption{\footnotesize M2' brane and M5'' orientations}
\label{tab:M2pM5ppOrientations}
\end{center}
\end{table}
The extra projection imposed by the M2' brane is
\be
\underline{M2'}: \quad  \epsilon = i R'\Gamma^{234}R'{}^{-1}\epsilon \,, 
\label{M2pProj}
\ee
with $R'= \exp\big[ \frac{\alpha}{2}\Gamma^{2,10} + \frac{\beta}{2}\Gamma^{47}  \big]$ and the factor $i$ is due to the fact that the M2' is a Euclidean brane. 
The analysis of the system of equations \eqref{BasicSetupProj} and \eqref{M2pProj} with angles tuned according to \eqref{AnglesConstr} reveals that two supercharges can be preserved if the angles $\alpha,\beta$ satisfy\footnote{We used Mathematica to solve for this system with a given choice of gamma matrices. Some relations simplifying the analysis are $R'\Gamma^{234}R'{}^{-1}= R'{}^2\Gamma^{234}$, $R'{}^2 = [\cos\alpha + \sin\alpha \, \Gamma^{2,10}][\cos\beta + \sin\beta \, \Gamma^{47}]$ and similar relations involving the $R$ rotation matrix in the M5' projection \eqref{BasicSetupProj}.} 
\be
\sin(\alpha-\beta) = 0 \,.
\ee
Which two supercharges are preserved depends on whether $\alpha-\beta=0$ or $\alpha-\beta=\pi$ modulo $2\pi$ \footnote{The two corresponding brane configurations are related to each other by reversing the brane orientation.}. The two types of configurations are equivalent and we need only consider the situation
\be
\alpha = \beta \,.
\label{AlphaBeta}
\ee
We can denote these branes M2$_{\alpha}$. As we compactify the $(x^2,x^{10})$ plane to a $\tau$-torus the angle $\alpha$ gets quantized, as depicted in Figure \ref{TorusSpan}. The M2$_{\alpha}$ brane wrapping the $r \, \scC_x + s \, \scC_y$ cycle is dualized to a $1_{(r,s)}$ string in type IIB. With the choice $\tau=i$ we have
\be
\tan\alpha = \frac{s}{r} \qquad (\tau=i)\,.
\label{AlphaQuantize}
\ee
We notice that the M2$_{\alpha}$ and M5$_{\theta}$ branes with angles given by \eqref{AlphaQuantize}, \eqref{ThetaQuantize} intersect in $|ps-qr|$ points on the torus (unless they are parallel),
\be
\text{M2}_\alpha \cap \text{M5}_\theta = |ps-qr| \,, \quad (r,s) \neq \pm (p,q) \,.
\label{IntersectionNumber}
\ee 
This will be important in later analyses.
\smallskip

Finally we can add M5'' branes oriented as in Table \ref{tab:M2pM5ppOrientations}, with angles $\gamma,\delta$ to be fixed so that two supercharges are preserved. After dualizing to type IIB they become D3' branes and are also responsible for chiral operator insertions in the 3d theory.  The projection imposed by the M5'' is
\be
\underline{M5''}: \quad  \epsilon = i R''\Gamma^{24567,10}R''{}^{-1}\epsilon \,,
\ee
with $R'' = \exp\big[ \frac{\gamma}{2}\Gamma^{58} + \frac{\delta}{2}\Gamma^{69}  \big]$. The orientations preserving two supercharges have
\be
\sin(\gamma - \delta)=0 \,.
\ee
Again, which two supercharges are preserved depends on whether $\gamma - \delta=0$ or $\gamma - \delta=\pi$ modulo $2\pi$, and we will focus on the configurations with
\be
\delta = \gamma \,,
\label{GammaDelta}
\ee
that we denote M5$_{\gamma}$. This choice is correlated with the choice \eqref{AlphaBeta}, because, remarkably, the two supercharges preserved by the M5$_{\gamma}$ brane are the same as those preserved by the M2$_{\alpha}$ brane. In particular they are independent of the value of $\theta, \alpha, \gamma$.

\medskip

To summarize the brane configurations realizing chiral operators in the 3d low-energy theory are given by the M2-M5-M5$_{\theta}$ system with additional M2$_{\alpha}$ and/or M5$_{\gamma}$ with the orientations of Table \ref{tab:BraneOrientationsFinal}, where we also indicated the corresponding type IIB dual configurations.
\begin{table}[h]
\begin{center}
\begin{tabular}{|c||c|c|c|c|c|c|c|c|c|c|c|}
  \hline
      & 0 & 1 & 2 & 3 & 4 & 5 & 6 & 7 & 8 & 9 & 10 \\  \hline
 M2  & X & X &  & X &   &   &   &   &   &  &  \\
M5 & X & X & X &   &  X & X  & X  &  &  &  & \\
  M5$_{\theta}$ & X & X & $\theta_{[2,10]}$  &   & $\theta_{[4,7]}$  & $\theta_{[5,8]}$  & $\theta_{[6,9]}$  & $\theta_{[4,7]}$  & $\theta_{[5,8]}$  & $\theta_{[6,9]}$ & $\theta_{[2,10]}$ \\ 
  M2$_{\alpha}$ &  &  & $\alpha_{[2,10]}$  &  X & $\alpha_{[4,7]}$  &   &   & $\alpha_{[4,7]}$  &  &  & $\alpha_{[2,10]}$ \\ 
   M5$_{\gamma}$ &  &  & X  &   & X  & $\gamma_{[5,8]}$  & $\gamma_{[6,9]}$  & X  &  $\gamma_{[5,8]}$  & $\gamma_{[6,9]}$  & X \\
   \hline
   \hline
  D3  & X & X & X & X &   &   &   &   &   &  & -  \\
  NS5 & X & X & X &   & X  & X  & X  &  &  &  & - \\
  $5_{(p,q)}$ & X & X & X &   & $\theta_{[4,7]}$  & $\theta_{[5,8]}$  & $\theta_{[6,9]}$  & $\theta_{[4,7]}$  & $\theta_{[5,8]}$  & $\theta_{[6,9]}$ & -  \\ 
$1_{(r,s)}$ &  &  &  &  X & $\alpha_{[4,7]}$  &   &   & $\alpha_{[4,7]}$  &  &  & -  \\ 
  D3$_{\gamma}$ &  &  &   &   & X  & $\gamma_{[5,8]}$  & $\gamma_{[6,9]}$  & X  &  $\gamma_{[5,8]}$  & $\gamma_{[6,9]}$  & - \\
   \hline
\end{tabular}
\caption{\footnotesize M-theory brane array and dual IIB brane array realizing chiral operator insertions in 3d $\N=3$ theories. Here $\tan\theta = q/p$, $\tan\alpha=s/r$ and $\gamma$ is arbitrary.}
\label{tab:BraneOrientationsFinal}
\end{center}
\end{table}
It is a rather unexpected result that the D3$_{\gamma}$ brane in the type IIB configuration can be at an arbitrary real angle $\gamma$ and still preserve the same two supercharges. These configurations will be related to the insertion of a linear combination of two chiral operators, however they will not play an central role. We will also use the name D3$_{(p,q)}$ to denote a D3$_{\gamma}$ brane with the angle $\gamma$ tuned so that it is in a Hanany-Witten orientation with the 5$_{(p,q)}$ brane, i.e. $\gamma = \frac{\pi}{2} - \frac{q}{p}$. This means that if a D3$_{(p,q)}$ crosses a 5$_{(p,q)}$ brane when moving along $x^3$ a 1$_{(p,q)}$ string is created, stretched between them. This follows from the Hanany-Witten D3 creation effect \cite{Hanany:1996ie} and S-duality in IIB string theory.

For the special configurations with orthogonal orientation $\theta = \frac{\pi}{2}$ preserving eight supercharges, we find that the addition of M2$_{\alpha}$ branes  and M5$_{\gamma}$ branes at $(\alpha,\gamma) \in \{(0, \frac{\pi}{2}), (\frac{\pi}{2},0) \}$ preserve four supercharges. With the choice of the square torus $\tau=i$, these configurations reduce to those studied in \cite{Assel:2017hck} with F1 strings, D1 branes and D3 branes inserting half-BPS operators.

Finally we notice that equivalent brane setups are obtained by $SO(3)$ rotations acting simultaneously on the $x^{4,5,6}$ and $x^{7,8,9}$ spaces. This reflects the choice of a chiral subalgebra in the $\N=3$ theory or equivalently a choice of $U(1)_R$ subgroup of the $SU(2)_R$ R-symmetry.


\section{Analysis in simple theories}
\label{sec:SimpleCases}

In this section the moduli space of vacua of some simple abelian $\N=4$ and $\N=3$ theories is obtained from the analysis of the type IIB brane setups related to chiral operators insertions and chiral ring relations.

\subsection{$U(1)_\kappa$ theory}
\label{ssec:U1k}

The simplest theory is the abelian $\N=3$ Chern-Simons theory at level $\kappa\neq 0$ without matter hypermultiplets. In the infrared limit, after integrating out the massive fermions and scalars, the theory becomes the pure abelian Chern-Simons theory at level $\kappa$ which has no propagating degrees of freedom. The UV theory has a single chiral operator $\varphi$, the complex scalar in the vector multiplet. There is no monopole operator since the bare monopole of magnetic charge $n\neq0$ has a gauge charge $-n\kappa$ which cannot be canceled by a matter dressing. In addition the scalar $\varphi$ is set to zero by the constraints \eqref{PotentialConstraints} with no matter, describing the unique vacuum of the theory.
 \begin{figure}[h!]
\centering
\includegraphics[scale=0.75]{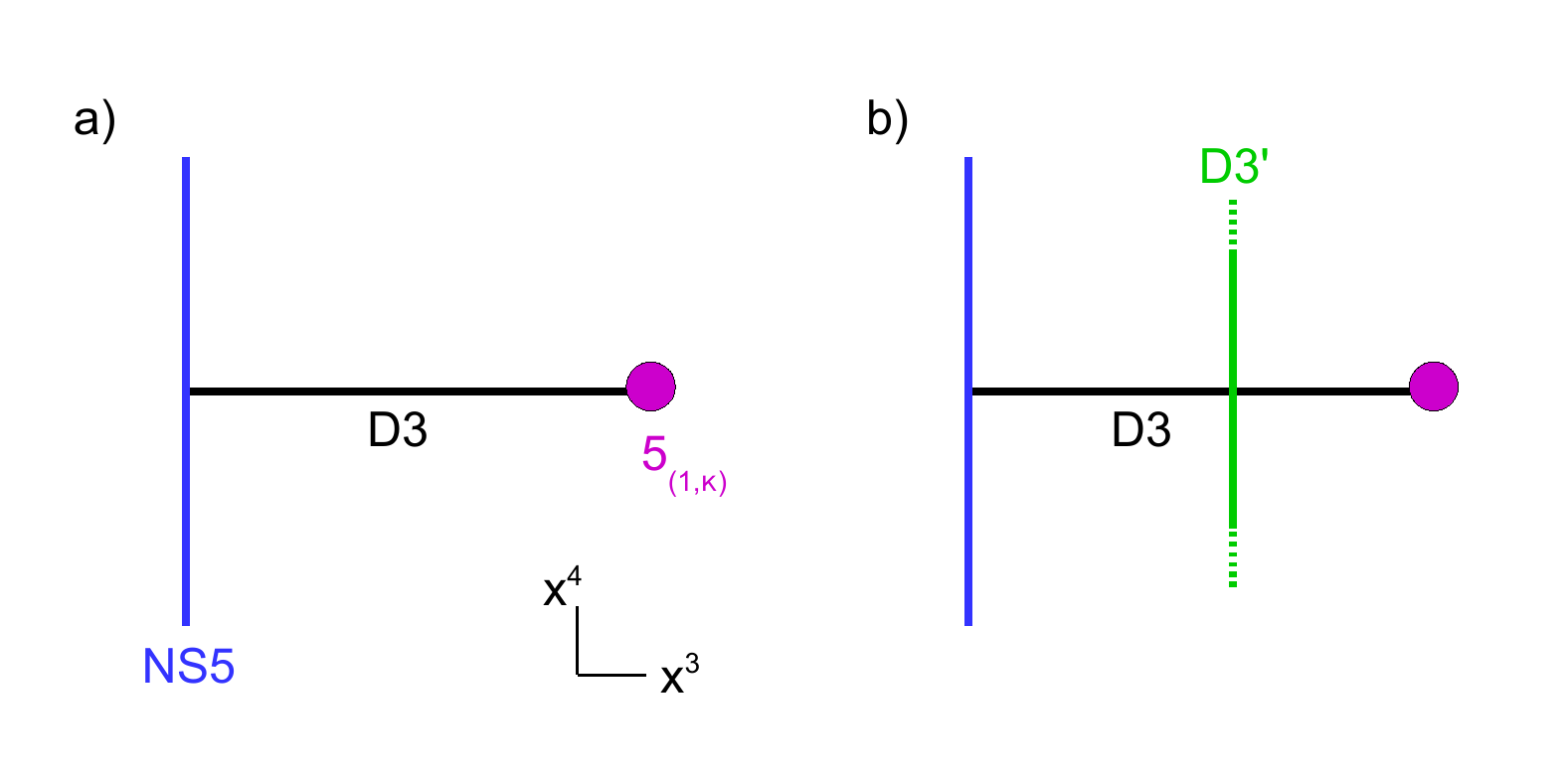} 
\vskip -0.5cm
\caption{\footnotesize a) Brane configuration realizing the $U(1)_\kappa$ theory, in the $(x^3,x^4)$ plane. b) The addition of a D3$_{\gamma}$ (= D3') brane inserts the chiral operator $\varphi$ vanishing in the vacuum.}
\label{U1kap}
\end{figure}

The brane configuration realizing the theory has a single D3 brane stretched between an NS5 and a $5_{(1,\kappa)}$, as in Figure \ref{U1kap}-a. We take the 5-branes to sit at the origin in all directions transverse to their worldvolume. The vev of the vector multiplet scalar $\varphi$ is associated with the position of the D3 segment along the $x^{5+i6}$ directions (this choice of transverse plane is correlated to a choice of complex structure on the moduli space of vacua). Since the D3 segment in this particular brane setup is fixed to the origin of the $x^{5+i6}$ plane by the fact that it is attached to the two 5 branes, the scalar $\varphi$ is set to zero in vacuum configurations, as indicated by the field theory F-term constraints.

Chiral operator insertions correspond to adding $1_{(r,s)}$ string or D3$_{\gamma}$ branes to the configuration. Since the two 5-branes have different orientations,  no $1_{(r,s)}$ string can be stretched between them. This leaves the possibility of adding a D3$_{\gamma}$ brane crossing the D3 segment, as in Figure \ref{U1kap}-b. 
The insertion due to the addition of the D3$_{\gamma}$ brane can be understood as arising from the light D3-D3$_{\gamma}$ open string modes. The specific orientations between the D3 and D3$_{\gamma}$ branes are such that they intersect at a point and the lowest open string mode is a single zero-dimensional fermion $\chi$, whose complex mass is the distance $\varphi$ between the two D3s along $x^{5+i6}$. Integrating out the fermion leads to the insertion of $\varphi$ in the path integral of the theory. Since the D3 segment is stuck at the origin in the $x^{5+i6}$ plane, the D3$_{\gamma}$ crosses exactly the D3 and the fermion $\chi$ is massless $\varphi=0$.

\medskip

It is also interesting to consider the $U(1)_\kappa$ theory with $N_f$ fundamental hypermultiplets $(Q^\alpha, \ti Q_\alpha)$, $\alpha=1, \cdots, N_f$. 
The chiral operators comprise, in addition to $\varphi$, the meson operators $Z^\alpha{}_\beta = Q^\alpha \ti Q_\beta$. There are still no monopole operators because of the constraint \eqref{MonopConstraint1} related to the fact that they cannot be dressed with matter fields charged under the $U(1)$ gauge group.

The vacuum equations \eqref{PotentialConstraints} are 
\be\ba
& \varphi Q^\alpha = 0 \,, \quad \ti Q_{\alpha}\varphi = 0 \,, \quad \sum_\alpha Q^\alpha \ti Q_\alpha = \kappa \varphi \,.
\ea\ee
From these equations we deduce 
\be 
\varphi=0 \,, \quad  \tr Z \equiv Q^\alpha \ti Q_\alpha = 0 \,,
\ee
 In addition there are relations $Z^\alpha{}_{\beta} Z^{\gamma}_{\delta} = Z^\alpha{}_{\beta} Z^{\gamma}_{\delta}$ following from the definition of $Z^\alpha{}_\beta$. 
 The space of vacua is then given by a Higgs branch
\be
\tr Z = 0 \,, \quad \text{rank}(Z) \le 1 \,,
\ee
with $\varphi$ vanishing. This identifies the Higgs branch with the closure of the $SL(N_f,\bC)$ nilpotent orbit $\scO_{(2,1^{N_f-2})}$ \cite{Gaiotto:2008ak}.  This is the identical to the Higgs branch of the $\N=4$ SQED theory with $N_f$ flavor hypermultiplets, which can be obtained by setting $\kappa=0$. 

\begin{figure}[h!]
\centering
\includegraphics[scale=0.75]{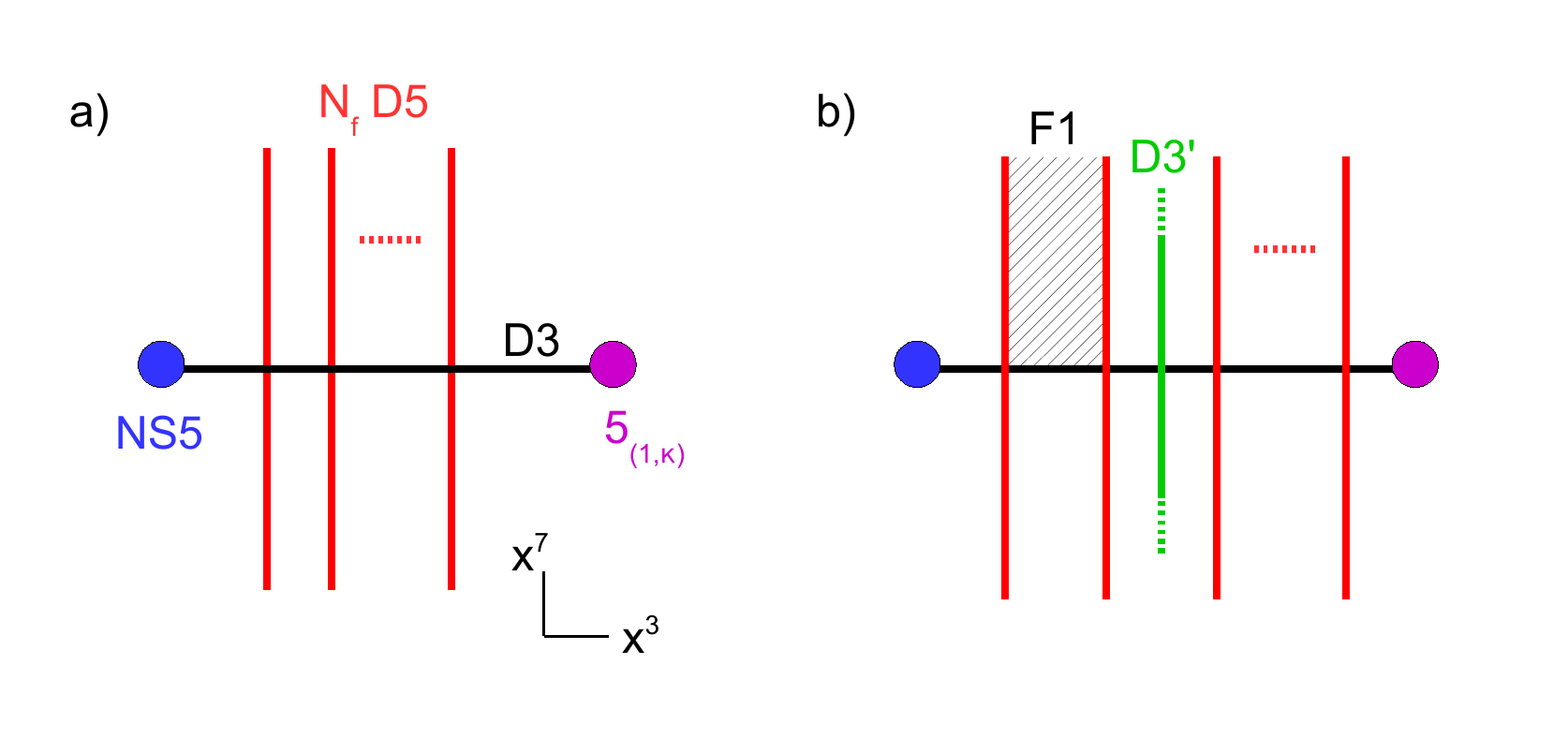} 
\vskip -0.5cm
\caption{\footnotesize a) Brane configuration realizing the $U(1)_\kappa$ with $N_f$ flavor hypermultiplets, in the $(x^3,x^7)$ plane. b) The addition of F1 strings stretched between two D5s and/or D3' branes crossing the D3 segment insert meson operators.}
\label{U1kapMatter}
\end{figure}

The brane configuration realizing the theory now has $N_f$ extra D5 branes oriented along $x^{012789}$, corresponding to 5$_{(0,1)}$ branes in Table \ref{tab:BraneOrientationsFinal}, as in Figure \ref{U1kapMatter}-a.
To realize chiral operator insertions we can add strings stretched between the D5 branes and ending on the D3 segment, which have to be F1 strings ($\alpha=\pi/2$) according to the orientations of Table \ref{tab:BraneOrientationsFinal}, and/or D3' branes, as in Figure \ref{U1kapMatter}-b. These configurations have been studied in detail in \cite{Assel:2017hck} (see Section 5.1.1 of that paper) in the context of $\N=4$ SQED, so we will not repeat the analysis. 
The short summary is that the F1 strings insert operators $Z^\alpha{}_{\beta}$ with $\alpha < \beta$ or $\alpha > \beta$ depending on whether they end on the D3 from above or from below. The D3'$\cong$ D3$_{\gamma}$ brane oriented along $x^{5,6}$ with $\gamma=0$  is responsible for the insertion of specific linear combinations of the remaining meson operators $Z^\alpha{}_\alpha$. The F-term constraint $\tr Z=0$ can be understood from D3' brane moves. The constraints $Z^\alpha{}_{\beta} Z^{\gamma}_{\delta} = Z^\alpha{}_{\beta} Z^{\gamma}_{\delta}$ follow from considering brane setups with a full F1 string stretched between two D5s and crossing the D3 segment, and reading the operator insertion in two different ways. We will describe the Higgs branch analysis in more details in other theories. 

When the D3$_{\gamma}$ brane is along $x^{89}$, namely when $\gamma=\pi/2$, the distance between the D3$_\gamma$ and the D3 segments is vanishing (those can move only along the D5 directions $x^{789}$ spanned by the D3$_{\frac{\pi}{2}}$). Integrating out the massless fermion living at the pointlike intersection of the D3s yields a zero that is the insertion of the vanishing scalar $\varphi$, consistently with the analysis in \cite{Assel:2017hck}. Other orientations of the D3$_{\gamma}$ do not yield independent chiral operator insertions.

There is no other string or brane that we can add for the insertion of other chiral operators, in agreement with the fact that the theory has only a Higgs branch.

\subsection{Chern-Simons Duals of $T[SU(2)]$}
\label{ssec:CSTSU2}

\begin{figure}[h!]
\centering
\includegraphics[scale=0.75]{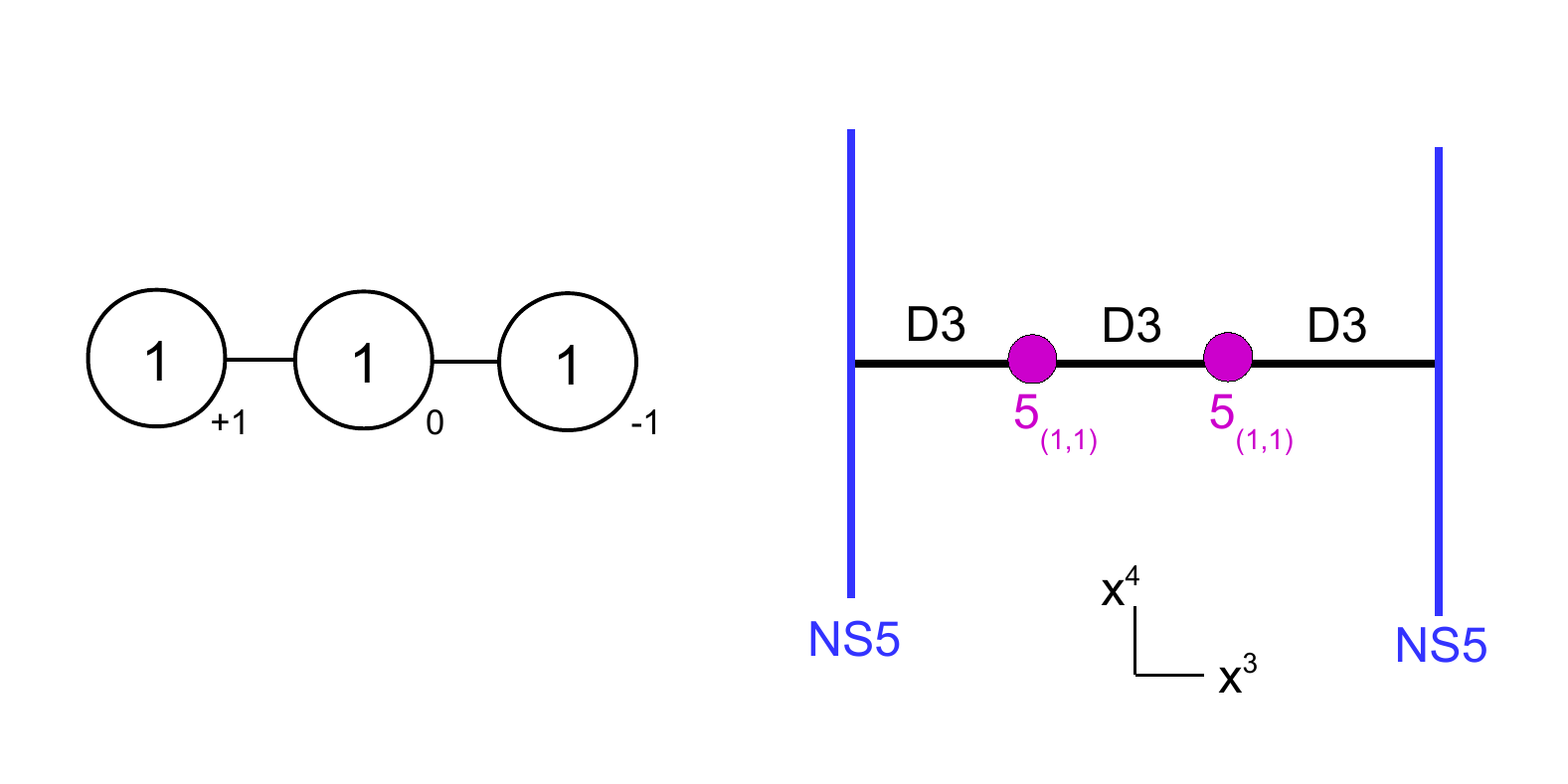} 
\vskip -0.5cm
\caption{\footnotesize Quiver diagram of the $T_A$ theory (subscripts indicate Chern-Simons levels) and associated brane configuration.}
\label{TA}
\end{figure}

We now move to more interesting situations. We consider first the abelian quiver of Figure \ref{TA}, which has gauge group $U(1)_{1}\times U(1)_0 \times U(1)_{-1}$, with indices indicating Chern-Simons levels ($0$ means pure Yang-Mills), and only bifundamental matter. Let us call this theory $T_A$. Its brane configuration has two NS5 branes and two 5$_{(1,1)}$ branes, ordered as shown in Figure \ref{TA}, with D3 segments stretched between them. Because the brane configuration has only two types of 5 branes we know that the $\N=3$ theory flows in the IR to an $\N=4$ fixed point (see Section \ref{sec:Branes}). 
Actually this theory is related by $\N=4$ mirror symmetry to the $U(1)$ SQED theory with two flavor hypermultiplets  \cite{Assel:2014awa,Cremonesi:2016nbo,Jafferis:2008em}, also called $T[SU(2)]$ theory. This can be seen from the brane realization by acting with the $SL(2,\bZ)$ duality in IIB string theory which leaves invariant the NS5 branes and transform 5$_{(1,1)}$ branes into D5 branes. This transformation is $-STS$ in the notation of \cite{Assel:2014awa}.\footnote{The $T$ action transforms (NS5,D5) into (5$_{(1,1)}$,D5) and the $S$ action transforms (NS5,D5) into (D5, -NS5). } This is an extension of the more familiar mirror symmetry \cite{Intriligator:1996ex} which is identified with the $S$ action in IIB string theory and trades D5s and NS5s. The action of $S$ and $T$ generates a larger orbit of equivalent brane configurations, whose low-energy 3d theories may or may not have  Lagrangian descriptions (see \cite{Gaiotto:2008ak,Assel:2014awa}).

The $T_A$ and $T[SU(2)]$ theories are dual in the sense that they flow to the same infrared theory at corresponding points on their vacuum space, and in particular they flow to the same fixed point at the origin of their vacuum space. In particular the theories must have isomorphic spaces of vacua described as algebraic varieties, since this is not affected by the RG flow. The space of vacua of the $T[SU(2)]$ theory has a Coulomb branch and a Higgs branch (exchanged under $S$-mirror symmetry), each of which being parametrized by three complex coordinates $x,y,z$ subject to the relation $xy=z^2$. This has been studied in detail in \cite{Assel:2017hck}. Mirror symmetry thus predicts that the $T_A$ theory also has two such branches.

The chiral operators of $T_A$ contain the three vector multiplet scalars $\varphi_i$, $i=1,2,3$, for the three nodes, and the two mesons $X_j = q_j \ti q_j$, $j=1,2$, with $(q_j,\ti q_j)$ denoting the two bifundamental scalars.
In addition there are bare monopole operators $\ol V_{n_1n_2 n_3}$ with magnetic charges $n_1,n_2$ and $n_3$ for the $U(1)_1$, $U(1)_0$ and $U(1)_{-1}$ gauge nodes respectively. According to \eqref{MonopElCharge} the bare monopole $\ol V_{n_1 n_2 n_3}$ has electric charge $(-n_1,0,n_3)$ and dimension $\frac 12 (|n_1-n_2|+|n_2-n_3|)$ according to \eqref{MonopDim}. The gauge invariant monopoles are therefore
\be
V_{n_1n_2} = \left\lbrace
\begin{array}{c}
\ol V_{n_1n_2n_1} (q_1)^{n_1} (q_2)^{n_1} \,, \quad n_1 \ge 0 \,, \cr
\ol V_{n_1n_2n_1} (\ti q_1)^{-n_1} (\ti q_2)^{-n_1} \,, \quad n_1 < 0 \,,
\end{array} \right.
\ee
with $(n_1, n_2) \in \bZ^2-\{ (0,0)\}$. $V_{n_1n_2}$ has dimension $|n_1-n_2| + |n_1|$. The monopoles $V_{n_1n_2}$ satisfy the constraint \eqref{MonopConstraint1} $n_1 - n_1=0$.

 It happens in 3d $\N=4$ Yang-Mills theories that the Coulomb branch is generated by a subset of the monopole operators, which have small dimensions \cite{Bullimore:2015lsa}. The other monopoles can be expressed as polynomials in the former monopoles, due to quantum chiral ring relations. We propose that a basis of monopole operators is given by the operators with smallest dimensions whose magnetic charges form a basis of the lattice of allowed monopole magnetic charges, in all $\N=3$ theories. We will verify this conjecture in examples, worked out from the brane analysis. We will refer to this basis of monopoles simply as the monopoles of a given theory, ignoring monopole operators of higher magnetic charges. We will comment on those redundant operators in Section \ref{ssec:HigherChargeMonop}.
 
 The monopoles in $T_A$ are thus $u^{\pm} = V_{0 \pm}$ and $v^{\pm}= V_{\pm \pm}$, with dimension one, the basis of chiral operators is $\varphi_i, X_j, u^\pm,v^\pm$. The vacuum equations \eqref{PotentialConstraints}, \eqref{MonopConstraint2} read
 \be\ba
& (\varphi_1-\varphi_2) q_1 = (\varphi_1-\varphi_2) \ti q_1 = (\varphi_2-\varphi_3) q_2 = (\varphi_2-\varphi_3) \ti q_2 = 0 \,,  \cr
&  q_1 \ti q_1 - \varphi_1 = -q_1 \ti q_1 + q_2 \ti q_2 = -q_2 \ti q_2 + \varphi_3 = 0 \,, \cr
& u^{\pm} q_j = u^{\pm} \ti q_j = 0 \,, \quad j=1,2\,.
\ea\ee
Note that the equations \eqref{MonopConstraint2} do not imply constraints involving $v^\pm$. 
In terms of gauge invariant operators we obtain
 \be\ba
& (\varphi_1-\varphi_2) X_1 = (\varphi_1-\varphi_2) v^{\pm}  = (\varphi_2-\varphi_3) X_2 = (\varphi_2-\varphi_3) v^\pm = 0 \,,  \cr
& X_1 - \varphi_1 = X_2 - X_1 = -X_2 + \varphi_3 = 0 \,, \cr
& u^{\pm} X_1  = 0 \,, \quad u^{\pm} X_2  = 0 \,, \quad u^{\pm} v^{\pm} = 0 \,.
\label{TAvac}
\ea\ee
The equations on second line allows us to fix  $\varphi_1= \varphi_3 = X_1 = X_2 \equiv X$.
We find two branches of vacua parametrized by free moduli
\be\ba
& \text{branch I:} \quad  v^\pm = X = 0 \,, \quad (u^{\pm}, \varphi_2) \ \text{free} \,, \cr
& \text{branch II:} \quad  u^\pm = 0 , \,  \varphi_2 = X \,, \quad (v^{\pm}, X ) \ \text{free} \,.
\ea\ee
On each branch the gauge group is broken to a $U(1)$ subgroup, $U(1) = U(1)_0$ for the branch I and $U(1) = $ diag$(U(1)_{+1}\times U(1)_0\times U(1)_{-1})$ for the branch II, therefore they are two Coulomb-like branches.
To match the two branches of the $T[SU(2)]$ theory this naive picture must be corrected by the addition of quantum relations on each branch, of the form
\be\ba
& \text{branch I:} \quad  u^+ u^- \sim (\varphi_2)^2 \,, \cr
& \text{branch II:} \quad  v^+ v^- \sim X^2  \,.
\label{TArelations}
\ea\ee
This is very reminiscent of what happens for the Coulomb branch of the $T[SU(2)]$ theory and it is possible that these relations can be obtained by adapting the CFT computation of \cite{Borokhov:2002cg}. 
Instead we will derive these relations from the brane analysis. 

\medskip

In the brane configuration with orientations as in Table \ref{tab:BraneOrientationsFinal} the vevs $\varphi_i$ of the three vector multiplet complex scalars correspond to the positions of the three D3$_i$ segments in the complex plane $x^{5+i6}$ and the vevs of the two bifundamental scalars $X_j$ correspond to the distance in the complex plane $x^{8+i9}$ between the 5$_{(1,1)j}$ brane and the reconnected D3 segments across the 5$_{(1,1)j}$ \cite{Hanany:1996ie,Gaiotto:2008sa}. When the D3 segments break (i.e. do not reconnect) on the 5$_{(1,1)j}$ brane the vev $X_j$ is vanishing. Here $i$ and $j$ label the D3 segments and the 5$_{(1,1)}$ branes respectively, from left to right in the brane configuration.
From the brane configuration of the $T_A$ theory (Figure \ref{TA}) we observe two possible motions of the D3 segments corresponding to the two branches of vacua: the motion of the D3$_2$ segment along the 5$_{(1,1)}$s directions, for branch I,  and the motion of the three reconnected D3 segments along the NS5s directions, for branch II. According to the discussion above we find the constraints $\varphi_1 = \varphi_3 = X_1 = X_2 = 0$ on the branch I and $\varphi_1 = \varphi_3 = X_1 = X_2 = \varphi_2$ on the branch II, matching the field theory description. 
To derive the constraints on the branch II we have used the fact that the reconnected D3 segment moves along the 5$_{(1,1)}$ branes, which are oriented at the angle $\theta = \pi/4$ in the $x^{58}$ and $x^{69}$ planes. 
These initial observations tell us about the branch splittings. They must be completed by studying the brane setups inserting chiral operators, in particular monopole operators.

\medskip

\begin{figure}[h!]
\centering
\includegraphics[scale=0.75]{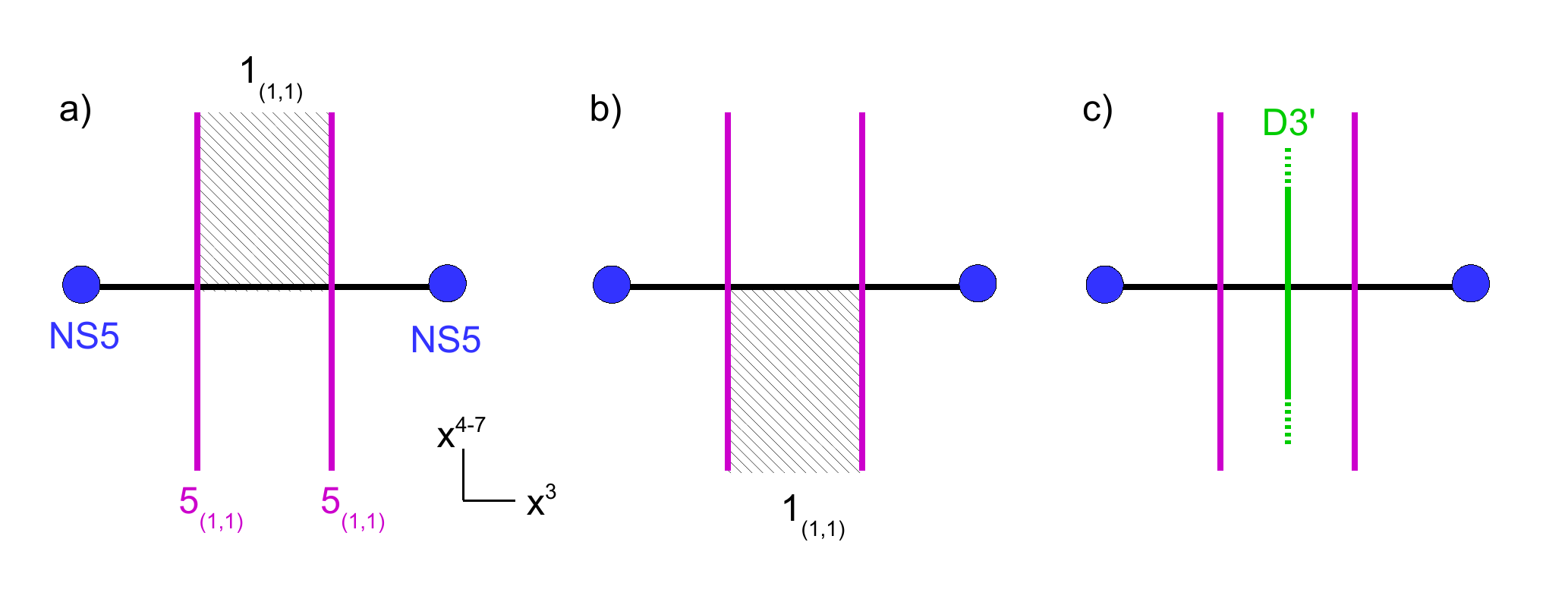} 
\vskip -0.5cm
\caption{\footnotesize Brane setups inserting: a) $u^+$, b) $u^-$,c) $\varphi_2$. The vertical axis corresponds to the direction spanned by the 5$_{(1,1)}$ brane in the $x^{47}$ plane.}
\label{TAOpI}
\end{figure}

The chiral operators are inserted in the theory by adding 1$_{(r,s)}$ strings and D3$_{\gamma}$ branes oriented as in Table \ref{tab:BraneOrientationsFinal} to the brane configuration of Figure \ref{TA}. 
We propose brane setups inserting the chiral operators on the two branches in close analogy with \cite{Assel:2017hck}.
For the branch I operators, we have:
\begin{itemize}
\item The monopole operators $u^+$ and $u^-$ are inserted by adding a 1$_{(1,1)}$ string stretched between the two 5$_{(1,1)}$s, ending on the middle D3 segment from above and from below respectively, as in Figure  \ref{TAOpI}-a,b.
\item The scalar operator $\varphi_2$ is inserted by adding a D3$_{(1,1)}$ brane between the two 5$_{(1,1)}$ branes, as in Figure \ref{TAOpI}-c.
\end{itemize}
We remind that a D3$_{(p,q)}$ brane is a D3$_{\gamma}$ brane with the angle $\gamma$ tuned so that it is in a Hanany-Witten orientation with the 5$_{(p,q)}$ brane, i.e. $\gamma = \frac{\pi}{2} - \frac{q}{p}$.

The justifications for these insertions are essentially the same as for the Coulomb branch operators in \cite{Assel:2017hck}. 
The semi-infinite 1$_{(1,1)}$ string ending on the D3 from above inserts a line defect carrying one unit of magnetic charge and one unit of electric charge in the D3 worldvolume SYM theory, namely it inserts a Wilson-'t Hooft loop or (1,1)-dyonic loop \cite{Kapustin:2005py} in the 4d $\N=4$ SYM theory defined on an interval along $x^3$. This line defect is extended along the interval between the two 5$_{(1,1)}$ branes and the boundary conditions at the ends of this interval \cite{Gaiotto:2008sd} are compatible with the line defect.
In the low-energy limit the theory becomes effectively three-dimensional and the insertion of the BPS dyonic loop due to the 1$_{(1,1)}$ string becomes the insertion of the local chiral monopole operator of magnetic charge one in the $U(1)_0$ node $u^+$.\footnote{The dyonic loop in 4d can be described as a half-BPS 't Hooft loop dressed with a half-BPS Wilson loop factor. In the 3d limit the Wilson loop yields, after a possible regularization, an overall factor multiplying the $u^+$ monopole insertion. We do not attempt to compute this factor, since it is not relevant to the 3d discussion.}  

Similarly the addition of the semi 1$_{(1,1)}$ string stretched between the 5$_{(1,1)}$s and ending on the middle D3 segment from below (Figure \ref{TAOpI}-b) inserts the monopole of negative charge $u^-$ in the low-energy $T_A$ theory.

The presence of the D3$_{(1,1)}$ brane in Figure \ref{TAOpI}-c introduces a fermionic mode $\chi$ living at the point-like intersection of the D3 segment and the D3$_{(1,1)}$\footnote{This follows from the orientations of the two D3s, which have eight Neumann-Dirichlet directions.}, with a complex mass $m_{\chi} = \Delta_{D3-D3_{(1,1)}} \propto \varphi_2$ given by the distance between the D3 and D3$_{(1,1)}$ in the $x^{5-8}+i x^{6-9}$ complex plane  spanned by the 5$_{(1,1)}$s.
Its zero-dimensional action is\footnote{The fermion action with a complex mass is not real. It can be thought as the dimensional reduction of a 2d Fermi multiplet theory coupled to a background vector field, Wick rotated to Euclidean signature.}
\be
S_{\rm 0d} \sim  \bar\chi\varphi_2\chi \,.
\ee
Integrating out the zero-dimensional fermion yields the insertion of the scalar operator $\varphi_2$, as proposed.

\medskip

\begin{figure}[h!]
\centering
\includegraphics[scale=0.75]{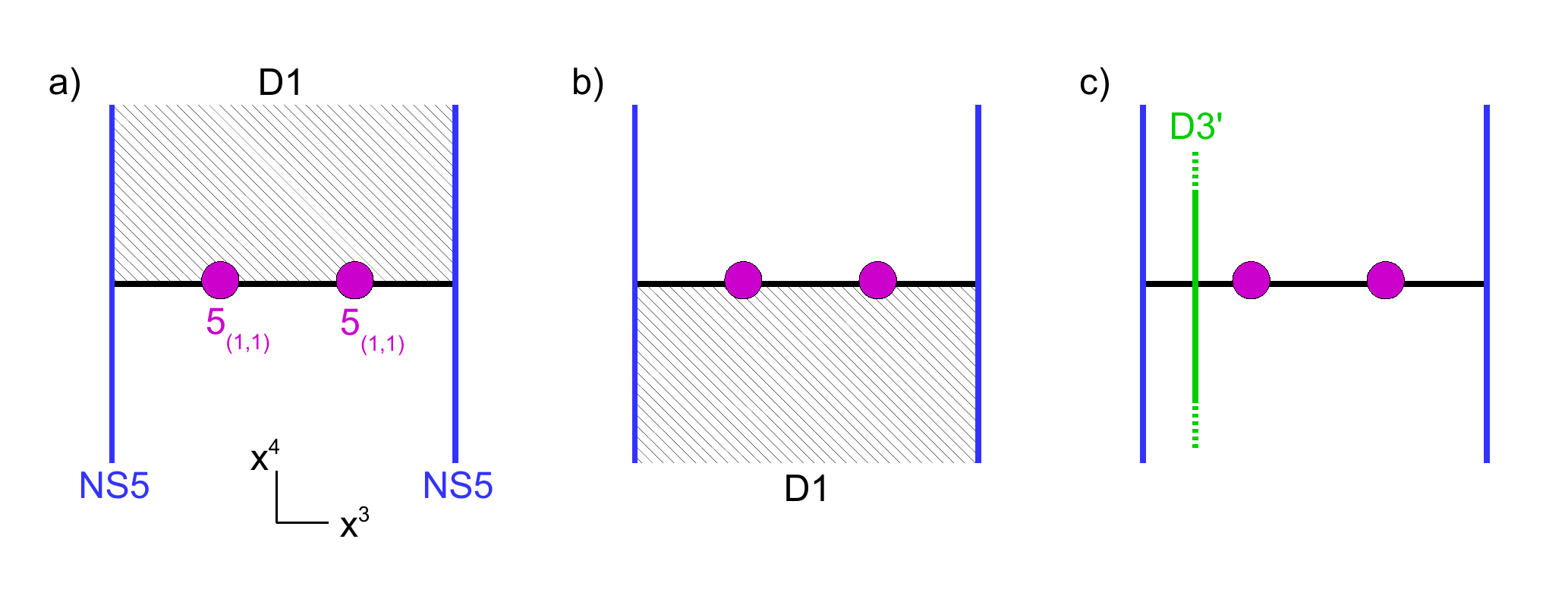} 
\vskip -0.5cm
\caption{\footnotesize Brane setups (in the $x^{3,4}$ plane) inserting: a) $v^+$, b) $v^-$,c) $\varphi_1 \equiv X$.}
\label{TAOpII}
\end{figure}

For the branch $II$ operators we propose:
\begin{itemize}
\item The monopole operators $v^+$ and $v^-$ are inserted by adding a D1 string stretched between the two NS5 branes, ending on the D3 segments from above and from below respectively, as in Figure \ref{TAOpII}-a,b.
\item The scalar operator $X$ is inserted by adding a D3$_{(1,0)}$ brane between the two NS5 branes (at any $x^3$ position between them), as in Figure \ref{TAOpII}-c.
\end{itemize}
The additional strings and branes have the orientation of Table \ref{tab:BraneOrientationsFinal} associated to the insertion of local chiral operators in the 3d theory. The justifications are analogous to those of the branch I operators.

The presence of the semi D1 string stretched between the NS5s and ending on the D3 segments (Figure \ref{TAOpII}-a) inserts a 't Hooft loop of magnetic charge one in each of the three 4d $\N=4$ SYM $U(1)$ theory defined on the three adjacent intervals. In the low energy limit this becomes the insertion of a monopole singularity of magnetic charge one in each of the $U(1)$ nodes, corresponding to the bare monopole operator $\ol V_{+++}$. In addition there are contributions associated to the D1 crossing the 5$_{(1,1)}$ branes.
It is generally hard to study the low-energy physics of brane systems involving 5$_{(p,q)}$ branes with $p\neq 0$. We postulate that the D3-5$_{(1,1)i}$-D1 intersection with a semi-D1 string crossing the 5$_{(1,1)i}$ and ending on the D3 from above is responsible for the insertion of a bifundamental scalar factor $q_i$, for $i=1,2$.
Later we will find generelization with general 1$_{(r,s)}$ strings and 5$_{(p,q)}$ branes crossings.
The total insertion is then $\ol V_{+++} q_1 q_2 = v^+$, as proposed.

Similarly the presence of the semi-D1 stretched between the two NS5s, ending on the D3 segments from below (Figure \ref{TAOpII}-b) inserts the bare monopole of charge $-1$ in each of the three nodes $\ol V_{---}$, together with bifundamental scalars coming from the D3-5$_{(1,1)}$-D1 crossings. We propose that the D3-5$_{(1,1)i}$-D1 intersection with a semi-D1 string crossing the 5$_{(1,1)i}$ and ending on the D3 from below is responsible for the insertion of a bifundamental scalar factor $\ti q_i$, for $i=1,2$. The total insertion is $\ol V_{---} \ti q_1 \ti q_2 = v^-$.

\begin{figure}[h!]
\centering
\includegraphics[scale=0.75]{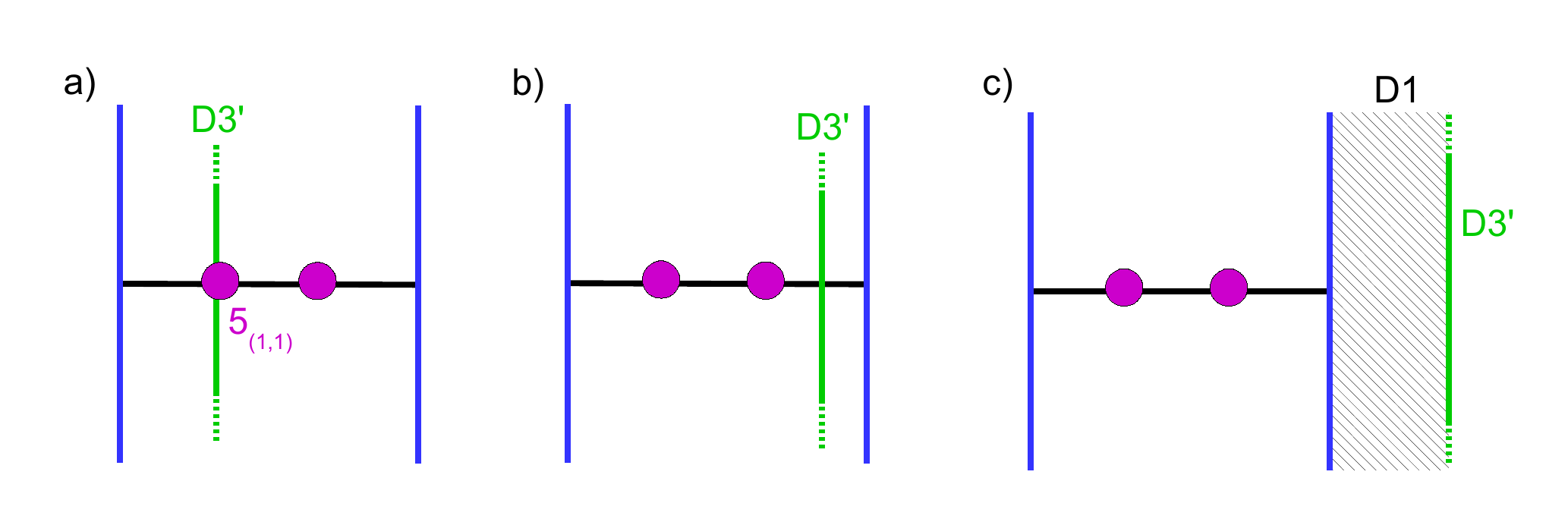} 
\vskip -0.5cm
\caption{\footnotesize Moving the D3'($\cong$ D3$_{(1,0)}$) along $x^3$ we obtain different operator insertions, which are identified on the branch II: a) $X_1$, b,c) $\varphi_3$. }
\label{TAD3move}
\end{figure}

Finally the D3$_{(1,0)}$ brane on Figure \ref{TAOpII}-c intersects the first D3 segment. Integrating out the zero-dimension fermion living at the intersection of the D3s yields the insertion of the complex scalar $\varphi_1$. 
It is commonly argued that the low-energy physics is unaffected by brane moves along the $x^3$ direction \cite{Hanany:1996ie}, therefore we may insert the same operator with the D3$_{(1,0)}$ brane moved to different $x^3$ positions. The insertion due to the D3$_{(0,1)}$ at different $x^3$ positions have been worked out in \cite{Assel:2017hck} (see Section 5.2.1) with the following results. If we move the D3$_{(1,0)}$ on top of the 5$_{(1,1)1}$ brane the D3-D3$_{(1,0)}$ open string modes insert the meson operator $X_1$. If the D3$_{(1,0)}$ stands between the two 5$_{(1,1)}$ branes, crossing the D3$_2$ segment, the operator insertion is $\varphi_2$. If we move the D3$_{(1,0)}$ on top of the 5$_{(1,1)2}$ brane we end up with the insertion of the meson operator $X_2$. If we move it so that it crosses the third D3 segment it results in the insertion of the complex scalar $\varphi_3$. This is partially depicted in Figure \ref{TAD3move}. 
The operator insertions related by D3$_{(1,0)}$ moves are equal on the branch II:
\be
X_1 = X_2 = \varphi_1 = \varphi_2 = \varphi_3 \equiv X \quad (\text{branch II}) \,.
\label{RelD3moves}
\ee
We might go further and move the D3$_{(1,0)}$ brane to the right of the brane configuration as in Figure \ref{TAD3move}-c. As the D3$_{(1,0)}$ crosses the right NS5, a D1 string is created stretched between them by the Hanany-Witten effect. Therefore this setup also corresponds to the insertion of the operator $\varphi_3 = X$. Similarly when the D3$_{(1,0)}$ is on the left of the configuration, with a D1 stretched between the D3$_{(1,0)}$ and the left NS5, the operator insertion is $\varphi_1 = X$. This is compatible with the results of \cite{Assel:2017hck}. 

Such brane setups with D3' branes therefore correspond to the insertion of chiral operators on branches of vacua. They are not useful to derive ring relations and we will mostly ignore them on the remainder of the paper.

\bigskip

\begin{figure}[h!]
\centering
\includegraphics[scale=0.75]{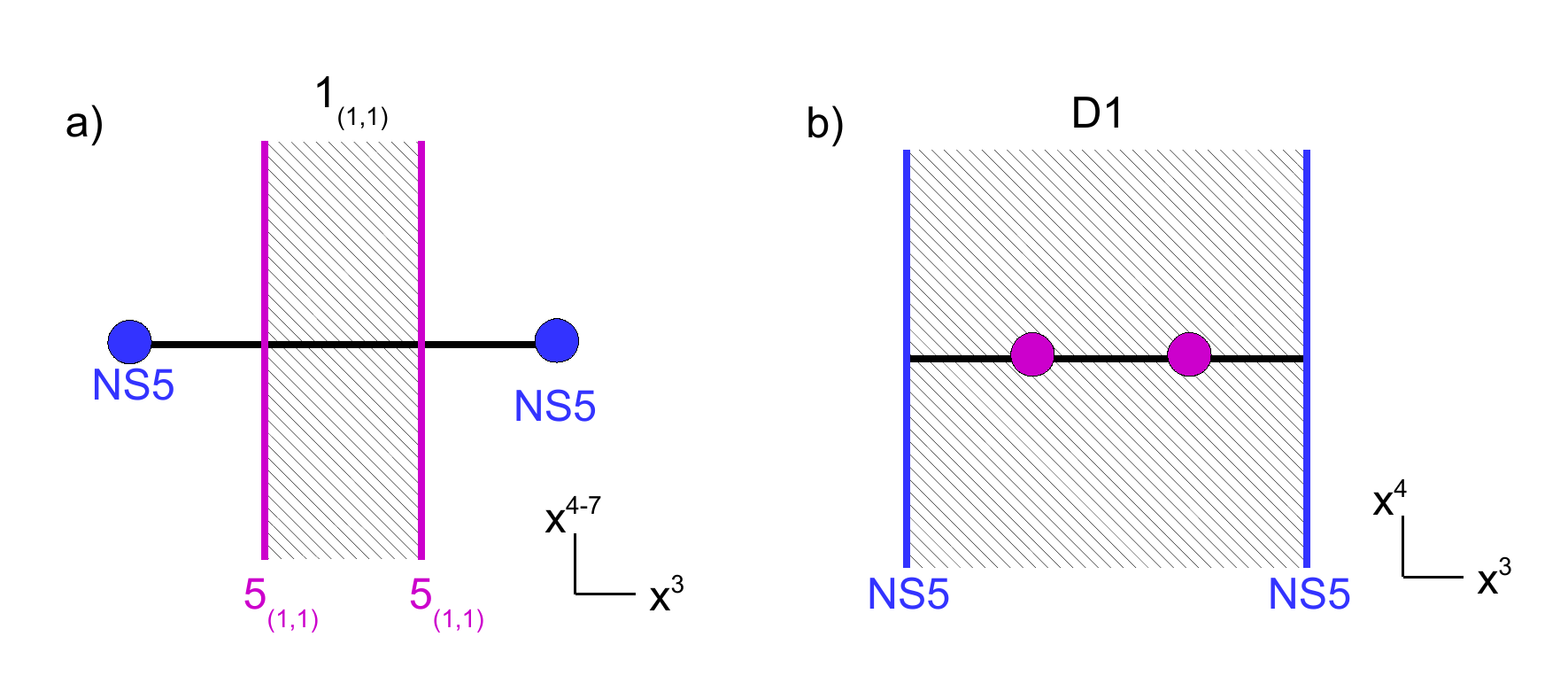} 
\vskip -0.5cm
\caption{\footnotesize Each of the two brane setups can be read in two ways, leading to the relations: a) $u^+u^- = (\varphi_2)^2$, b) $v^+v^-=X^2$.}
\label{TARel}
\end{figure}

We have found the brane setups realizing the chiral operators. Now we show that the relations \eqref{TArelations} predicted by mirror symmetry can be derived from other brane setups.
For this purpose we consider the two brane setups of Figure \ref{TARel}.

In Figure \ref{TARel}-a there is a full 1$_{(1,1)}$ string stretched between the two 5$_{(1,1)}$ branes. We can think of it as two semi infinite 1$_{(1,1)}$ strings ending on the middle D3 segment from above and from below respectively, inserting the product of monopole operators $u^+u^-$. Alternatively we can think of it as a single 1$_{(1,1)}$ string crossing the D3 segment. In that case the chiral operator insertion comes from integrating out light open string modes. 
There are several light modes: 2d modes living on the 1$_{(1,1)}$ worldvolume and D3$_{i}-1_{(1,1)}$ modes. 
In the brane configuration we choose the 1$^{(1,1)}$ string to cross exactly the D3$_2$ segment (so that we can interpret it as a product of monopoles insertion), therefore the D3$_{2}$-1$_{(1,1)}$ modes are massless and integrating them out yields a constant. Similiarly integrating out the 1$_{(1,1)}$ worldvolume theory yields a constant. Remain the D3$_{1/3}$-1$_{(1,1)}$ modes, which are excitations of small fundamental strings stretched across one or the other 5$_{(1,1)}$ brane. This is analogous to D3$_{1/3}$-D1 modes of strings stretched across an NS5.\footnote{The two configurations are actually related by the $T$ transformation of $SL(2,\bZ)$.} In that case it was found in \cite{Assel:2017hck} that the light mode is once again a zero-dimensional fermion, with complex mass given by the distance between the D1 and the D3. Applied to the present setup it implies that the D3$_{1/3}$-1$_{(1,1)}$ modes are  two 0d fermion $\chi_1,\chi_3$ with complex mass given by the distance between D3$_{1/3}$ and the 1$_{(1,1)}$ in the transverse plane $x^{5-8}+ix^{6-9}$, namely $m_1 \propto \varphi_1 - \varphi_2 = -\varphi_2$ and $m_3 \propto \varphi_3 - \varphi_2=-\varphi_2$. Here we have set $\varphi_1=\varphi_3=0$, since the D3$_1$ and D3$_3$ segments cannot be displaced in transverse space due to the presence of the 1$_{(1,1)}$ string ending on D3$_2$. This is the sign that we are dealing with the branch I of vacua. Integrating out the two fermions we find the insertion of the operator $(\varphi_2)^2$.

Identifying the two ways of reading the operator insertion we arrive at the chiral ring relation
\be
u^+ u^- = (\varphi_2)^2 \,,
\ee
which arises on the branch with $\varphi_1 = \varphi_3=0$, namely on the branch I. The brane analysis does not fix the coefficients of operator insertions but these can be absorbed (in the present case) into redefinitions of the chiral operators and are therefore immaterial.

\medskip

In Figure \ref{TARel}-b we see a full D1 stretched between the two NS5s and we insist on the fact that it exactly crosses all D3 segments. This implies that the three segments reconnect and in particular that $\varphi_1=\varphi_2=\varphi_3$, so that we are on the branch II.
The brane setup can again be given two interpretations. First it shows two semi-infinite D1s ending on the D3s, inserting the product of monopole operators $v^+ v^-$, according to the earlier discussion. Alternatively it can be seen as a single D1 crossing the D3s inserting a chiral operator of vanishing monopole charge. The massless excitations on the D1 worldvolume and from D3-D1 strings integrate to an overall factor (which may need regularization). The only contribution relevant to the chiral operator insertion in the 3d theory are associated to light modes from the D1-5$_{(1,1)}$ strings.\footnote{These are small 1$_{(1,1)}$ strings that can stretch btween the D1 and  the 5$_{(1,1)}$. They are $SL(2,\bZ)$ dual to fundamental strings stretched between D5 and D1.} We do not know how to study these localized modes, however we have already assumed that the half-D1 ending on the D3 from above or from below and crossing 5$_{(1,1)i}$ carries the insertion of the bifundamental scalar $q_i$ or $\ti q_i$ respectively. Therefore the full D1 crossing 5$_{(1,1)i}$ carries the insertion of the product $q_i\ti q_i = X_i$. From the two D1-5$_{(1,1)}$ crossings we obtain the insertion of $X_1 X_2 = X^2$ on the branch II.
Identifying the two insertions we get the ring relation
\be
v^+ v^- = X^2 \,,
\ee
on the branch II.

The final moduli space that we obtain from the brane analysis has the two branches
\be\ba
& \text{branch I:} \quad  u^+ u^- =  (\varphi_2)^2 \quad [ \, v^\pm = X = 0 \, ] \,,  \cr
& \text{branch II:}  \quad v^+ v^- = X^2   \quad [ \, u^\pm = 0 , \,  \varphi_2 = X  \,] \,.
\label{TAbranchesFinal}
\ea\ee
It corresponds to the union $(\bC^2/\bZ_2) \cup (\bC^2/\bZ_2)$ with the origins of the two branches identified. The 3d infrared SCFT at the origin has $\N=4$ supersymmetry with $SU(2)_I \times SU(2)_{II}$ R-symmetry, each $SU(2)$ factor acting on a single branch.

\begin{figure}[h!]
\centering
\includegraphics[scale=0.75]{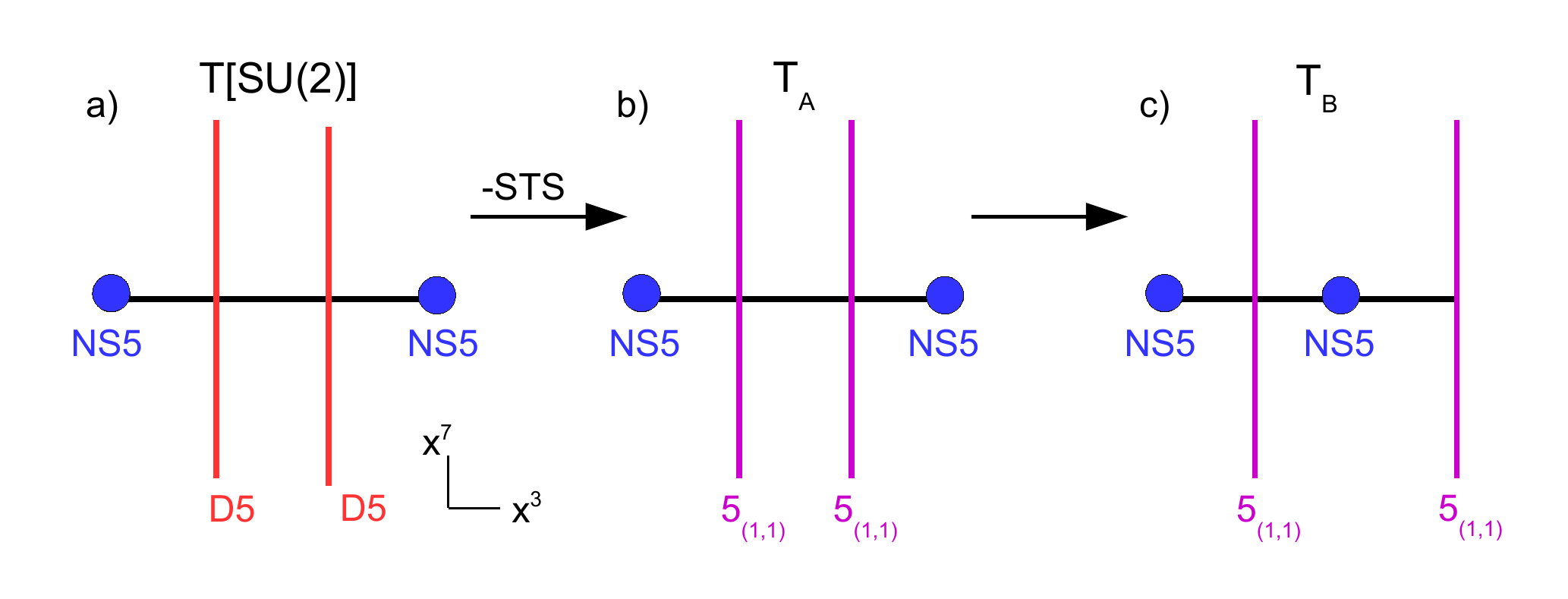} 
\vskip -0.5cm
\caption{\footnotesize. The brane realization of $T[SU(2)]$ (a) becomes the brane realization of $T_A$ (b) after an $SL(2,\bZ)$ transformation. Exchanging the right 5$_{(1,1)}$ and NS5 (with Hanany-Witten D3 creation effect) leads to the brane realization of $T_B$ (c).}
\label{Dualities}
\end{figure}

This is compatible with the prediction of mirror symmetry which follows from the $SL(2,\bZ)$ type IIB action on the brane configuration (see Figure \ref{Dualities}). More precisely, starting from the brane realization of $T[SU(2)]$ (Figure \ref{Dualities}-a), one must act with the $-STS \in SL(2,\bZ)$ transformation obtaining the configuration in non-zero axion background where the NS5 and 5$_{(1,1)}$ are orthogonal, preserving manifestly $\N=4$ supersymmetry. This is a brane realization of the $T_A$ theory, equivalent in the low-energy limit to the $\N=3$ brane configuration with non-orthogonal branes and vanishing axion background.
The dualities actually involve four theories: $T_A^{(1)}, T_A^{(2)}, T[SU(2)]^{(1)}, T[SU(2)]^{(2)}$, where $T_A^{(1)}$ and $ T_A^{(2)}$ are related by the exchange of the branches I and II, and $T[SU(2)]^{(1)}$ and $T[SU(2)]^{(2)}$ are related by the exchange of their Higgs and Coulomb branches. These self-dualities follow from $SL(2,\bZ)$ transformations which exchange the two types of 5-branes and map the brane configuration of the theory back to itself.
\medskip

To this list of dual theories one can add a third theory $T_B$ (or a third couple of theories $T_B^{(1)}$, $T_B^{(2)}$), with a brane realization related to that of $T_A$ by a 5 brane move, as shown in Figure \ref{Dualities}-b,c.
The theory $T_B$ is a $U(1)_{1}\times U(1)_{-1}\times U(1)_1$ linear quiver theory with only bifundamental matter, which we study now as our next example.
We will directly extract the relevant information from the brane analysis, bypassing any initial gauge theory analysis. 

Let us call $\varphi'_i$, $i=1,2,3$ the vector multiplet complex scalars of the three $U(1)$ nodes. 
We also denote $(q'_j,\ti q'_j)$, $j=1,2$, the bifundamental hypermultiplets and $X'_j = q'_j \ti q'_j$ the two chiral meson operators. 
As in the $T_A$ theory, the D3 motions are constrained in the brane configuration of Figure \ref{Dualities}-c, reflecting the presence of two branches of vacua. 
The constraints on each branch are read with the rules explained above.
We find that one branch, say the branch I, has $\varphi'_1=\varphi'_2=X'_1$ and $\varphi'_3=X'_2=0$, corresponding to the reconnection of the D3$_1$ and D3$_2$ segments with motion along the NS5 directions $x^5+ix^6$, and the second branch, say branch II, has $\varphi'_2 = \varphi'_3 = -X'_2$ and $\varphi'_1=X'_1=0$, corresponding to the reconnection of the D3$_2$ and D3$_3$ segments with motion along the 5$_{(1,1)}$ directions $x^{5-8}+ix^{6-9}$ (this can be read from Figure \ref{XandVarphi} of the next section with $\kappa=1$).

\begin{figure}[h!]
\centering
\includegraphics[scale=0.75]{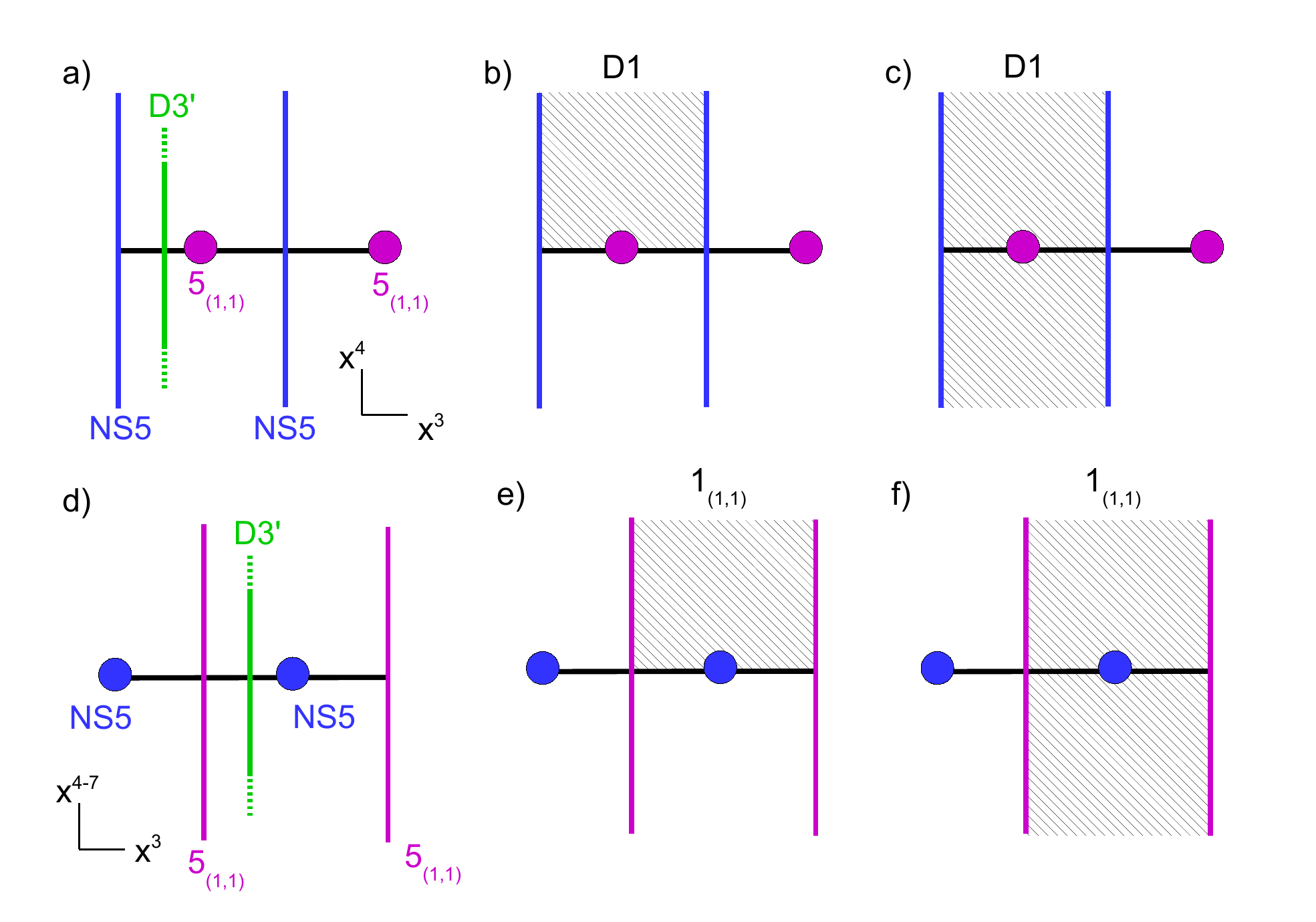} 
\vskip -0.5cm
\caption{\footnotesize Brane setups inserting chiral operators: a) $X'_I$, b) $v^+_{I}$, d) $X_{II}$, e) $v^+_{II}$, and brane setups leading to the ring relations $v_{I}^+ v_{I}^- = X_{I}^2$ (c) and $v_{II}^+ v_{II}^- = X_{II}^2$ (f).}
\label{TBOpRel}
\end{figure}

The operator $X_I \equiv \varphi_1=\varphi_2=X'_1$ on branch I  is realized by adding a $D3_{(1,0)}$ brane standing between the two NS5s (Figure \ref{TBOpRel}-a). The operator $X_{II} \equiv -\varphi_2=-\varphi_3=X'_2$ on branch II  is realized by adding a $D3_{(1,1)}$ brane standing between the two 5$_{(1,1)}$s (Figure \ref{TBOpRel}-d).
In addition there are chiral monopole operators, realized by adding semi-infinite D1 and 1$_{(1,1)}$ strings stretched between the two NS5s and between the two 5$_{(1,1)}$ respectively (Figure \ref{TBOpRel}-b,e). These are the chiral monopoles $v_I^{+}=\ol V'_{++0}q'_1$ and $v_{I}^- = \ol V'_{--0} \ti q'_1$ from the semi D1s, which have magnetic charges $(\pm,\pm,0)$, and $v_{II}^{+}=\ol V'_{0++}\ti q'_2$ and $v_{II}^- = \ol V'_{0--} q'_2$ from the semi 1$_{(1,1)}$s, which have magnetic charges $(0,\pm,\pm)$. The definition of these monopoles is directly read from the brane setups as we did for the $T_A$ theory, with similar assumptions on the matter scalar insertion due to the D3-NS5-1$_{(1,1)}$ intersections. They turn out to be the gauge invariant chiral monopole operators of minimal dimension $\Delta=1$ of the $T_B$ theory,\footnote{The existence of these chiral monopole operators of dimension one is tied to the existence of conserved currents which enhance the $U(1)^2$ topological symmetry to $SU(2)\times SU(2)$ in the infrared SCFT \cite{Gaiotto:2008ak, Bashkirov:2010hj}. This  phenomenon also arises in the $T_A$ and $T[SU(2)]$ theories.}  as we expect. 
We see that $v_I^{\pm}$ can take non-zero vev only on the branch I, because they cannot be realized when the D3$_1$ and D3$_2$ segments are disconnected (corresponding to moving on the branch II). Similarly $v_{II}^\pm$ can take non-zero vev only on the branch II, because they cannot be realized when the D3$_2$ and D3$_3$ segments are disconnected. 

Finally we read the quantum chiral ring relations $v_{I}^+v_{I}^-=X_{I}^2$ and $v_{II}^+v_{II}^-=X_{II}^2$ from the brane setups of Figure \ref{TBOpRel}-c,f. The configuration \ref{TBOpRel}-c , for instance, can be read as the insertion of the product of monopoles $v_{I}^+v_{I}^-$ or as the insertion of $X_{I}^2$ coming from integrating out two fermionic modes of mass $X_{I}$, associated to the D1 crossing the middle 5$_{(1,1)}$ and the D1-D3$_3$ light mode.
We obtain the $(\bC^2/\bZ_2) \cup (\bC^2/\bZ_2)$ moduli space with two branches meeting at their origin
\be\ba
& \text{branch I:} \quad  v_{I}^+ v_{I}^- = X_{I}^2 \quad [\, v_{II}^\pm = X_{II} = 0 \,]  \,, \cr
& \text{branch II:}  \quad  v_{II}^+ v_{II}^- = X_{II}^2  \quad [\, v_{I}^\pm = X_{I} = 0 \,] \,,
\label{TBbranchesFinal}
\ea\ee
isomorphic to the moduli space of the $T_A$ and $T[SU(2)]$ theories.

\begin{figure}[h!]
\centering
\includegraphics[scale=0.75]{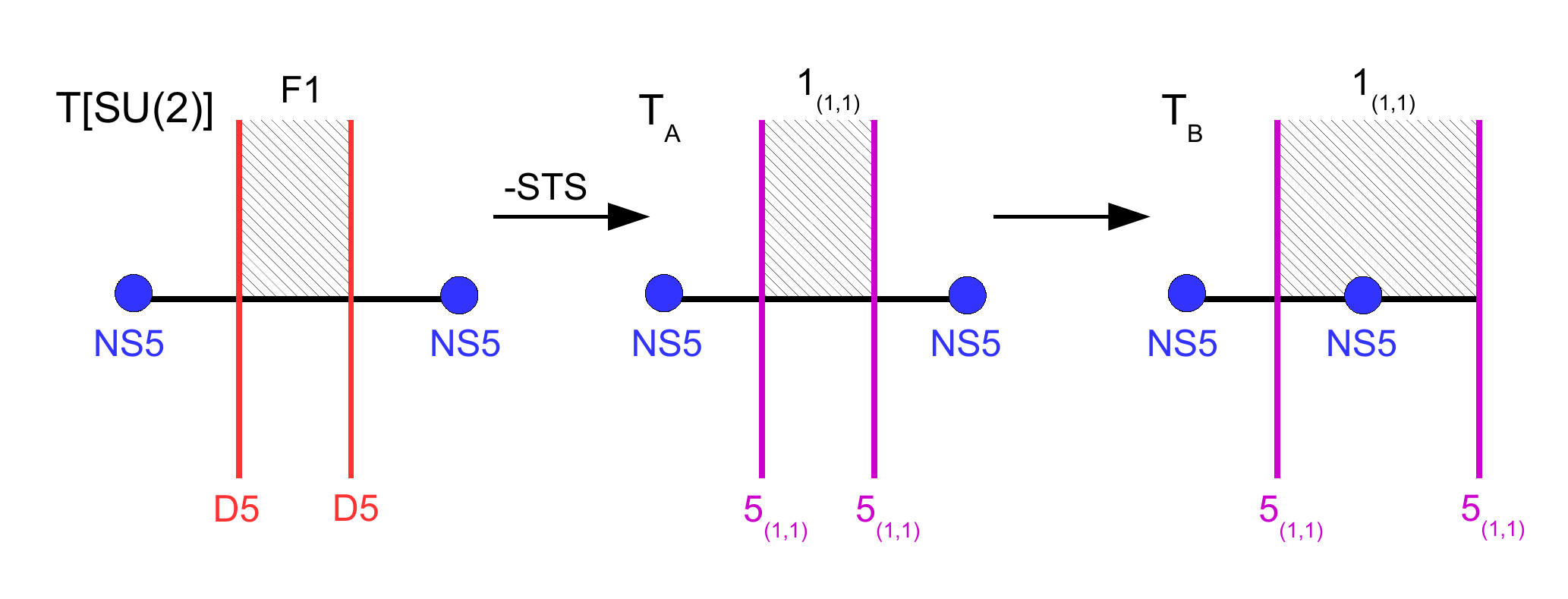} 
\vskip -0.5cm
\caption{\footnotesize The brane realization of the meson operator $Z^2{}_1$ in $T[SU(2)]$ becomes the setup for $u^+$ in $T_A$ and the setup for $v^+_{II}$ in $T_B$, by $SL(2,\bZ)$ action and 5 brane moves.}
\label{MirrorMap}
\end{figure}

 From the $SL(2,\bZ)$ action we can also relate the brane setups for the operator insertions in dual theories and find the mirror map of operators. For instance the insertion of the monopole operator $u^+$ in $T_A$ is realized by the setup of Figure \ref{TAOpI}-a. It is related by 5 brane moves to the setup of Figure \ref{TBOpRel}-e inserting the monopole operator $v^+_{II}$ in $T_B$. It is also related by $SL(2,\bZ)$ transformation to the brane setup inserting a meson operator $Z^2{}_{1}$ in the $T[SU(2)]$ theory (see \cite{Assel:2017hck}). This is illustrated in Figure \ref{MirrorMap}. We therefore get the mirror map $u^+ \leftrightarrow v^+_{II} \leftrightarrow Z^2{}_1$, and more generally the mirror map of Table \ref{tab:MirrorMap}.
\begin{table}[h]
\begin{center}
\setlength\extrarowheight{3pt}
\begin{tabular}{|c|c|c|}
\hline
  $T[SU(2)]$  &  $T_A$ & $T_B$  \\    
\hline
\hline
 $\varphi_{T[SU(2)]}$   & $X$ & $X_{I}$  \\ [2pt]
\hline
 $u^+_{T[SU(2)]}$   & $v^+$ & $v^+_{I}$  \\ [2pt]
\hline
 $u^-_{T[SU(2)]}$   & $v^-$ & $v^-_{I}$   \\ [2pt]
\hline
\hline
 $Z^2{}_2$   & $\varphi_2$ & $X_{II}$  \\ [2pt]
\hline
 $Z^2{}_1$   & $u^+$ & $v^+_{II}$  \\ [2pt]
\hline
  $-Z^1{}_2$   & $u^-$ & $v^-_{II}$   \\ [2pt]
\hline
\end{tabular}
\caption{\footnotesize Mirror map of operators for the $T[SU(2)]$ (in the notation of \cite{Assel:2017hck}), $T_A$ and $T_B$ theories. Each theory is also self-dual under a mirror symmetry exchanging the two branches of vacua (first three rows exchanged with last three rows).}
\label{tab:MirrorMap}
\end{center}
\end{table}

\subsection{$U(1)_{\kappa} \times U(1)_{-\kappa} \times U(1)_{\kappa}$ theory }
\label{ssec:U1cube}

\begin{figure}[h!]
\centering
\includegraphics[scale=0.7]{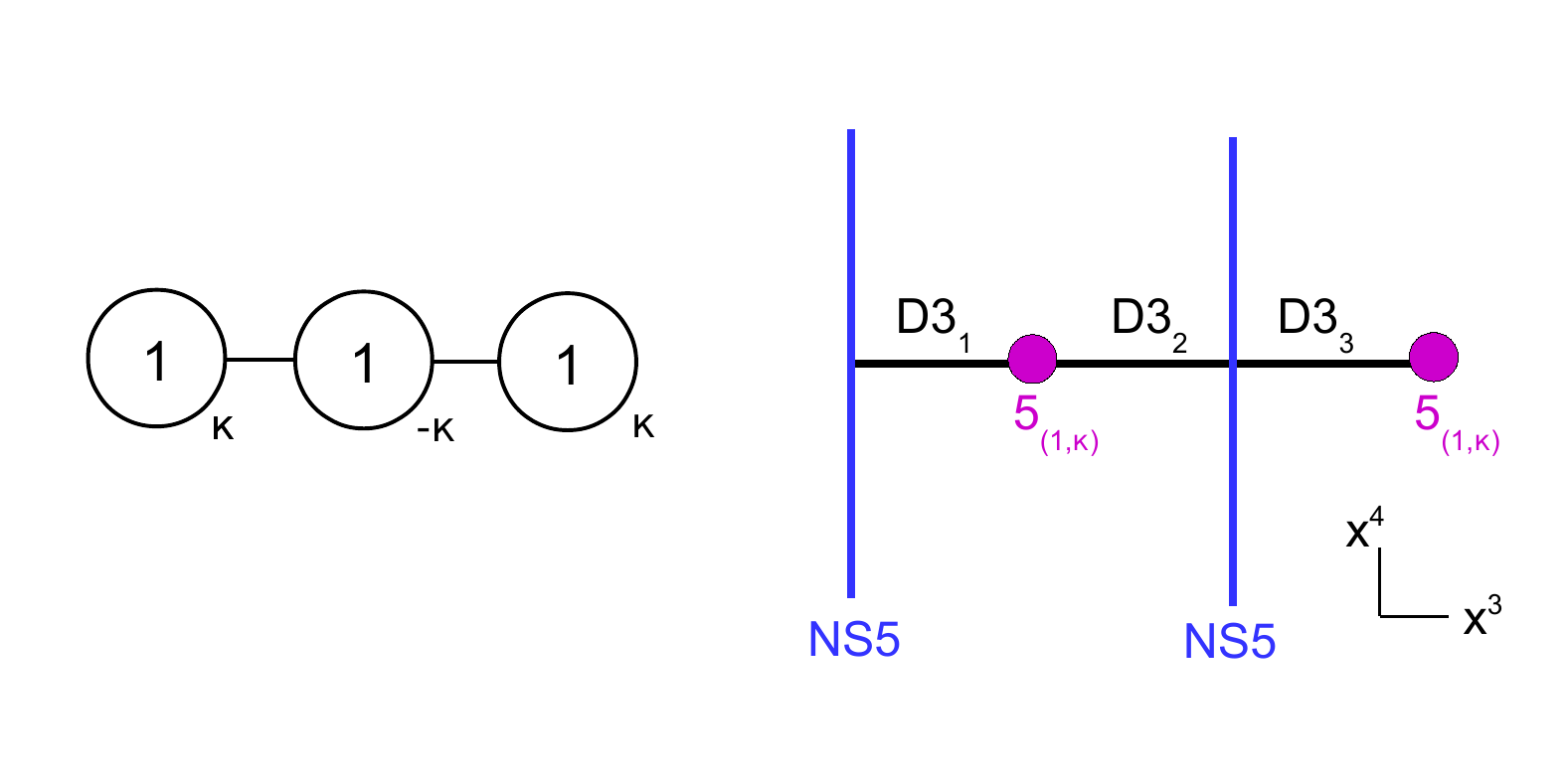} 
\vskip -0.5cm
\caption{\footnotesize Quiver diagram and brane realization of the $U(1)_{\kappa} \times U(1)_{-\kappa} \times U(1)_{\kappa}$ theory.}
\label{TBk}
\end{figure}

A simple generalization of the $T_B$ theory is the  $U(1)_{\kappa} \times U(1)_{-\kappa} \times U(1)_{\kappa}$ theory with only bifundamental matter, with $\kappa\neq 0$. The theory is realized with two NS5 and two 5$_{(1,\kappa)}$ branes ordered as in Figure \ref{TBk}. Since there are only two types of 5 branes, the infrared 3d SCFT has $\N=4$ supersymmetry. The moduli space of this theory was studied in \cite{Jafferis:2008em} by using an infrared dual theory obtained by integrating out the diagonal $U(1)$ component of the gauge multiplet. By studying the one-loop corrections to the moduli space metric it was found for $\kappa>0$ that the moduli space of vacua has two $(\bC^2/\bZ_{\kappa+1})$ branches meeting at the origin. We will recover this result easily (without computations) from the brane analysis.
We denote again $\varphi_i$ the three vector multiplet complex scalars and $(q_j,\ti q_j)$ the hypermultiplets, with the two mesons $X_j = q_j \ti q_j$, $j=1,2$, as well as $\ol V_{n_1n_2n_3}$ the bare monopole operators of magnetic charges $(n_1,n_2,n_3)$.

\begin{figure}[h!]
\centering
\includegraphics[scale=0.75]{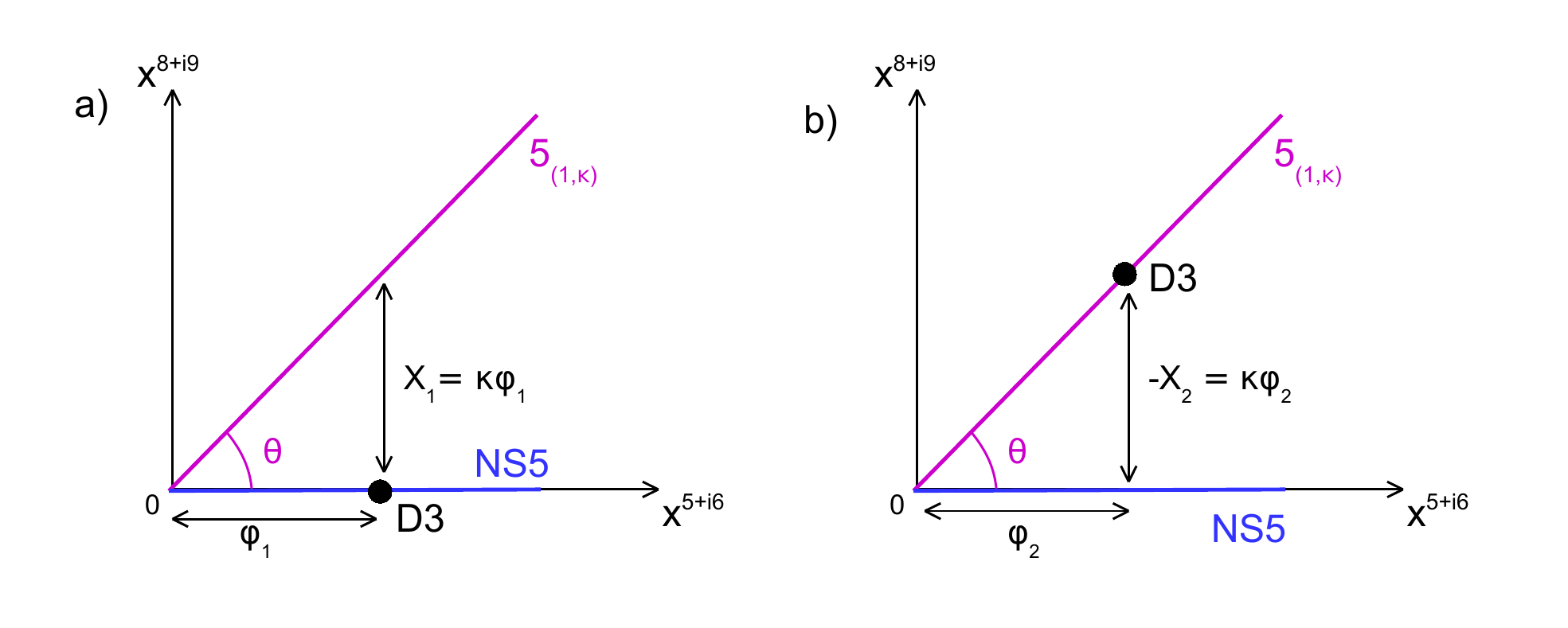} 
\vskip -0.5cm
\caption{\footnotesize Brane setups with the reconnected D3 segment displaced (a) along $x^{5+i6}\equiv x^{5}+i x^{6}$ for the branch I, $(b)$ along $x^{5-8}+i x^{6-9}$ for the branch II. Here $\tan\theta = \kappa$}
\label{XandVarphi}
\end{figure}

As for the $T_B$ theory there are two branches of vacua. The branch I can be associated to the reconnection of the D3$_1$ and D3$_2$ segments with motions along the NS5 directions, leading to the constraints $X_1 = \kappa \varphi_1 = \kappa \varphi_2 \equiv X_I$ and $X_2=\varphi_3=0$.
The branch II can be associated to the reconnection of the D3$_2$ and D3$_3$ segments with motions along the 5$_{(1,\kappa)}$ directions, leading to the constraints $X_2 = -\kappa \varphi_2 = -\kappa \varphi_3\equiv X_{II}$ and $X_1=\varphi_1=0$.
The constraints on the two branches are read from the rules described in the previous section and which are summarized in Figure \ref{XandVarphi}-a,b.
It is a simple exercise to check that these two branches agree with the F-term constraints \eqref{PotentialConstraints}.

The brane setups inserting chiral operators are essentially the same as in the $T_B$ theory (Figure \ref{TBOpRel}-a,b,d,e), with the difference that for the branch II operators one uses 1$_{(1,\kappa)}$ strings stretched between the 5$_{(1,\kappa)}$ branes and D3$_{(1,\kappa)}$ branes standing between the 5$_{(1,\kappa)}$ branes. Therefore we still have six operators: $v^\pm_{I}, X_I$ for the branch I and $v^\pm_{II}, X_{II}$ for the branch II.

The insertion of the scalar $X_I$ by the D3$_{(1,0)}$ brane (as in Figure \ref{TBOpRel}-a for $\kappa=1$) and the insertion of the scalar $X_{II}$ by the D3$_{(1,\kappa)}$ brane (as in Figure \ref{TBOpRel}-d for $\kappa=1$) are as in the $T_B$ theory.

The semi D1 string stretched between the NS5s (as in Figure \ref{TBOpRel}-b for $\kappa=1$) still inserts the bare monopole $\ol V_{++0}$ but the contribution from the D3-5$_{(1,\kappa)}$-D1 crossing now depends on $\kappa$. When $\kappa=1$, we argued that a bifundamental scalar $q_1$ was inserted. To understand the insertion for arbitrary $\kappa$ it is useful to look at the M-theory uplift of this configuration studied in Section \ref{sec:Branes}. In M-theory the 5$_{(1,\kappa)}$ uplifts to an M5$_{\theta}$, with $\tan\theta=\kappa$, and the D1 string uplifts to an M2$_{\alpha}$ brane with $\alpha=0$, as indicated in Table \ref{tab:BraneOrientationsFinal}. The D1-5$_{(1,\kappa)}$ crossing becomes an M2$_{\alpha}$-M5$_{\theta}$ crossing as schematically depicted in Figure \ref{TorusSpan}. The two branes intersect $|\kappa|$ times, therefore we propose that for $\kappa>0$ the semi D3-5$_{(1,\kappa)}$-D1 crossing with the D1 ending on the D3 segments from above inserts $\kappa$ times the bifundamental scalar $q_1$, namely the factor $(q_1)^{\kappa}$. 
For $\kappa <0$ we propose that the insertion is $(\ti q_1)^{-\kappa}$.\footnote{The D5-charge conjugation changing the 5$_{(1,\kappa)}$ into the 5$_{(1,-\kappa)}$ can be associated to the hypermultiplet scalar exchange $q \leftrightarrow \ti q$.} 
Repeating the argument for the semi D1 ending on the D3s from below and for the semi 1$_{(1,\kappa)}$ string ending on the D3s from above or from below (the NS5-1$_{(1,\kappa)}$ system uplifts to an M2-M5 system with again $|\kappa|$ intersections), with similar assumptions, we find the insertions of $v^+_I,  v^-_I, v^+_{II},  v^-_{II}$ due to the four setups, with, for $\kappa >0$,
\be 
v^+_{I} = \ol V_{++0} (q_1)^\kappa \,, \quad v^-_{I} = \ol V_{--0} (\ti q_1)^\kappa \,, \quad v^+_{II} = \ol V_{0++} (\ti q_2)^\kappa \,, \quad v^-_{II} = \ol V_{0--} (q_2)^\kappa \,,
\ee
and for $\kappa<0$,
\be 
v^+_{I} = \ol V_{++0} (\ti q_1)^{-\kappa} \,, \quad v^-_{I} = \ol V_{--0} (q_1)^{-\kappa} \,, \quad v^+_{II} = \ol V_{0++} (q_2)^{-\kappa} \,, \quad v^-_{II} = \ol V_{0--} (\ti q_2)^{-\kappa} \,.
\ee
One can easily check, using \eqref{MonopElCharge} and \eqref{MonopDim}, that they are the gauge invariant monopole operators of minimal dimension $\Delta=\frac{|\kappa|+1}{2}$. They satisfy the constraint \eqref{MonopConstraint1}. 

\begin{figure}[h!]
\centering
\includegraphics[scale=0.7]{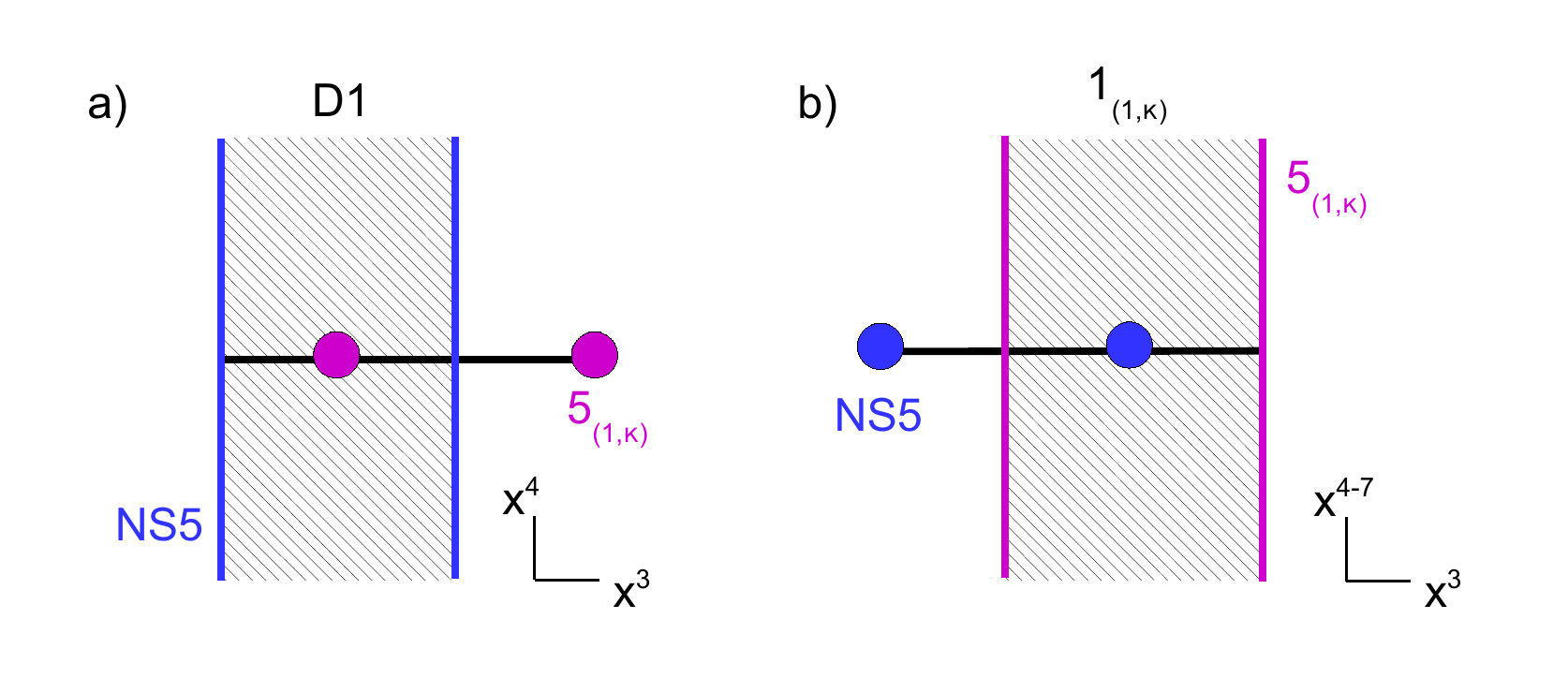} 
\vskip -0.5cm
\caption{\footnotesize Brane setups associated to the relations: a) $v^+_{I}v^-_{I}= (X_{I})^{|\kappa|+1}$, b) $v^+_{II}v^-_{II}= (X_{II})^{|\kappa|+1}$.}
\label{TBkRel}
\end{figure}

After having identified the chiral operators one must find the chiral ring relations. Here again the relevant setups are as in Figure \ref{TBOpRel}-c,f with the 5$_{(1,1)}$s and 1$_{(1,1)}$ replaced by 5$_{(1,\kappa)}$s and 1$_{(1,\kappa)}$.  These setups are reproduced in Figure \ref{TBkRel}-a,b for clarity.

The setup of Figure \ref{TBkRel}-a, related to the branch I, has one interpretation as the monopole insertion $v^+_{I}v^-_{I}$ and a second interpretation as the insertion of $(X_{I})^{|\kappa|+1}$. In the second interpretation, which comes from integrating out light modes in the configuration, assuming $\kappa>0$, there is a contribution $(q_1)^{\kappa} (\ti q_1)^\kappa =  (X_1)^\kappa \equiv (X_{I})^\kappa$ from the the full D1 crossing the 5$_{(1,\kappa)}$ (combining the contributions of two semi D1s) and a contribution $\varphi_3-\varphi_2 \equiv -\kappa^{-1} X_I \propto X_I$ from the D1-D3$_3$ fermionic mode. The reasoning is similar for $\kappa<0$. This leads to the relation $v^+_{I}v^-_{I}= (X_{I})^{|\kappa|+1}$ on the branch I, where, here again, coefficients have been absorbed in operator redefinitions.

\begin{figure}[h!]
\centering
\includegraphics[scale=0.8]{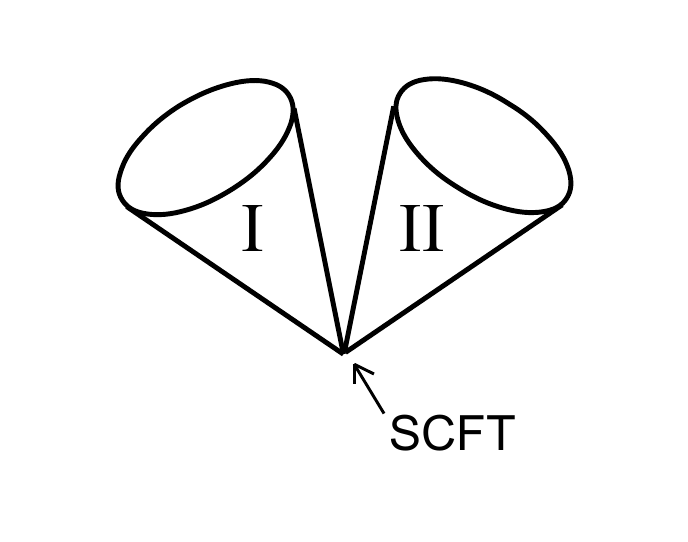} 
\vskip -0.5cm
\caption{\footnotesize Moduli space of the $U(1)_{\kappa}\times U(1)_{-\kappa}\times U(1)_\kappa$ theory. There are two $(\bC^2/\bZ_{|\kappa|+1})$  branches meeting at the origin (SCFT point).}
\label{TBkModSpace}
\end{figure}

The setup of Figure \ref{TBkRel}-b, related to the branch II, identifies the insertion of $ v^+_{II}v^-_{II}$ and $(X_{II})^{|\kappa|+1}$. In the second interpretation, e.g. for $\kappa>0$, one factor $(q_2)^{\kappa} (\ti q_2)^\kappa = (X_2)^\kappa \equiv (X_{II})^{\kappa}$ comes from integrating out lights modes at the 1$_{(1,\kappa)}$-NS5 crossing (combining the contributions of two semi 1$_{(1,\kappa)}$s) and one factor $\varphi_1-\varphi_2  \propto X_{II}$ comes from the 1$_{(1,\kappa)}$-D3$_1$ fermionic mode. We obtain $v^+_{II}v^-_{II}= (X_{II})^{|\kappa|+1}$ on the branch II, and in total two branches with $A_{|\kappa|}$ type singularities
\be\ba
& \underline{\text{branch I:}} \quad  v_{I}^+ v_{I}^- = (X_{I})^{|\kappa|+1}   \quad [\, v_{II}^\pm = X_{II} = 0  \,] \,, \cr
& \underline{\text{branch II:}}\quad   v_{II}^+ v_{II}^- = (X_{II})^{|\kappa|+1} \quad  [\,  v_{I}^\pm = X_{I} = 0  \,]  \,,
\label{TBkbranchesFinal}
\ea\ee
describing the geometry $(\bC^2/\bZ_{|\kappa|+1}) \cup (\bC^2/\bZ_{|\kappa|+1})$ with the two branches meeting at their origins,  in agreement with the result of \cite{Jafferis:2008em}. This is represented in Figure \ref{TBkModSpace}.

\subsection{$T[3]$ theory}
\label{ssec:T3}

We can go further and add fundamental matter to the $U(1)_{\kappa}\times U(1)_{-\kappa}\times U(1)_\kappa$ theory, with $M_i$ fundamental hypers in the $i$th node. In the brane realization we must add $M_i$ D5 branes (= 5$_{(0,1)}$ branes) oriented as in Table \ref{tab:BraneOrientationsFinal} and crossing the D3$_i$ segment, for each $i$. Since we have three kinds of 5 branes now, the supersymmetry is not enhanced in the infrared limit and the theory has only $\N=3$ supersymmetry. 

The moduli space of vacua has a new branch, which is a Higgs branch, emanating from the origin of the two-branches space of Figure \ref{TBkModSpace}, and where fundamental matter scalars can acquire vevs. In the brane picture the new branch is associated to the motion of D3 segments stretched between D5s and moving along the D5 worldvolume directions $x^{789}$.

\begin{figure}[h!]
\centering
\includegraphics[scale=0.75]{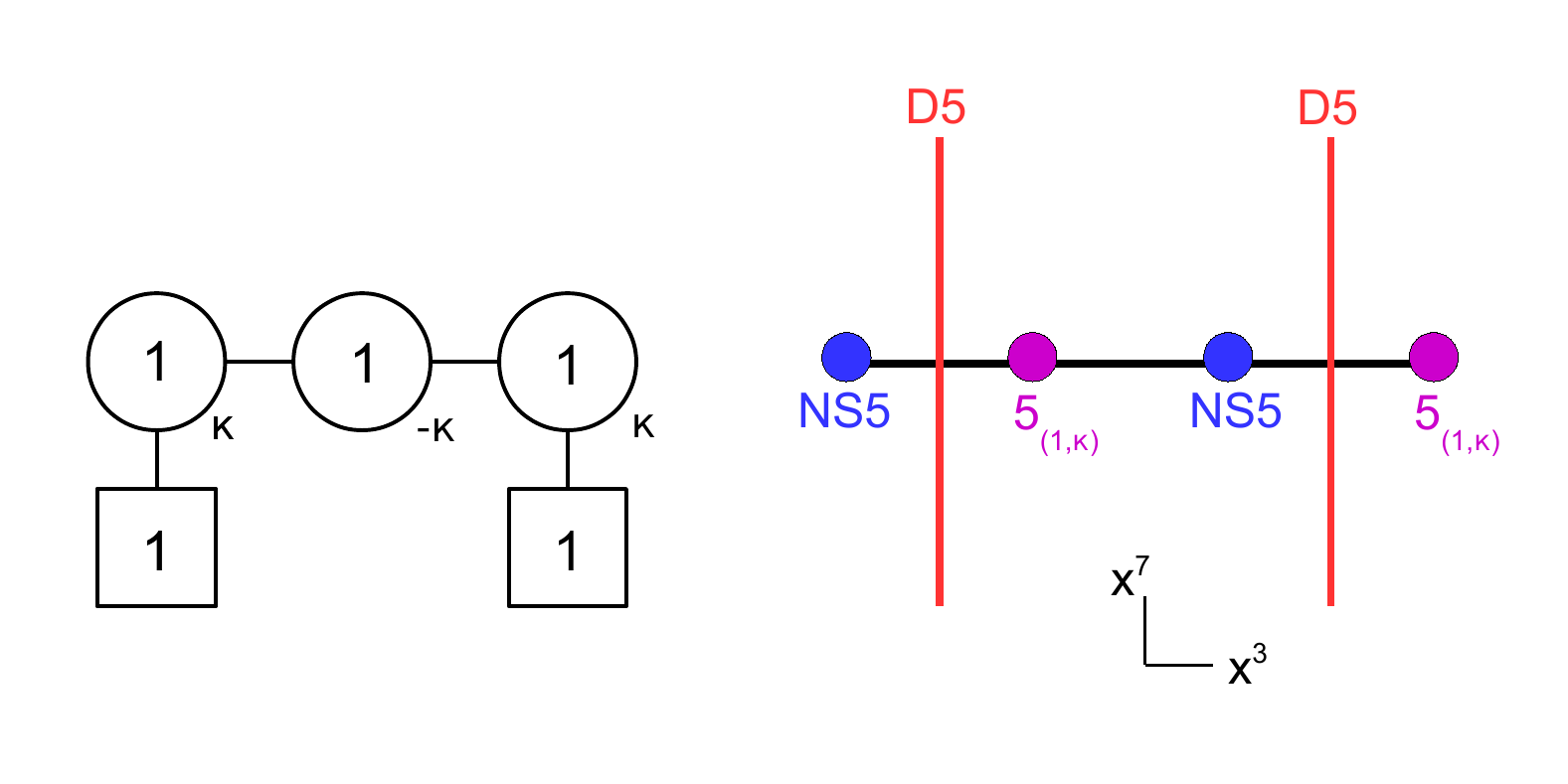} 
\vskip -0.5cm
\caption{\footnotesize Quiver and brane realization of the $T[3]$ theory in the $x^{7,3}$ plane.}
\label{T3}
\end{figure}

Let us study what happens in the simple case with only two fundamental hypermultiplets, say with one in the first node and one in the third node. The quiver diagram and the brane realization are those of Figure \ref{T3} and we call this theory $T[3]$, referring to the three branches that we will find.

We denote $(Q_\alpha,\ti Q_\alpha)$ the hypermultiplet scalars sourced by the D5$_\alpha$ branes, with $\alpha=1,2$ labeling the D5s from left to right, and $Z_1 = Q_1 \ti Q_1$, $Z_2 = Q_2 \ti Q_2$, $Z^+ = Q_1 \ti q_1 \ti q_2 \ti Q_2$, $Z^- = \ti Q_1 q_1 q_2 Q_2$  are the four meson operators.

From the brane construction we identify three branches corresponding to motions of D3 segments along the three kinds of 5 branes (NS5s, 5$_{1,\kappa}$s and D5s respectively),
\be\ba
& \text{branch I:} \quad \kappa\varphi_1 = \kappa\varphi_2 = X_1 \equiv X_I \,, \quad \varphi_3 = X_2=Z_1=Z_2=0 \,, \cr
& \text{branch II:} \quad -\kappa\varphi_2 =  -\kappa\varphi_3 = X_2 \equiv X_{II} \,, \quad \varphi_1 = X_1=Z_1=Z_2=0 \,, \cr
& \text{branch III:} \quad  -Z_1=Z_2 = X_1 = X_2 \equiv X_{III} \,, \quad \varphi_1= \varphi_2 = \varphi_3 =0 \,. \cr
\ea\ee
The constraints on the operators $\varphi_i$ ($i=1,2,3$) and $X_j$ ($j=1,2$) defining the branches I end II are as in the previous subsection. In addition the meson operators $Z_\alpha$ take vevs when the D3 segment crossing the D5$_\alpha$ breaks in two pieces on the brane. The vev of $Z_\alpha$ then corresponds to the separation between the two D3 segments in the $x^{8+i9}$ directions \cite{Gaiotto:2008sa}.
These three branches arise in the field theory from the F-term constraints \eqref{PotentialConstraints}
\be\ba
& Q^1\varphi_1 = \ti Q_1 \varphi_1 =  Q^2\varphi_3 = \ti Q_2 \varphi_3 = 0 \,, \cr
& q_1(\varphi_1-\varphi_2) = \ti q_1(\varphi_1-\varphi_2) = q_2(\varphi_2-\varphi_3) = \ti q_2(\varphi_2-\varphi_3) = 0 \,, \cr
& \ti Q_1 Q^1 + \ti q_1 q_1 = \kappa \varphi_1 \,, \quad  -\ti q_1 q_1 + \ti q_2 q_2 = -\kappa \varphi_2 \,, \quad  -\ti q_2 q_2 + \ti Q_1 Q^2 = \kappa \varphi_3 \,.
\ea\ee

The three operators $X_I, X_{II}, X_{III}$ can be inserted by the brane setups with a D3$_{(1,0)}$, D3$_{(1,\kappa)}$ and D3$_{(0,1)}$ placed between the two NS5s, the two 5$_{(1,\kappa)}$s and the two D5s respectively.

\begin{figure}[h!]
\centering
\includegraphics[scale=0.75]{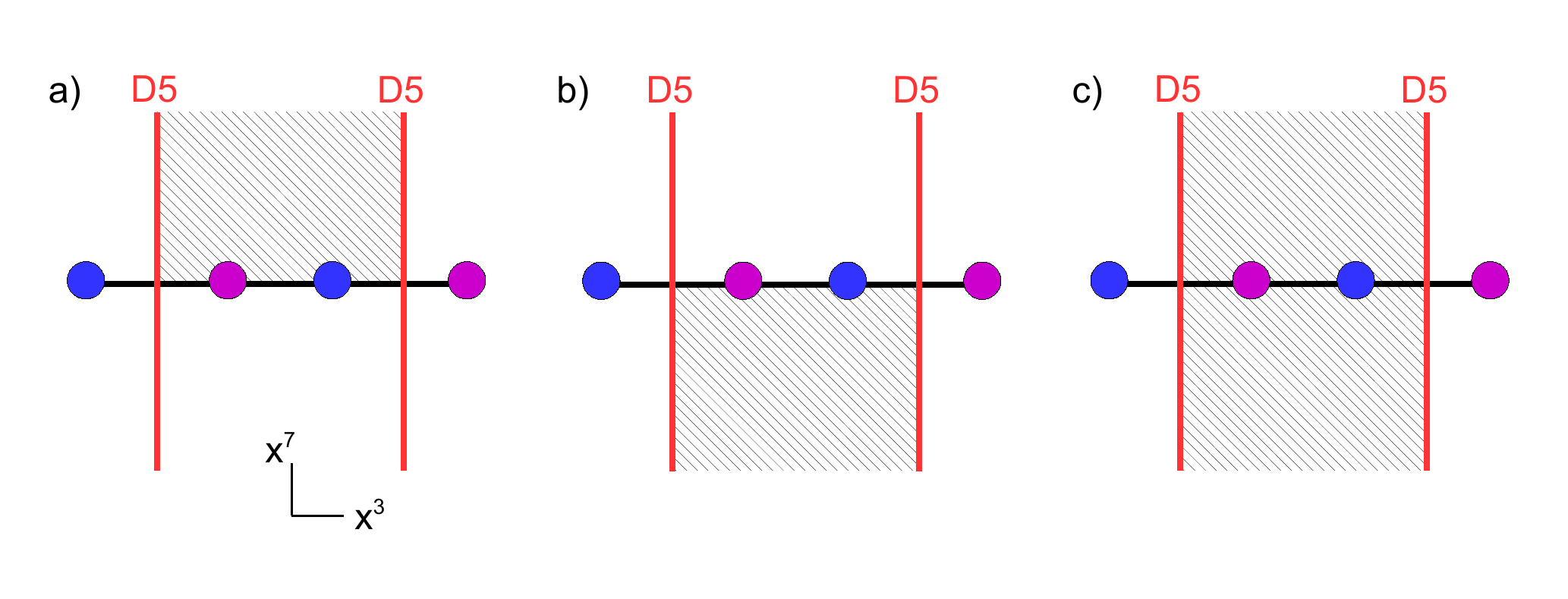} 
\vskip -0.5cm
\caption{\footnotesize Brane setups inserting: a) $Z^+$, b) $-Z^-$. From the setup c) we read $-Z^+ Z^- = -Z_1 X_1 X_2 Z_2  = (X_{III})^{4}$.}
\label{T3Mesons}
\end{figure}

Each branch has two more chiral operators which can take vevs: $v^\pm_{I}$ on the branch I, $v^\pm_{II}$ on the branch II and $Z\pm$ on the branch III, which are inserted by adding a semi-infinite D1, 1$_{(1\kappa)}$ and F1 string stretched between the two NS5s, the two 5$_{(1,\kappa)}$s and the two D5s respectively, ending on the D3 segments from above or from below. This is as before for $v^\pm_{I}$ and $v^\pm_{II}$, namely the presence of the D5s does not modify the monopole operator insertion, however it affects the dimensions of the monopoles \eqref{MonopDim} which become $\Delta(v^\pm_{I})=\Delta(v^\pm_{II})=\frac{\kappa}{2}+1$. For the insertion of $\pm Z^\pm$ the brane setups are shown in Figure \ref{T3Mesons}-a,b, as shown in \cite{Assel:2017hck}.\footnote{The brane setups with F1 strings inserting $\pm Z^\pm$ are actually permuted in comparison with \cite{Assel:2017hck}, in order to obtain consistency with the general insertion for a 1$_{(r,s)}$ string later. This does not affect the analysis in \cite{Assel:2017hck}.}

\begin{figure}[h!]
\centering
\includegraphics[scale=0.75]{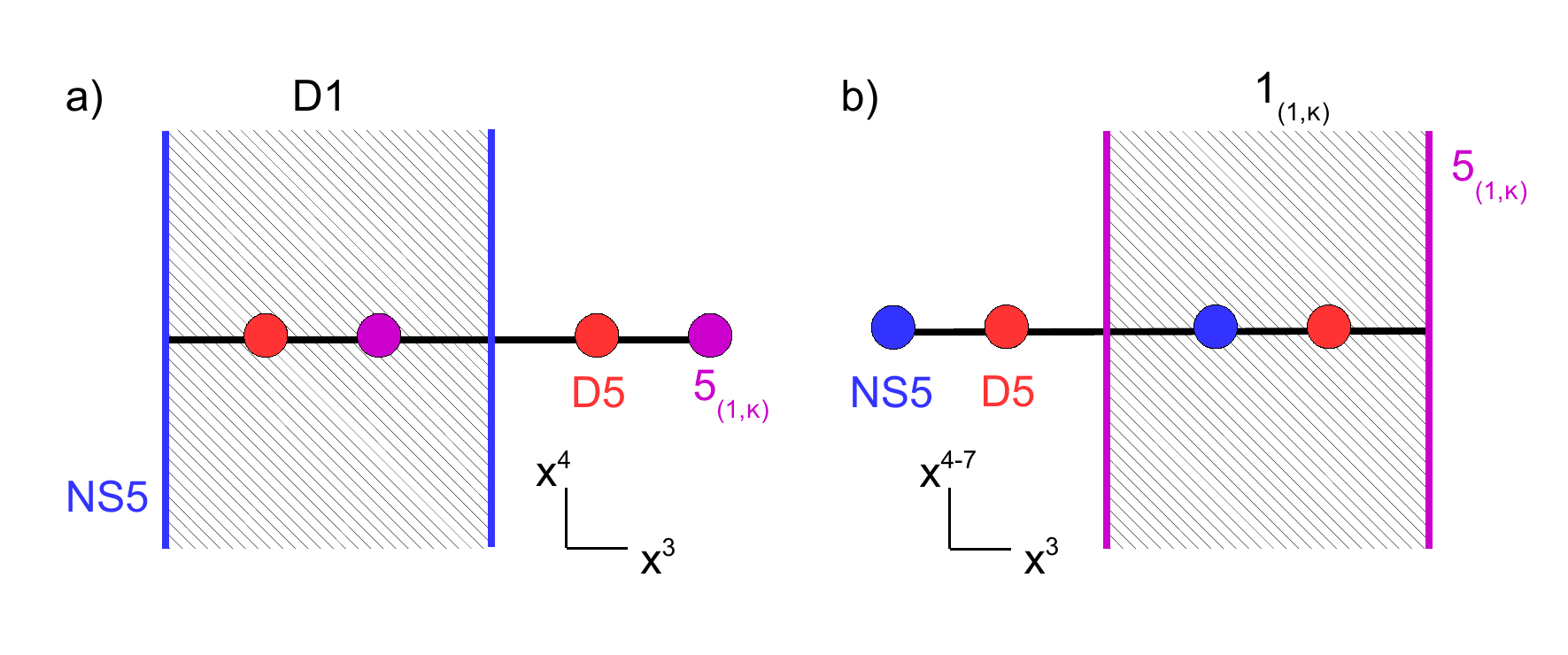} 
\vskip -0.5cm
\caption{\footnotesize Brane setups with full D1 and 1$_{(1,\kappa)}$ strings, from which we read the relation (a) on the branch I and (b) on the branch II, respectively .}
\label{T3Rel}
\end{figure}

The quantum relations involving the monopole operators on the branches I and II are read as before, however the presence of the D5s introduce additional light modes. For instance in the setup of Figure \ref{T3Rel}-a with a full D1 string, the D1-5$_{(1,\kappa)1}$ modes insert a factor $(X_1)^{|\kappa|}$ and the D1-D3$_3$ modes insert a factor $\varphi_3-\varphi_2 \propto X_I$, as before. In addition the D1-D5$_1$ open strings source a zero dimensional fermion of complex mass $-\varphi_1 \propto X_I$ (distance between D5$_1$ and D1 along $x^{5+i6}$), which introduces an extra factor of $X_I$. This leads to the relation $v^+_I v^-_I = (X_I)^{|\kappa|+2}$. 
Similarly the setup with the full 1$_{(1,\kappa)}$ string (Figure \ref{T3Rel}-b) has 1$_{(1,\kappa)}$-NS5$_2$ modes inserting a factor $(X_2)^{|\kappa|}$ and 1$_{(1,\kappa)}$-D3$_1$ modes inserting a factor $(\varphi_1 - \varphi_2) \propto X_{II}$, as before. In addition there is now a 1$_{(1,\kappa)}$-D5$_2$ fermionic mode inserting  a extra factor $-\varphi_3 \propto X_{II}$. This leads to the relation $v^+_{II} v^-_{II} = (X_{II})^{|\kappa|+2}$.

Finally the relation on the branch III (or Higgs branch) can be read from the setup of Figure \ref{T3Mesons}-c, where a full F1 string is stretched between the two D5s. One interpretation is the insertion of the product of mesons $-Z^+Z^-$ corresponding to the insertion of two semi-F1 strings ending on the D3 segments from above and from below. The other interpretation is obtained from the analysis of \cite{Assel:2017hck}, with a contribution from the full F1 string ending on the D5$_1$ from the right inserting a factor $Z_1$, a contribution from the full F1 string ending on the D5$_2$ from the left inserting a factor $-Z_2$, a contribution from the full F1 string crossing the NS5$_2$ inserting a factor $X_2$. The only novelty is the contribution from  the F1 crossing the 5$_{(1,\kappa)}$. Since the F1 intersects the 5$_{(1,\kappa)}$ a single time (as follows from the analysis of the M-theory uplift of Section \ref{sec:Branes}), we propose that the insertion is as for a full F1 crossing an NS5 brane, corresponding in this case to a factor $X_1$. The total insertion is therefore $-Z_1Z_2 X_1 X_2 = (X_{III})^4$ and the relation is $-Z^+Z^- = -Z_1 Z_2 X_1 X_2 = (X_{III})^4$. This is nothing but the trivial relation following the definition of the meson operators in terms of elementary scalar fields. This can serve as a check of our brane algorithm.

\begin{figure}[h!]
\centering
\includegraphics[scale=0.8]{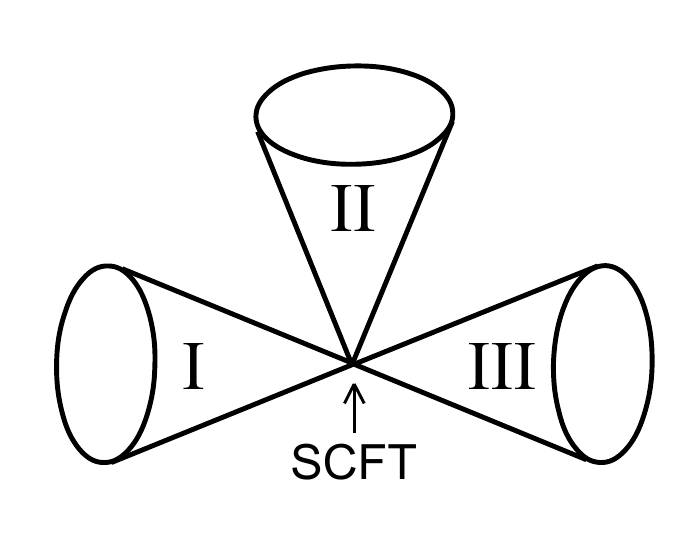} 
\vskip -0.5cm
\caption{\footnotesize Moduli space of the $T[3]$ theory. There are three branches meeting at the origin (SCFT point).}
\label{T3ModSpace}
\end{figure}

We thus obtain a moduli space with three branches parametrized by three chiral operators each and satisfying the ring relations
\be\ba
& \text{branch I:} \quad v^+_I v^-_I = (X_I)^{|\kappa|+2} \,, \cr
& \text{branch II:} \quad v^+_{II} v^-_{II} = (X_{II})^{|\kappa|+2} \,, \cr
& \text{branch III:} \quad -Z^+Z^- = (X_{III})^4 \,. 
\label{T3BranchesFinal}
\ea\ee
The moduli space is therefore $(\bC^2/\bZ_{|\kappa|+2}) \cup (\bC^2/\bZ_{|\kappa|+2}) \cup (\bC^2/\bZ_{4})$, with the three branches meeting at their origins, as shown in Figure \ref{T3ModSpace}.

\medskip

We must comment on the distinction between ring relations on the geometric branches of vacua and chiral ring relations in the theory. In our analysis we have first distinguished three branches and then derived the algebraic ring relations between the complex coordinates on each branch, which are relations between the vevs of chiral operators. However the relations one obtains by reading the brane setups are really the chiral ring relations in the theory and are valid without referring to a given branch of vacua. From the three brane setups of Figure \ref{T3Mesons}-c, \ref{T3Rel}-a,b, we have obtained the chiral ring relations
\be\ba
& Z^+Z^- = Z_1 Z_2 X_1 X_2 \,, \cr
&  v^+_I v^-_I = \varphi_1(\varphi_3-\varphi_2) (X_1)^{|\kappa|} \,, \cr
& v^+_{II} v^-_{II} = (\varphi_1 - \varphi_2)\varphi_3 (X_2)^{|\kappa|} \,,
\ea\ee
up to irrelevant coefficients. On each branch two out of the three relations trivially vanish and one relation remains, leading to \eqref{T3BranchesFinal}.

\subsection{Deformations}
\label{ssec:Deformations}

Three-dimensional $\N \ge 3$ theories do not admit continuous marginal deformations (preserving that amount of supersymmetry) \cite{Cordova:2016xhm}, however they admit massive deformations which arise from weakly gauging a Cartan subalgebra of the global symmetry algebra. In abelian linear quiver theories the global symmetries are $U(1)$ flavor symmetries acting on each fundamental hypermultiplet, corresponding to a triplet of mass deformations $\vec m$, and $U(1)$ topological symmetries acting on monopole operators, corresponding to a triplet of Fayet-Iliopoulos (FI) deformations $\vec\xi$. In $\N=3$ theories, the FI parameters $\vec\xi$ and the mass parameters $\vec m$ both transform as triplets of the $SU(2)_R$ R-symmetry.

In $\N=4$ theories turning on mass parameters has the effect of lifting the Higgs branch and deforming the Coulomb branch, while turning on FI parameters has the opposite effect of lifting the Coulomb branch and deforming the Higgs branch. The two effects are related by mirror symmetry which exchanges Coulomb and Higgs branches \cite{Intriligator:1996ex}. The mass deformations are understood in the brane picture as displacements of the D5 branes along the $x^{456}$ directions, and the FI deformations as displacements of the NS5 branes along the $x^{789}$ directions \cite{Hanany:1996ie}. In a chosen complex structure the triplets split into a real and a complex deformation $(m_\bR,m_\bC)$ or $(\xi_\bR,\xi_\bC)$. The complex parameters $m_\bC$ or $\xi_\bC$ then appear in the Coulomb or Higgs chiral ring relations of the $\N=2$ subalgebra associated to that complex structure. The real parameters are related to resolutions of the singularities \cite{Bullimore:2015lsa}.

 In $\N=3$ theories, turning on a massive deformation still has the effect of lifting some branches of vacua and deforming the surviving branches. In the brane picture the triplets of FI parameters correspond to displacements of the 5$_{(1,\kappa)}$ branes along $x^{789}$ (and the mass parameters to the displacements of D5s along $x^{456}$). The deformation of the moduli space of vacua can be found from the analysis of the deformed brane configuration.
 \medskip
 
 In the explicit example of the $T[3]$ theory, the gauge theory has three $U(1)$ nodes, with three FI deformations $\vec\eta_i=(\eta_{\bR,i}, \eta_{\bC,i})_{i=1,2,3}$,  and two fundamental hypermultiplets with masses $\vec m_\alpha = (m_{\bR,\alpha},m_{\bC,\alpha})_{\alpha=1,2}$, in the complex structure that we choose to describe the moduli space. For simplicity we will set the real parameters to zero\footnote{It is not clear to us how to incorporate their effect.} $\eta_{\bR,i} = m_{\bR,\alpha} =0$ and we recast $\eta_{\bC,i} = \eta_i$, $m_{\bC,\alpha} = m_\alpha$.

The correspondence between the complex parameters and the brane configuration is as follows. $m_1$ and $m_2$ correspond to the position of the D5$_1$ and D5$_2$ branes along the complex plane $x^{5+i6}$. The positions $\xi_i$ of the NS5s and $\zeta_i$ of the 5$_{(1,\kappa)}$ branes along the complex plane $x^{8+i9}$, ordered from left to right ($\xi_1$ for NS5$_1$, $\xi_2$ for NS5$_2$, ... etc), are related to the FI parameters through $\eta_1 = \xi_1 - \zeta_1$, $\eta_2 = \zeta_1 - \xi_2$, $\eta_3 = \xi_2 - \zeta_2$.

When $m_1,m_2$ are turned on, the D5 branes are displaced and the D3 segments cannot break and move along the D5 directions anymore. This means that the branch III is lifted. In the field theory the fundamental hypers aquired a mass, lifting the Higgs branch. The other branches get deformed. 
Similarly when $\zeta_1,\zeta_2$ are turned on, the 5$_{(1,\kappa)}$s are displaced along $x^{8+i9}$, the branch II is lifted and the other branches get deformed. And when $\xi_1,\xi_2$ are turned on, the NS5s are displaced along $x^{8+i9}$, the branch I is lifted and the other branches get deformed. 

\begin{figure}[h!]
\centering
\includegraphics[scale=0.75]{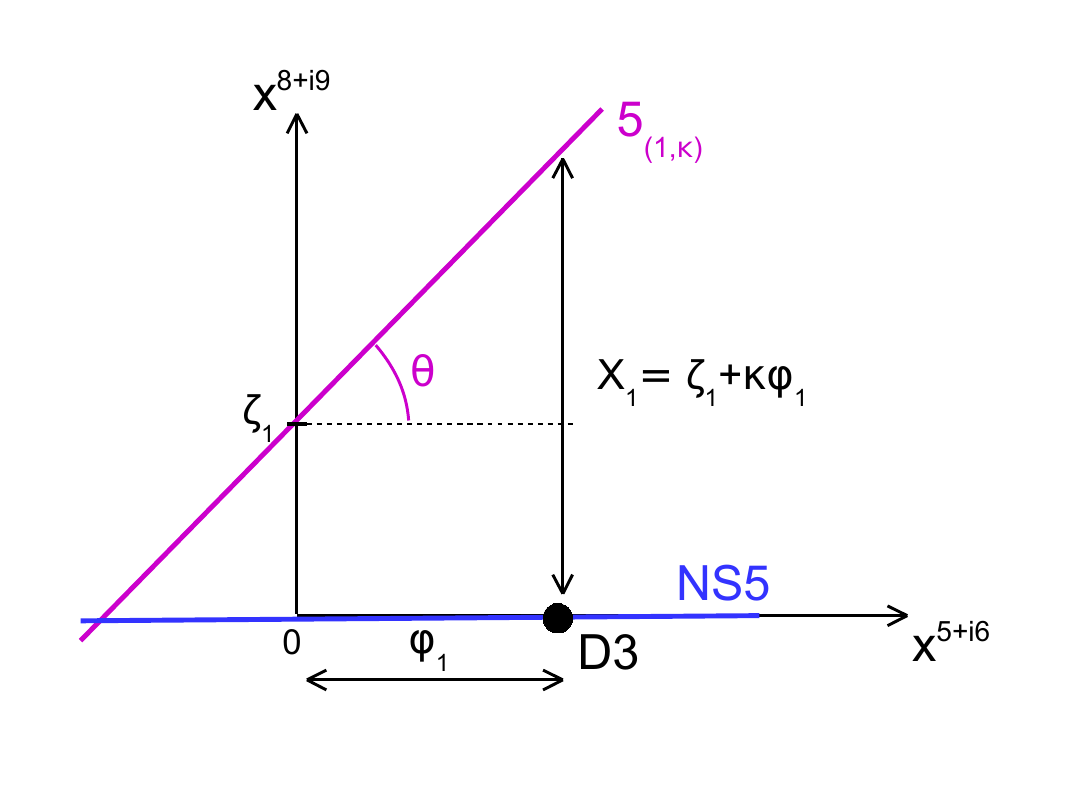} 
\vskip -0.5cm
\caption{\footnotesize The 5$_{(1,\kappa)1}$ brane is moved to the position $\zeta_1$ along $x^{8+i9}$, leading to the constraint $X_1 = \zeta_1+\kappa\varphi_1 (=\zeta_1+\kappa\varphi_2)$.}
\label{XandVarphiDeform}
\end{figure}

Let us analyze the situation when $m_\alpha$ and $\zeta_i$ are turned on (but $\xi_i=0$). Only the branch I survives, corresponding to D3 motion along the NS5s. The constrained movements of the D3 segments in this configuration translate into constraints on the branch I operators, however we observe that there are two posibilities for the positions of the D3s between the NS5$_2$ and the 5$_{(1,\kappa)2}$: either a D3 segment between NS5$_2$ and D5$_2$ , at position $m_2$ along $x^{5+i6}$, and another D3 segment between the D5$_2$ and the 5$_{(1,\kappa)2}$, or a single D3 segment between the NS5$_2$ and the 5$_{(1,\kappa)2}$, at position $-\frac{\zeta_2}{\kappa}$ along $x^{5+i6}$. This means that the branch I splits into two disjoint branches I-1 and I-2,
 \be\ba
& \text{branch I-1:} \quad \kappa\varphi_1 = \kappa\varphi_2 = X_1-\zeta_1 \equiv X_I \,, \quad \varphi_3 = m_2 \,, \cr
& \text{branch I-2:} \quad \kappa\varphi_1 = \kappa\varphi_2 = X_1-\zeta_1 \equiv X_I \,, \quad \varphi_3 = -\frac{\zeta_2}{\kappa}  \,.
\ea\ee
The modification $X_1 \to X_1-\zeta_1$ is due to the displacement of the 5$_{(1,\kappa)1}$ brane along $x^{8+i9}$, as shown in Figure \ref{XandVarphiDeform}.

The chiral ring relation on the branch I is also affected by the motion of the 5 branes. In the brane setup of Figure \ref{T3Rel}-a the 0d fermion sourced by the D5$_1$-D1 open strings has now complex mass $\varphi_1-m_1$, therefore the chiral ring relation becomes
\be
v^+_I v^-_I = (\varphi_1-m_1)(\varphi_3-\varphi_2) (X_1)^{|\kappa|} \,.
\ee
Evaluated on the two disjoint branches I-1 and I-2, we obtain
\be\ba
& \text{branch I-1:} \quad  v^+_I v^-_I = (X_I- \kappa m_1)(X_I- \kappa m_2) (X_I + \zeta_1)^{|\kappa|}  \,, \cr
& \text{branch I-2:} \quad  v^+_I v^-_I = (X_I- \kappa m_1)(X_I + \zeta_2) (X_I + \zeta_1)^{|\kappa|} \,,
\ea\ee
where some coefficients have been absorbed in coordinate redefinitions.

The analysis of the (surviving) branch II when $m_\alpha,\xi_i$ are turned on (but $\zeta_i=0$) proceed in the same way and leads to a splitting into two disjoint branches II-1 and II-2, with
\be\ba
& \text{branch II-1:} \quad  v^+_{II} v^-_{II} = (X_{II}-\kappa m_1)(X_{II}-\kappa m_2) (X_{II}+\xi_2)^{|\kappa|}  \,, \cr
& \text{branch II-2:} \quad  v^+_{II} v^-_{II}  = (X_{II}+\xi_1)(X_{II}-\kappa m_2) (X_{II}+\xi_2)^{|\kappa|}  \,.
\ea\ee

When $\xi_i$ and $\zeta_i$ are turned on (but $m_\alpha=0$) only the Higgs branch survives. This time there is no splitting into two branches, since the D3 segments, outside the D5s, have a single allowed position. The modified constraints read
\be\ba
& \text{branch III:} \quad  -Z_1-\xi_1= Z_2 - \zeta_2 = X_1-\zeta_1 = X_2-\xi_2 \equiv X_{III} \,, \quad \varphi_1= \varphi_2 = \varphi_3 =0 \,. 
\ea\ee
The chiral ring relation, read from the brane setup of Figure \ref{T3Mesons}-c, is unaffected by the NS5 and 5$_{(1,\kappa)}$ motions, so it is still $Z^+Z^- = Z_1 Z_2 X_1 X_2$, however, after plugging in the modified constraints, we obtain the deformed ring relation on the Higgs branch:
\be\ba
& \text{branch III:} \quad  -Z^+Z^- = (X_{III}+\xi_1)(X_{III}+\xi_2) (X_{III}+\zeta_1)(X_{III}+\zeta_2)  \,. 
\ea\ee

When all the FI and mass parameters are turned on, with generic values, there is no configuration of D3 segments preserving $\N=3$ supersymmetry. This the sign of spontaneous supersymmetry breaking, which is harder to study.
\medskip

We notice that the branch I and the branch II are isomorphic, even in the deformed case, provided we exchange the FI deformations $\xi_1,\xi_2$ with $\zeta_2,\zeta_1$. This is the sign of a self-duality associated with the $SL(2,\bZ)$ action which exchanges NS5 and 5$_{(1,\kappa)}$ branes, leaving the D5s invariant. The brane configuration goes back to itself after this action and a spacetime reflection $x^3\to -x^3$. In the process, the branches I and II have been mapped one into the other.

\subsection{Higher charge monopoles}
\label{ssec:HigherChargeMonop}

\begin{figure}[h!]
\centering
\includegraphics[scale=0.7]{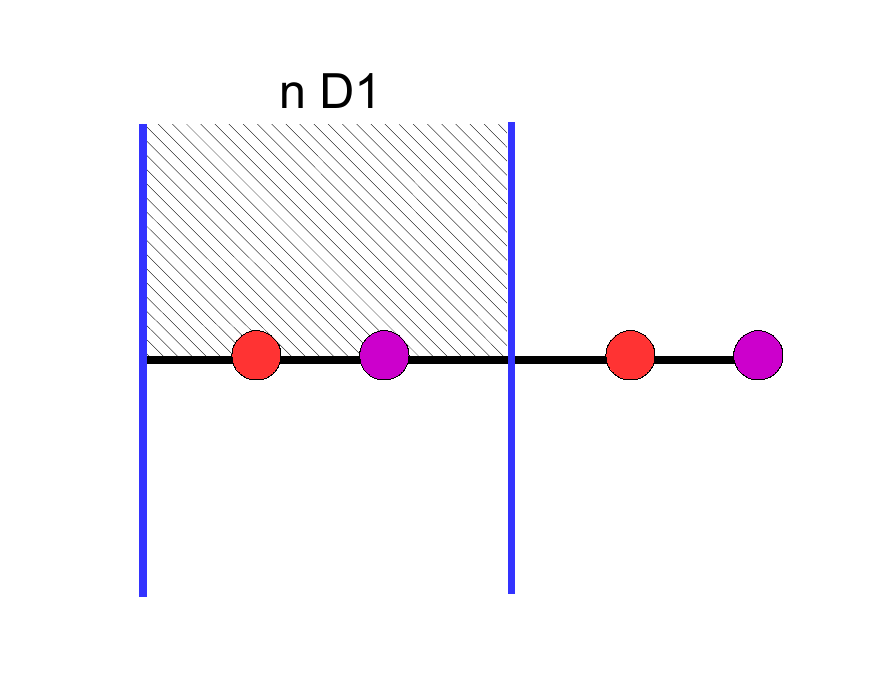} 
\vskip -1cm
\caption{\footnotesize Setup with $n$ superposed D1 strings, leading to the relation $v^n_I = (v^+_I)^n$.}
\label{HigherMonop}
\end{figure}

The brane analysis furnishes a basis of chiral operators parametrizing the branches of the moduli space of vacua. Except for the Higgs branch, these are monopole operators of minimal dimension, with magnetic charges spanning the lattice of allowed monopole magnetic charges in the theory. The monopole operators with higher magnetic charges (and higher dimension) also have a realization in the brane picture. They correspond to having several  strings, possibly superposed, ending on the D3 segments. For instance in the case of the $T[3]$ theory, we can consider the brane setup with $n>0$ semi-infinite D1 branes stretched between the two NS5s, ending on the D3 segments (see Figure \ref{HigherMonop}). The operator inserted has magnetic charge $n$ and, since there is no other brane setup inserting this monopole charge, we propose that it corresponds to the (gauge invariant) monopole operator $v^n_{I} = \ol V_{nn0} (q_1)^{\kappa n}$. Since we can interpret the same brane setup as $n$ times the insertion of $v^+_I$, we obtain the chiral ring relation
\be
v^n_I = (v^+_I)^n \,,
\ee
which can be understood as the relation between bare monopoles $ \ol V_{nn0} = ( \ol V_{++0})^n$. This relation can be used to eliminate $v^n_I$ from the chiral ring. 
All higher magnetic charge monopoles can be eliminated by this procedure, leaving only the operators realized with a single string in the basis of the chiral ring.
%


\section{Abelian linear quivers}
\label{sec:LinQuiv}

\begin{figure}[h!]
\centering
\includegraphics[scale=0.8]{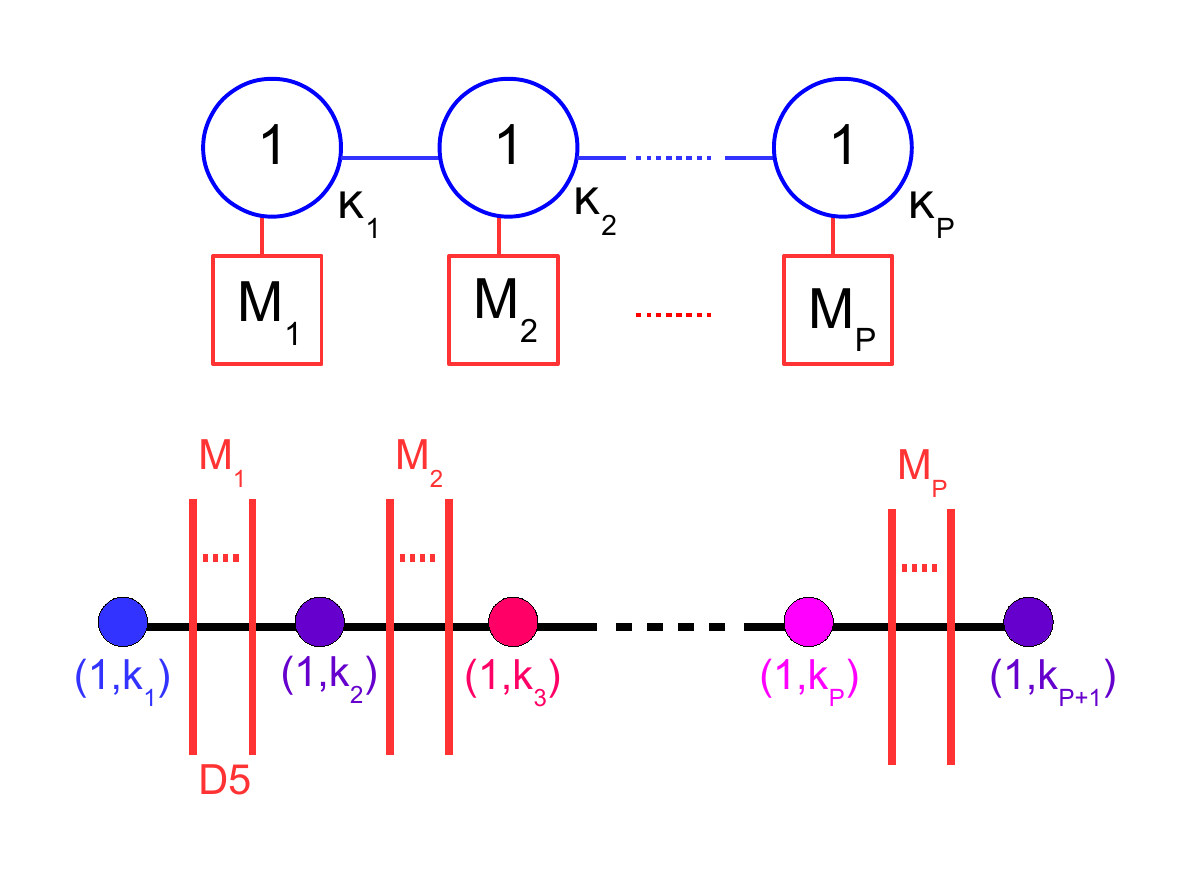} 
\vskip -0.5cm
\caption{\footnotesize Quiver and brane realization of an arbitrary abelian linear quiver $\N=3$ theory, with $\kappa_i = k_{i+1}-k_i$.}
\label{Ablinquiv}
\end{figure}

It is rather straightforward to generalize the analysis to an arbitrary abelian linear quiver theory with gauge group $\prod_{i=1}^P U(1)_{\kappa_i}$, bifundamental hypermultiplets and $M_i$ fundamental hypermultiplets in the $U(1)_{\kappa_i}$ node. The quiver diagram and the brane realization of the generic theory are displayed in Figure \ref{Ablinquiv}.
The brane configuration is given by a succession of 5$_{(1,k_i)}$ branes with a single D3 brane stretched between consecutive 5 branes and the relation to the gauge theory CS levels is $\kappa_i = k_{i+1}-k_i$, with by convention $k_1=0$ (the first 5 brane is an NS5). 

From the analysis of the previous section we can easily understand the general structure of the moduli space of vacua. 
The essential features are the following:
\begin{enumerate}
\item The space of vacua is a union of branches meeting on positive codimension subspaces. Each branch is a product of $L$ factors, with each factor associated to a type of 5 brane present in the type IIB realization of the theory, and $L$ is the number of different 5 brane types, 
\be \ba
\scM &= \bigcup_{n=1}^B \, \scB_n \,, \quad
\scB_n = \prod_{\ell=1}^L \scM^{(\ell)}_n \,,
\label{TotModSp}
\ea\ee
where $B$ is the total number of branches. The moduli parametrizing a factor $\scM^{(\ell)}_n$ are related to certain D3 segment motions along the directions spanned by the 5 branes of type $\ell$. The $\scM^{(\ell)}_n$ are hyperk\"ahler manifolds. Some factors $\scM^{(\ell)}_n$ are single points (and are usually dropped from the products).
\item For a given $\ell$, the factors $\scM^{(\ell)}_n$ obey an inclusion relation $\scM^{(\ell)}_{n_{\ell,1}} \subset \scM^{(\ell)}_{n_{\ell,2}} \subset \cdots \subset \scM^{(\ell)}_{n_{\ell,B}}$ for some permutation $(n_{\ell,1}, \cdots, n_{\ell,B}) \in S_B$ . We call the larger factor $\scM^{(\ell)}_{n_{\ell,B}}$ the {\it maximal branch of type $\ell$}.\footnote{This is a small abuse of language since the factor generically does not correspond to any branch $\scB_n$.}  If the 5 branes are D5 branes, the maximal branch is a Higgs branch, otherwise it is a Coulomb-like branch. There are $L$ maximal branches in total.
\item Two branches $\scB_{n_1}$ and $\scB_{n_2}$ intersect on a submanifold as follows
\be\ba
& \scB_{n_1} \cap \scB_{n_2} = \prod_{\ell=1}^L \scM^{(\ell)}_{n_{\ell \rm min}} \,, \cr
& n_{\ell \rm min} = \left\lbrace
\begin{array}{c}
n_1 \quad \text{if} \quad  \scM^{(\ell)}_{n_1} \subset \scM^{(\ell)}_{n_2} \,, \cr
n_2 \quad \text{if} \quad  \scM^{(\ell)}_{n_2} \subset \scM^{(\ell)}_{n_1} \,.
\end{array}\right.
\label{CapBranches}
\ea\ee
\item When the number of 5 branes of type $\ell$ is smaller than two all the factors $\scM^{(\ell)}_n$ are trivial (equal to a point), otherwise some factors are non-trivial. In particular when the brane configuration does not have any two 5 branes of the same type the entire moduli space of vacua is trivial  (a single point). This ties up the existence of branches of vacua to the presence of sequences $k_i, k_{i+1}, \cdots, k_j$ of Chern-Simons levels in the quiver chain satisfying $\sum_{m=i}^j k_m = 0$.
\item   The maximal branch of type $\ell$ has an algebraic description in terms of $(K_\ell)^2 - 1$ complex generators and a number of ring relations between them, which can be algorithmically extracted from the knowledge of $(K_\ell -1)^2$ pre-relations (see below), where $K_\ell$ is the number of 5 branes of type $\ell$ in the brane realization of the theory. It has quaternionic dimension $K_\ell-1$. The generators and the pre-relations can be read from brane setups.
\label{Claims}
\end{enumerate}
Let us explain each of these properties one by one.

The first point stems from the fact that the moduli parametrizing the space of vacua are associated (in part) with the motion of D3 segments along 5 branes. The possible motions of D3 segments in the brane configuration decompose into a certain number $B$ of mutually exclusive possibilities, each corresponding to a branch $\scB_n$. On the branch $\scB_n$ a certain number $d^{(\ell)}_n$ of D3 segments are displaced along the 5 branes of type $\ell$, parametrizing (in part) the factor $\scM^{(\ell)}_n$ of the branch. The quaternionic dimension of $\scM^{(\ell)}_n$ is dim$_\bH \, \scM^{(\ell)}_n = d^{(\ell)}_n$. Note that on each branch there are also D3 segments stretched between different types of 5 branes, stuck at the origin in space, which do not carry moduli.  This leads to the form \eqref{TotModSp} of the moduli space.

The second and third points follows from the observation that moving from one branch $\scB_n$ to another branch $\scB_m$ is achieved by letting a D3 segment of type $\ell$ (i.e. moving along 5 branes of type $\ell$) go to the origin in $x^{456789}$ space and possibly recombine with other D3 segments. This allows for other D3 segments to move away from the origin along 5 branes of different types. 
In the process the factor $\scM^{(\ell)}_n$ of the initial branch has shrunk to a factor $\scM^{(\ell)}_m \subset \scM^{(\ell)}_n$, and in the meantime some other factors $\scM^{(\ell')}_n$ have been enlarged to $\scM^{(\ell')}_m \supset \scM^{(\ell')}_n$.
In general one needs to repeat this process several times (moving several D3 segments) to go from $\scB_n$ to $\scB_m$, but the inclusion relations between the factors remain valid. It follows that at a given $\ell$, the factors $\scM^{(\ell)}_n$ obey an inclusion relation as described in point 2 above, and that the intersection between two branches is described by \eqref{CapBranches}.

The fourth point simply reflects the fact that there must be at least two branes of type $\ell$ to have a D3 segment moving along 5 branes of type $\ell$. In the gauge theory, this translates into the fact that there is sequence of Chern-Simons levels $k_i, k_{i+1}, \cdots, k_j$ such that $\sum_{m=i}^j k_m=0$, except if the 5 branes in question are D5s, in which case it simply means that there are at least two fundamental hypermultiplets in the theory.

The fifth point deserves a longer explanation and is more easily explained by studying a generic example. This will be the most elaborate theory we will consider in this paper.

\subsection{T[4] theory}
\label{ssec:T4}

\begin{figure}[h!]
\centering
\includegraphics[scale=0.75]{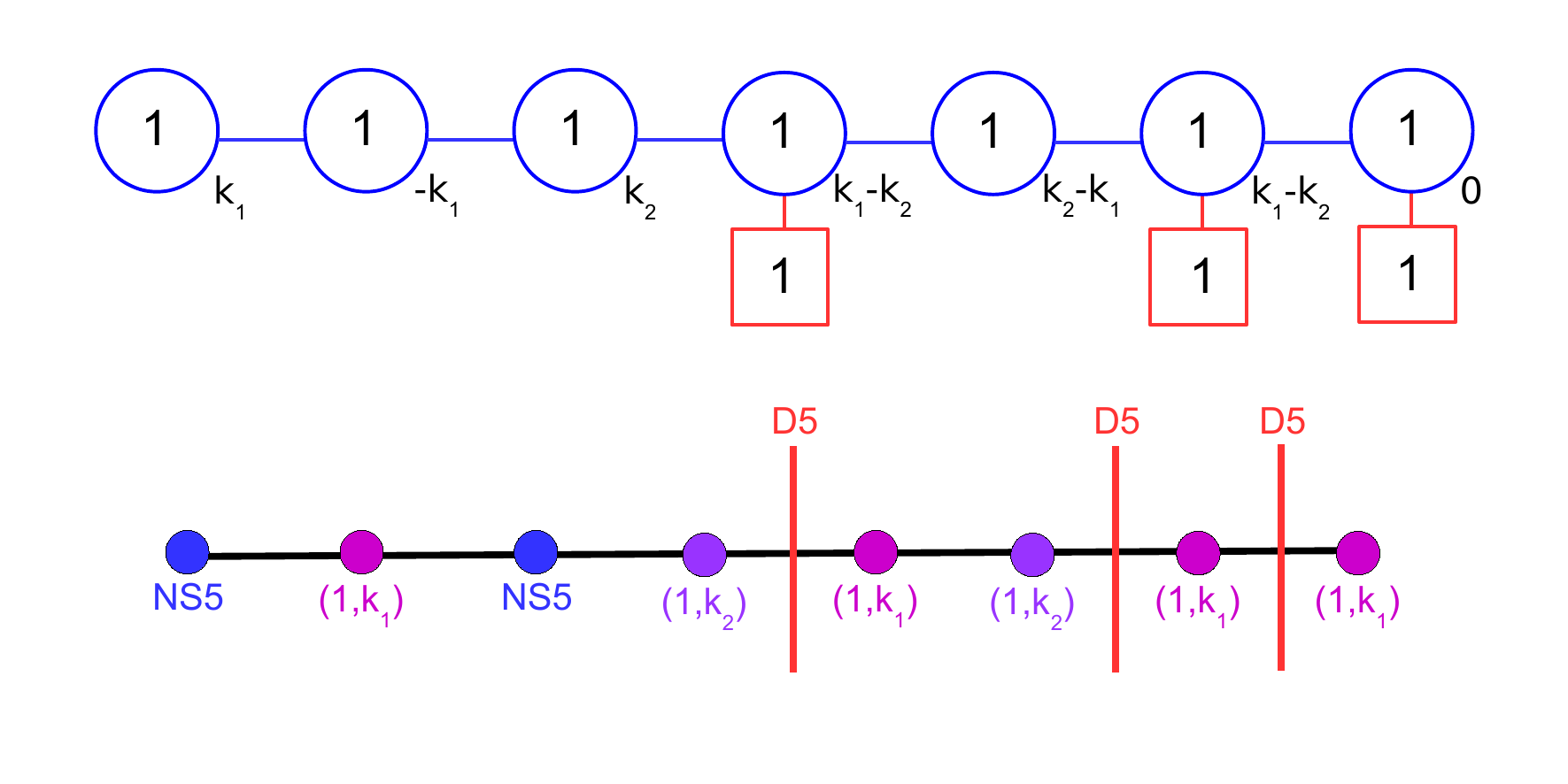} 
\vskip -1cm
\caption{\footnotesize Quiver and brane realization of the $T[4]$ theory in the $x^{3,7}$ plane.}
\label{T4}
\end{figure}

We consider an abelian linear quiver with gauge group $U(1)_{k_1} \times U(1)_{-k_1} \times U(1)_{k_2} \times U(1)_{k_1-k_2} \times U(1)_{k_2-k_1} \times U(1)_{k_1-k_2} \times U(1)_{0}$, with one fundamental hypermultiplet in the fourth, sixth and seventh nodes. We take $k_1k_2\neq 0$ and $k_1 \neq k_2$. The quiver diagram and brane realization are shown in Figure \ref{T4}. We will call this theory $T[4]$ in relation to the four maximal branches that we will find. We choose this theory because it illustrates all the points mentioned above and it is sufficiently generic to convince oneself that the method works for any abelian linear quiver. Despite the relative complexity of the theory, we will see that its space of vacua can be easily understood from the study of a few brane setups.

The brane realization of the theory has four types of 5 branes signaling four maximal branches, $\scM_I,\scM_{II},\scM_{III},\scM_{IV}$, which we associate with the 5 branes of different types by
\be
\scM_I \leftrightarrow \text{NS5} \,, \quad \scM_{II} \leftrightarrow 5_{(1,k_2)} \,, \quad \scM_{III} \leftrightarrow \text{D5}   \,, \quad \scM_{IV} \leftrightarrow  5_{(1,k_1)} \,,
\ee
where the ordering $I$ to $IV$ corresponds to increasing dimensions (or equivalently increasing numbers of 5 branes of a given type).
Each maximal branch is associated to the motion of D3 segments along a single type of 5 branes.

According to the discussion above, the actual branches $\scB_n$ of the theory are of the form
\be
\scB_n = \scM^{(I)}_n \times \scM^{(II)}_n \times \scM^{(III)}_n \times \scM^{(IV)}_n \,,
\ee
with $\scM^{(I)}_n \subset \scM_I$, \dots , $\scM^{(IV)}_n \subset \scM_{IV}$, some factors $\scM^{(...)}_n$ being trivial (single points).

We label the 5 branes of each type from left to right NS5$_1$, NS5$_2$, 5$_{(1,k_2)1}$,5$_{(1,k_2)2}$ , 5$_{(1,k_1)1}$, \dots , 5$_{(1,k_1)4}$, D5$_1$, \dots , D5$_3$, and the initial D3 segments D3$_1$, \dots, D3$_7$. We denote $\varphi_{i=1,\cdots,7}$ the vector multiplet scalars, $(q_j,\ti q_j)_{j=1,\cdots,6}$ the bifundamental hyper scalars, with the mesons $X_j = q_j \ti q_j$, and $(Q_\alpha, \ti Q_\alpha)_{\alpha = 1,2,3}$, the three fundamental hyper scalars, where the indices $i,j,\alpha$ follow the usual left to right ordering in the quiver picture. From the fundamental scalars one builds the meson operators $Z_\alpha$ and $Z^{\pm}_{[\alpha\beta]}$, $1 \le \alpha < \beta \le 3$,
\be\ba
& Z_\alpha = Q_\alpha \ti Q_\alpha \,, \quad \alpha =1,2,3\,, \cr
& Z^+_{[1,2]} = Q_1 \ti q_4 \ti q_5 \ti Q_2 \,, \hspace{0.8cm}    Z^-_{[1,2]} = Q_2 q_4 q_5 \ti Q_1 \,, \cr
& Z^+_{[2,3]} = Q_2 \ti q_6 \ti Q_3 \,, \hspace{1.1cm}  Z^-_{[2,3]} = Q_3  q_6 \ti Q_2 \,, \cr
& Z^+_{[1,3]} = Q_1 \ti q_4 \ti q_5 \ti q_6 \ti Q_3 \,, \quad  Z^-_{[1,3]} = Q_3 q_4 q_5 q_6 \ti Q_1 \,.
\ea\ee
The bare monopole operators of magnetic charge $\vec n$ are are $\ol V_{n_1n_2\cdots n_7}$.
We study first the maximal branches. 
\medskip

The maximal branches $\scM_I$ and $\scM_{II}$ are analogous to the branches of vacua of the theories studied in the previous section, since they are related to the presence of only two NS5s or two 5$_{(1,k_2)}$s in the brane configuration.

\begin{figure}[h!]
\centering
\includegraphics[scale=0.75]{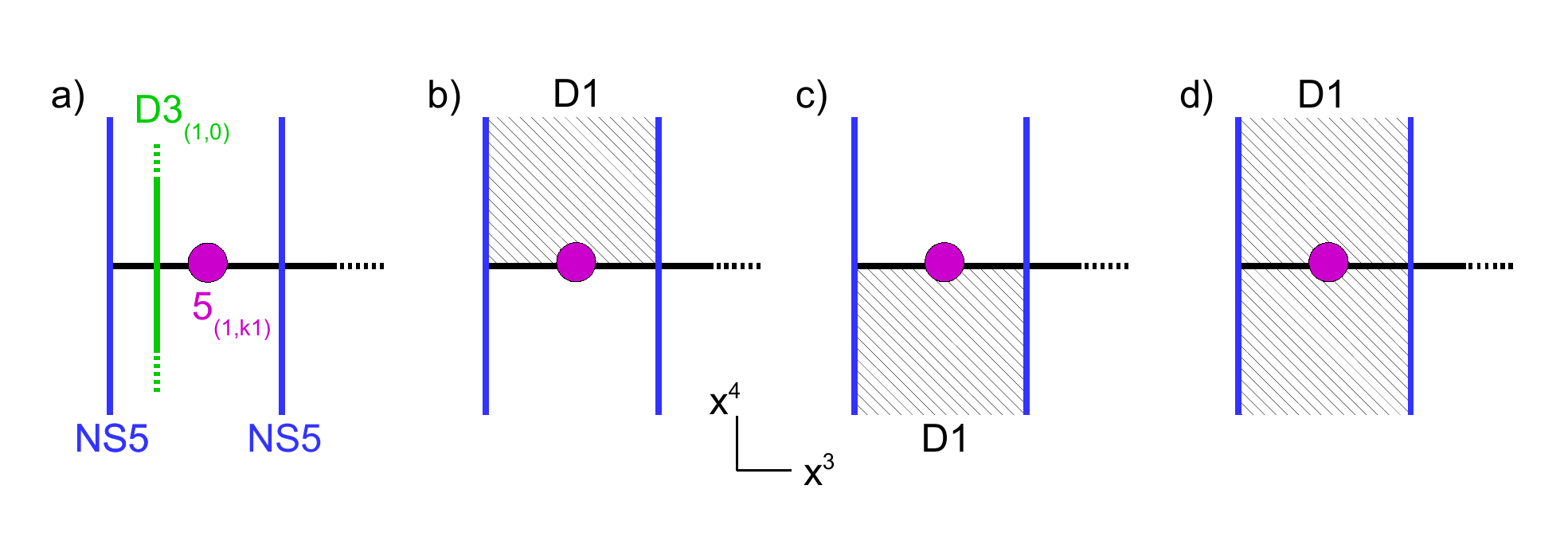} 
\vskip -0.5cm
\caption{\footnotesize Setups realizing the insertion of the branch I operators: (a) $X_I$, (b) $v^+_{I}$, (c) $v^-_{I}$, and setup revealing the relation  $v^+_I v^-_I = (X_I)^{|k_1|+1}$. Only the relevant part of the brane setup is shown.}
\label{T4branchI}
\end{figure}

The space $\scM_I$ is associated to motions of the D3 segment recombined across the 5$_{(1,k_1)}1$ along the NS5s. From the usual rules we read constraints solving the F-term equations on this branch:
\be
\underline{\scM_I}: \quad  k_1\varphi_1 = k_1\varphi_2 =  X_1 \equiv X_I \,, \quad \varphi_{3} = X_{2} = X_3 = 0\,.
\label{MI}
\ee
The other operators ($\varphi_{4},X_4, \cdots$) are not constrained by the D3 motion along the NS5s, so we do not specify them. We can think of them as being set to zero, if we wish, for the purpose of studying the factor $\scM_I$, which will appear in several branches $\scB_n$.
The space $\scM_I$ is parametrized by $X_I$ and the two monopole operators $v^\pm_I$, realized respectively by adding a D3$_{(1,0)}$ brane between the two NS5s and adding a semi D1 stretched between the NS5s, ending on the D3 segment from above and from below, as in Figure \ref{T4branchI}-a,b,c.
From the brane setups we read the definition of the monopoles:
\be
v^+_I = \ol V_{++00000} (q_1)^{k_1} \,, \quad v^-_I = \ol V_{--00000} (\ti q_1)^{k_1} \,, 
\ee
where we assumed $k_1>0$ (if $k_1<0$ one should replace $(q_1)^{k_1}$ by $\ti q_1^{-k_1}$ and $\ti q_1^{k_1}$ by $q_1^{-k_1}$). The factors $\ol V$ arise from the monopole charge insertion in the two first nodes from the semi D1 ending on the D3 segments, and the factors $(q_1)^{k_1}, (\ti q_1)^{k_1}$ arise from the semi D1-5$_{(1,k_1)}$ $k_1$-fold intersection. Let us remind here that a 1$_{(r,s)}$ string and a 5$_{(p,q)}$ brane intersect $|ps-qr|$ times according to the analysis of Section \ref{sec:Branes}.

There is one ring relation corresponding to the setup of Figure \ref{T4branchI}-d with a full D1. The setup relates the insertion of the product of monopoles $v^+_I v^-_I$ to the insertion of the operator $(X_1)^{|k_1|}(\varphi_3-\varphi_2)$, where the factor $(X_1)^{|k_1|}$
arises from $|k_1|$ 0d fermions of mass $X_1$ sourced by D1-5$_{(1,k_1)}$ strings, and the factor $(\varphi_3-\varphi_2)$ arises from a 0d fermion of mass $(\varphi_3-\varphi_2)$ sourced by D1-D3$_3$ strings. We obtain the chiral ring relation
\be
v^+_I v^-_I = (X_1)^{|k_1|}(\varphi_3-\varphi_2) \,,
\ee
which translate, up to unimportant coefficients, to the relation on the maximal branch I,
\be
\underline{\scM_I}:   \quad v^+_I v^-_I = (X_I)^{|k_1|+1} \,, \quad  \big[ \, k_1\varphi_1 = k_1\varphi_2 =  X_1 \equiv X_I \,, \ \varphi_3 = X_2= X_3 = 0 \, \big] \,.
\label{MIfinal}
\ee

\begin{figure}[h!]
\centering
\includegraphics[scale=0.75]{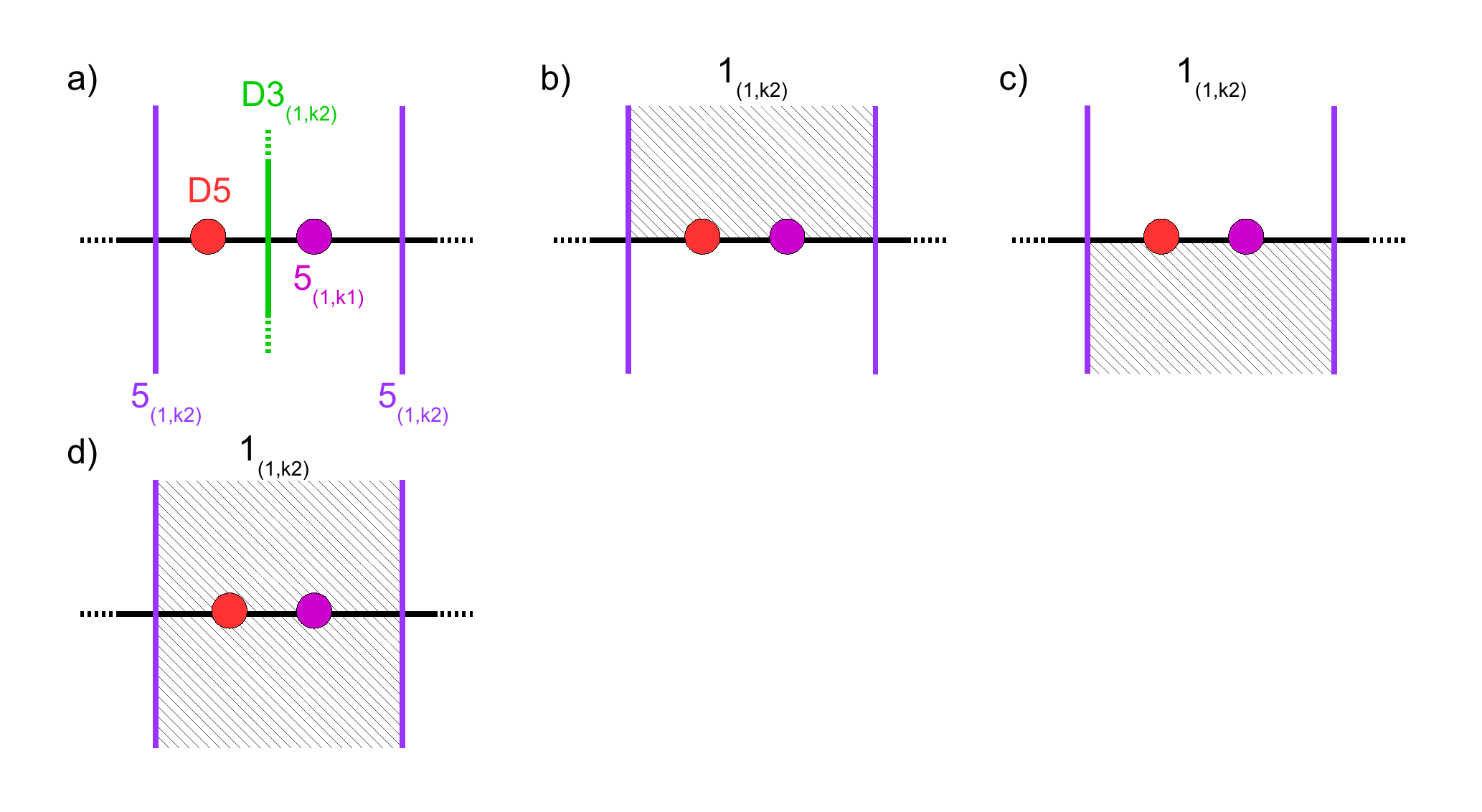} 
\vskip -0.5cm
\caption{\footnotesize Setups realizing the insertion of the branch II operators: (a) $X_{II}$, (b) $v^+_{II}$, (c) $v^-_{II}$, and setup revealing the relation  $v^+_{II} v^-_{II} = (X_{II})^{|k_1-k_2|+3}$. Only the relevant part of the brane setup is shown. The vertical direction is that spanned by the 5$_{(1,k_2)}$ branes in the $x^{47}$ plane.}
\label{T4branchII}
\end{figure}

The maximal branch II is similar. It is associated to the motion of the single reconnected D3 segment along the 5$_{(1,k_2)}$ branes, leading to the constraints
\be
\underline{\scM_{II}}: \quad  (k_1-k_2)\varphi_4 = (k_1-k_2)\varphi_5 =  X_4 \equiv X_{II} \,, \quad \varphi_3 = \varphi_6 = Z^1{}_1 =  X_2 = X_3 = X_5 = 0\,.
\label{MII}
\ee
The space $\scM_{II}$ is parametrized by $X_{II}$ and two monopole operators $v^\pm_{II}$. The brane setups for these three operators on $\scM_{II}$ are realized by the brane setups of Figure \ref{T4branchII}-a,b,c. The monopole operators have magnetic charge $\pm 1$ in the 4th and 5th nodes and are dressed with $|k_1-k_2|$ bifundamental scalar factors from the semi 1$_{(1,k_2)}$-5$_{(1,k_1)}$ $|k_1-k_2|$-fold intersection. Explicitly for $k_1 > k_2$ we read
\be
v^+_{II} = \ol V_{000++00} (q_4)^{k_1-k_2} \,, \quad v^-_{II} = \ol V_{000--00} (\ti q_4)^{k_1-k_2} \,.
\ee
There is a single ring relation on $\scM^{II}$, obtained from the setup with a full 1$^{(1,k_2)}$ stretched between the two 5$_{(1,k_2)}$s as in Figure \ref{T4branchII}-d, from which we read the chiral ring relation 
\be
v^+_{II}v^-_{II} = (X_4)^{|k_1-k_2|} \varphi_4 (\varphi_3-\varphi_4)(\varphi_6-\varphi_5)  \,,
\ee
where the factor $(X_4)^{|k_1-k_2|}$ comes from the 1$_{(1,k_2)}$-5$_{(1,k_1)}$ modes, the factor $\varphi_4$ comes from the 1$_{(1,k_2)}$-D5 mode, the factor $\varphi_3 - \varphi_4$ comes from the 1$_{(1,k_2)}$-D3$_3$ mode and the factor $\varphi_6-\varphi_5$ comes from the 1$_{(1,k_2)}$-D3$_6$ mode. 
This leads to
\be\ba
&\underline{\scM_{II}}:   \quad v^+_{II}v^-_{II} = (X_{II})^{|k_1-k_2|+3} \,, \cr
&  \big[ \, (k_1-k_2)\varphi_4 = (k_1-k_2)\varphi_5 =  X_4 \equiv X_{II} \,, \quad \varphi_3 = \varphi_6 = Z^1{}_1 =  X_2 = X_3 = X_5 = 0 \, \big] \,.
\label{MIIfinal}
\ea\ee

\begin{figure}[h!]
\centering
\includegraphics[scale=0.75]{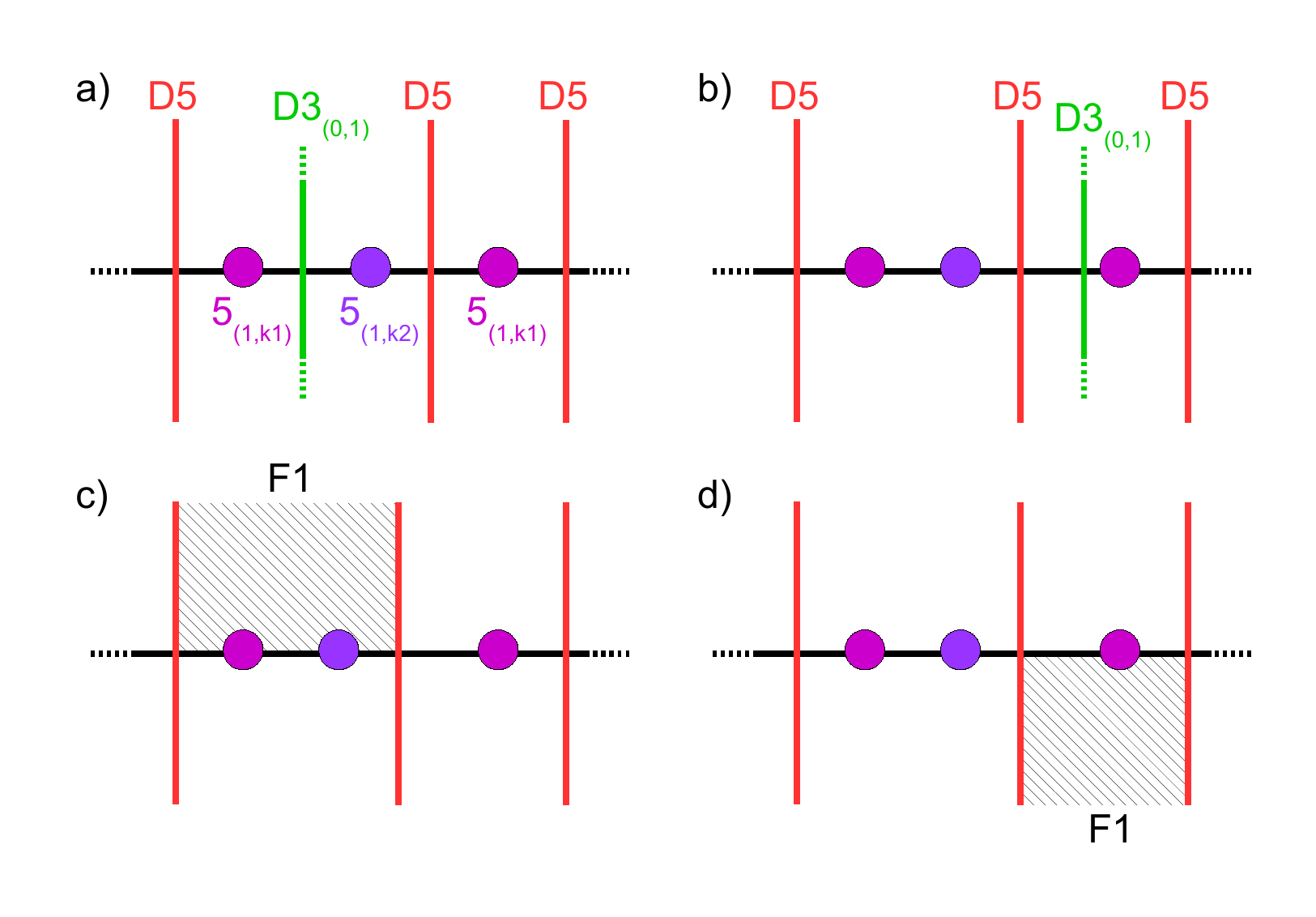} 
\vskip -1cm
\caption{\footnotesize Setups realizing the insertion of the branch III operators: (a) $X_{III,1}$, (b) $X_{III,2}$, (c) $Z^+_{[12]}$, (d) $-Z^-_{[23]}$. Only the relevant part of the brane setup is shown (in the $x^{34}$ plane).}
\label{T4branchIII}
\end{figure}

The maximal branch III, or Higgs branch, is associated to the motion of two reconnected D3 segments along the three D5 branes. We read the constraints
\be\ba
& \underline{\scM_{III}}: \quad  X_4 = X_5 = -Z_1 \equiv X_{III,1} \,, \quad X_6 = Z_3 \equiv X_{III,2} \,, \cr
& \quad   Z_2 = - Z_1 - Z_3 = X_{III,1} - X_{III,2}  \,, \quad \varphi_{4,5,6,7} =  X_3 = 0\,.
\label{MIII}
\ea\ee
To analyse this branch we will rely on the results of \cite{Assel:2017hck}.
The Higgs branch $\scM_{III}$ is parametrized by $X_{III,1},X_{III,2}$ and the mesons $Z^\pm_{[\alpha\beta]}$ with $\alpha < \beta$, which play the role of the monopole operators of other maximal branches. The operators $X_{III,1}$ and $X_{III,2}$ are realized by adding a D3$_{(0,1)}$ brane between the D5$_1$ and D5$_2$ or between the D5$_2$ and D5$_3$ respectively, as in Figure \ref{T4branchIII}-a,b. Let us give the explanations once more. In Figure \ref{T4branchIII}-a with the D3$_{(0,1)}$ at a generic $x^3$ position between D5$_1$ and D5$_2$ there is a D3-D3$_{(0,1)}$ fermionic mode of mass $X_{III,1}$, which corresponds to the distance between the two D3s along $x^{8+i9}$ (not visible in the figure). 
 Similarly In Figure \ref{T4branchIII}-a with the D3$_{(0,1)}$ at a generic $x^3$ position between D5$_2$ and D5$_3$ there is a D3-D3$_{(0,1)}$ fermionic mode of mass $X_{III,2}$, which corresponds to the distance between the two D3s along $x^{8+i9}$. 

The operators $Z^+_{[\alpha \, \alpha+1]}$ and $-Z^-_{[\alpha \, \alpha+1]}$ with $\alpha=1,2$ are realized by adding a semi F1 string stretched between the D5$_\alpha$ and D5$_{\alpha+1}$ branes and ending on the D3 segments from above and from below respectively. The insertions of $Z^+_{[12]}$ and $-Z^-_{[23]}$ are shown in Figure \ref{T4branchIII}-c,d as examples.
Here we introduced some minus signs to obtain the correct ring relations without having to redefine the meson operators, which have conventional definitions in terms of elementary fields.

\begin{figure}[h!]
\centering
\includegraphics[scale=0.75]{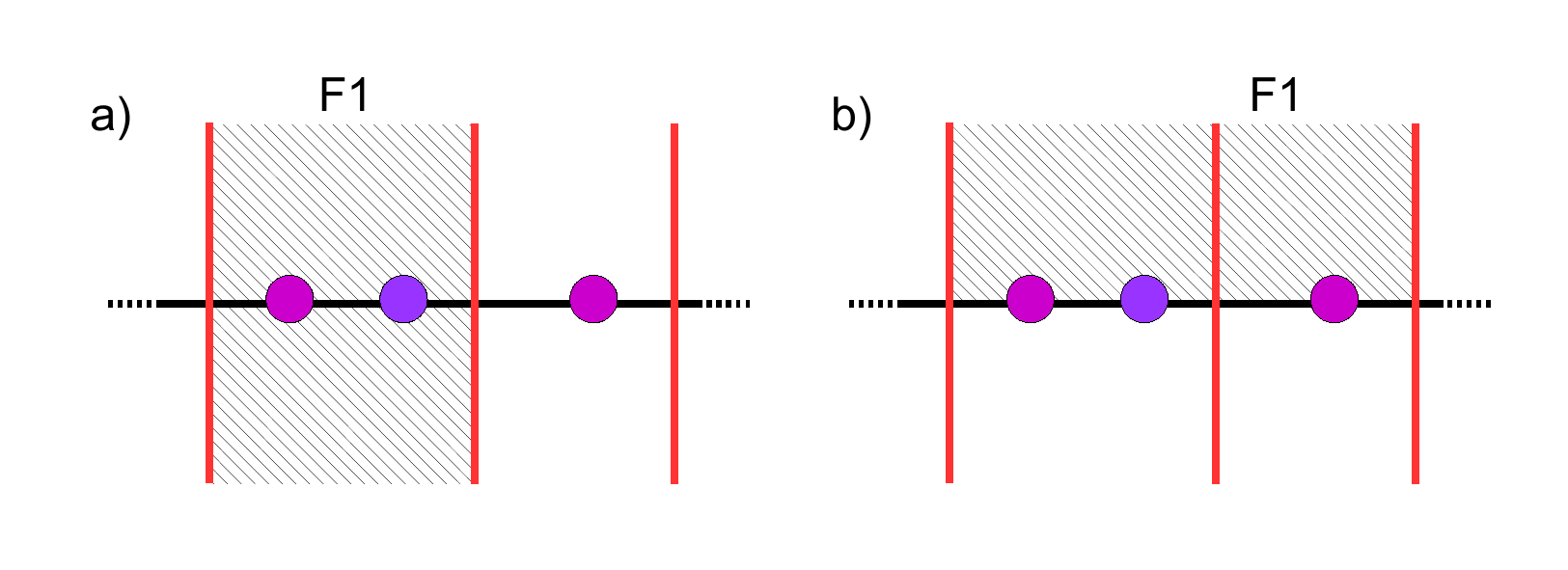} 
\vskip -1cm
\caption{\footnotesize Setups leading to the chiral ring relations: (a) $Z^+_{[12]} Z^-_{[12]} =  Z_1Z_2X_4X_5$, (b) $Z^+_{[12]} Z^+_{[23]} = Z^+_{[13]} Z_2$.}
\label{T4brIIIRel}
\end{figure}

This leaves us with the mesons $Z^\pm_{[\alpha \beta]}$ with $|\alpha-\beta | \le 2$, namely $Z^\pm_{[13]}$, for which we do not have brane setups inserting the operator alone. The relevant brane setups for these two operators have a semi F1 string stretched between the D5$_1$ and the D5$_3$, ending on the D3 segments from above and from below respectively. In the setup of Figure \ref{T4brIIIRel}-b we have the insertions of the fundamental scalars $Q_1$ and $\ti Q_3$ from the F1 ending on the D3-D5$_1$ corner and D3-D5$_3$ corners, and the insertions of the bifundamental scalars $\ti q_4$, $\ti q_5$ and $\ti q_6$ from the F1 crossing the 5$_{(1,k_1)2}$,5$_{(1,k_2)2}$ and 5$_{(1,k_1)3}$ branes, leading to the meson insertion $Z^+_{[13]} = Q_1\ti q_4 \ti q_5 \ti q_6 \ti Q_3$. However there is an extra contribution from the F1 crossing the D5$_2$ brane, which yields the extra insertion of $Q_2\ti Q_2 = Z_2$. The total insertion is therefore $Z^+_{[13]} Z_2$. This was already explained in \cite{Assel:2017hck}.
The same brane setup has also the interpretation of two semi F1 strings stretched between D5$_1$-D5$_2$ and D5$_2$-D5$_3$, inserting the product of mesons $Z^+_{[12]} Z^+_{[23]}$. Therefore we see that the setup where $Z^+_{[13]}$ appears is related to the chiral ring relation
\be 
Z^+_{[12]} Z^+_{[23]} = Z^+_{[13]} Z_2 \,,
\ee
which can be easily checked by replacing the mesons with their definition in terms of elementary Lagrangian fields.
Similarly the brane setup with the F1 ending on the D3 segments from below leads to the relation
\be
Z^-_{[12]} Z^-_{[23]} = Z^-_{[13]} Z_2 \,.
\ee
To complete the study we need to include the brane setups of the type presented in Figure \ref{T4brIIIRel}-a, which have a full F1 string stretched between a D5$_\alpha$ and D5$_{\alpha+1}$ branes. There are two brane setups, for $\alpha=1,2$, and two readings for each setup. As two semi F1 strings ending on D3 segments the insertion is $-Z^+_{[\alpha \, \alpha+1]}Z^-_{[\alpha \, \alpha+1]}$. As a full F1 string the insertion has a factor $-Z_{\alpha}Z_{\alpha+1}$ from the string ending on the D5$_{\alpha}$ from the left and on the D5$_{\alpha+1}$ from the right, and extra factors $X_j$ from the string crossing (only once) the 5branes-D3 intersections. They lead to the chiral ring relation $Z^+_{[12]} Z^-_{[12]} =Z_1 Z_2 X_4 X_5$ and $Z^+_{[23]} Z^-_{[23]} =Z_2 Z_3 X_6$. In total we obtain the four chiral ring relations
\be\ba
& Z^+_{[12]} Z^-_{[12]} =Z_1 Z_2 X_4 X_5 \,, \quad  Z^+_{[12]} Z^+_{[23]} = Z^+_{[13]} Z_2 \,, \cr
& Z^+_{[23]} Z^-_{[23]} =Z_2 Z_3 X_6 \,, \quad  Z^-_{[12]} Z^-_{[23]} = Z^-_{[13]} Z_2 \,.
\ea\ee
As explained in \cite{Assel:2017hck} these relations are not directly the full set of chiral ring relations of the Higgs branch, but corresponds instead to a minimal set of {\it pre-relations} from which the full set of chiral ring relations can be derived by the following recipe: one should include any relation that is obtained from the pre-relations by dividing by $Z_\alpha$ operators. In the case at hand we build
\be\ba
& Z^+_{[13]} Z_2 Z^-_{[23]}= Z^+_{[12]} Z^+_{[23]} Z^-_{[23]} = Z^+_{[12]} Z_2 Z_3 X_6 \quad \Rightarrow \quad  Z^+_{[13]} Z^-_{[23]} = Z^+_{[12]} Z_3 X_6 \,, \cr 
& Z^+_{[13]}  Z_2 Z^-_{[12]} = Z^+_{[12]} Z^+_{[23]}  Z^-_{[12]}  = Z^+_{[23]}  Z_1 Z_2 X_4 X_5 \quad \Rightarrow \quad  Z^+_{[13]} Z^-_{[12]} = Z^+_{[23]}  Z_1 X_4 X_5 \,, \cr
& Z^+_{[12]} Z_2 Z^-_{[13]}= Z^+_{[12]} Z^-_{[12]} Z^-_{[23]} = Z_1 Z_2 X_4 X_5 Z^-_{[23]} \quad \Rightarrow \quad  Z^+_{[12]} Z^-_{[13]} = Z^-_{[23]} Z_1 X_4 X_5 \,, \cr
& Z^+_{[23]} Z_2 Z^-_{[13]}= Z^+_{[23]} Z^-_{[12]} Z^-_{[23]} = Z_2 Z_3 X_6 Z^-_{[12]} \quad \Rightarrow \quad  Z^+_{[23]} Z^-_{[13]} =  Z^-_{[12]} Z_3 X_6 \,, \cr
& Z^+_{[13]} Z^-_{[13]}  (Z_2)^2= Z^+_{[23]} Z^+_{[12]} Z^-_{[12]} Z^-_{[23]} = Z_3 (Z_2)^2 Z_1 X_4 X_5 X_6   \ \Rightarrow \  Z^+_{[13]} Z^-_{[13]} = Z_1 Z_3 X_4 X_5 X_6 \,,
\ea\ee
where the left-hand-side relations follow from the pre-relations and the right-hand-side relations are obtained by dividing by $Z_2$. 
This leads to the full set of nine chiral ring relations. It is easy to check that in the gauge theory language these relations simply follow from the definition of the mesons in terms of elementary matter fields.
The important point is that the full set of relations can be worked out unambiguously from the pre-relations. The Higgs branch differs from  the algebraic space described by the pre-relations only by the lift of branches where a $Z_\alpha$ vanishes (where $Z_2$ vanishes in the current case). In the following we will only provide the generators and pre-relations to describe a given maximal branch.
 
The ring (pre)relations evaluated on the Higgs branch are thus
 \be\ba
& \underline{\scM_{III} \ \text{(pre-relations)} :} \cr
& Z^+_{[12]} Z^-_{[12]} = -(X_{III,1})^3 (X_{III,1} - X_{III,2}) \,, \quad  Z^+_{[23]} Z^-_{[23]} = (X_{III,1}-X_{III,2}) (X_{III,2})^2 \,, \cr
& Z^\pm_{[12]} Z^\pm_{[23]}= Z^\pm_{[1,3]}(X_{III,1}-X_{III,2}) \,.
\label{MIIIfinal}
\ea\ee
The maximal branch III has quaternionic dimension two, corresponding to having two D3 segments moving along the D5s. It admits two hyperk\"ahler subspaces of dimension one which can appear as factors $\scM^{(III)}_n$ of moduli space branches:\footnote{There is (at least) another dimension one hyperk\"ahler submanifold corresponding to the reconnection of the two D3 segments ($X_{III,1}=X_{III,2}$), but it does not corresponds to openning directions along other branches and therefore plays no role.}
\be\ba
 & \underline{\scM_{III,-1}:} \quad  Z^+_{[23]} Z^-_{[23]} = - (X_{III,2})^3  \qquad [ Z^\pm_{[12]} = Z^\pm_{[13]} = X_{III,1} =0 ]   \,, \cr
 & \underline{\scM_{III,-2}:} \quad  Z^+_{[12]} Z^-_{[12]} = - (X_{III,1})^4  \quad  [ Z^\pm_{[23]} = Z^\pm_{[13]} = X_{III,2} =0 ]  \,,
 \label{MIIIsubspaces}
\ea\ee
corresponding to having one or the other D3 segment fixed at the origin. 
\medskip

Finally we turn to the maximal branch IV. It is associated with the displacement of D3 segments between the four 5$_{(1,k_1)}$ branes, whose constrained motions translate into
\be\ba
& \underline{\scM_{IV}}: \cr
&  \varphi_2 = \varphi_3 = \varphi_4 = - \frac{X_2}{k_1}  = \frac{X_3}{k_2 -k_1}  \equiv X_{IV,1} \,, \cr
&   \varphi_5 = \varphi_6 = \frac{X_5}{k_2 -k_1} \equiv  X_{IV,2} \,, \quad   \varphi_7 \equiv  X_{IV,3} \,, \cr
& \varphi_1 = X_1 = X_4 = X_6 = Z^\alpha{}_\alpha = 0 \,.
\label{MIV}
\ea\ee
We repeat that all these constraints follows from reading the vevs of the operators from the setting with D3 segments displaced along 5$_{(1,k_1)}$ branes, with the vevs of $\varphi_i$ corresponding to the positions of the D3$_i$ segment in the complex plane $x^{5+i6}$, the vevs of $X_j$ corresponding to the distance (with sign) between the 5 brane and the reconnected D3 segment along $x^{8+i9}$ (see Figure \ref{XandVarphi}), and the vevs of $Z^\alpha{}_\alpha$ corresponds to the distance (with sign) between the two D3 segments breaking on the D5$_{\alpha}$ brane along $x^{8+i9}$.

\begin{figure}[h!]
\centering
\includegraphics[scale=0.75]{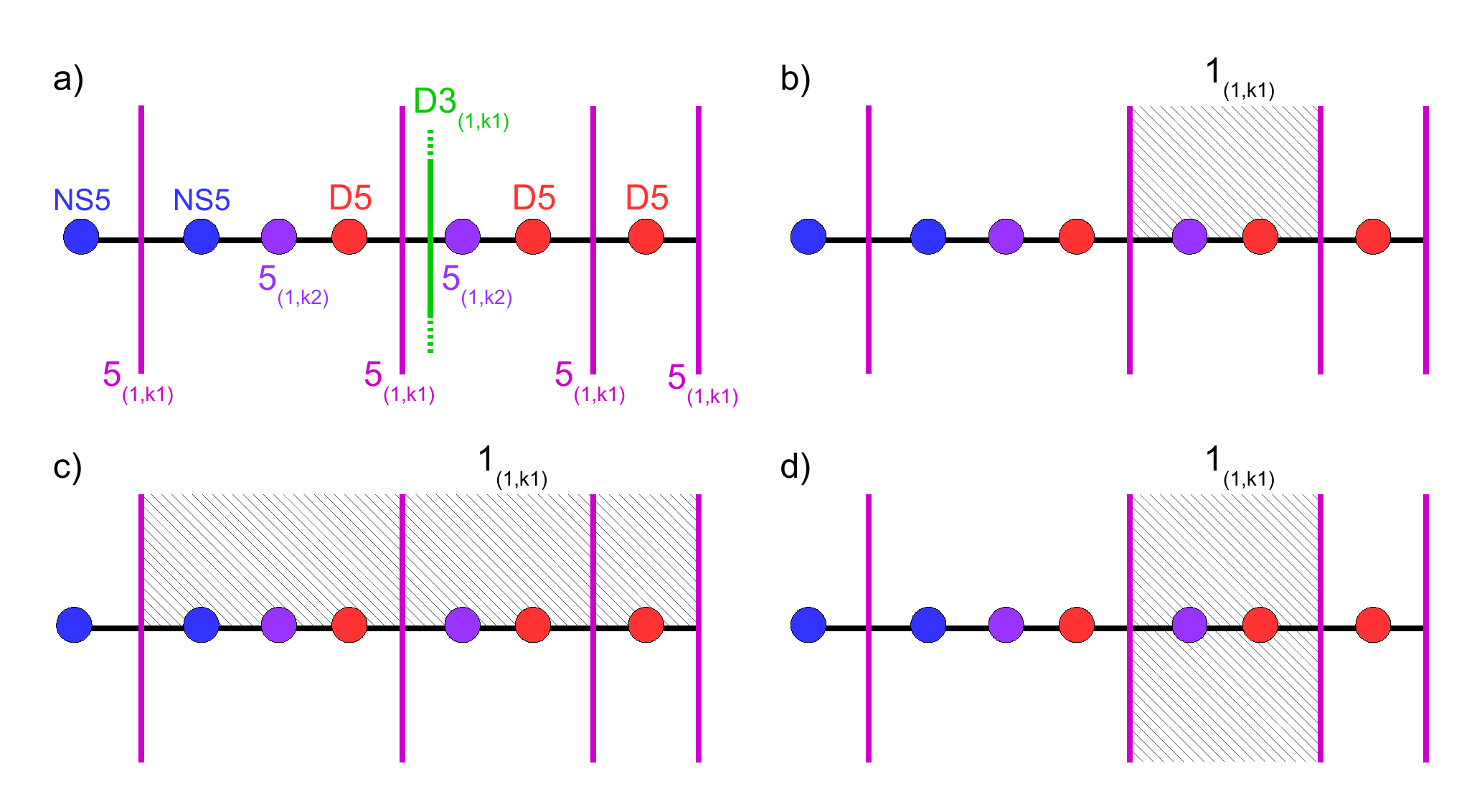} 
\vskip -0.5cm
\caption{\footnotesize Setups realizing the insertion of some branch IV operators: (a) $X_{IV,2}$, (b) $v^+_{IV,2} (\equiv V^+_{[2,3]})$, and setups for the chiral ring relations (c) $V^+_{[1,2]}V^+_{[2,3]}V^+_{[3,4]}=V^+_{[1,4]}(\varphi_4-\varphi_5)(\varphi_6-\varphi_7)$, (d) $V^+_{[2,3]}V^-_{[2,3]}=(\varphi_4-\varphi_5)(X_5)^{|k_1-k_2|}\varphi_6(\varphi_7-\varphi_6)$.}
\label{T4branchIV}
\end{figure}

The maximal branch $\scM_{IV}$ is parametrized by the $X_{IV,i}$ realized by brane setups with a D3$_{(1,k_1)}$ brane between the 5$_{(1,k_1)i}$ and 5$_{(1,k_1)i+1}$ branes, for $i=1,2,3$, as in Figure \ref{T4branchIV}-a for $i=2$, and some monopole operators. Part of these monopoles are the $v^\pm_{IV,i}$, $i=1,2,3$, realized by brane setups with a semi 1$_{(1,k_1)}$ string stretched between 5$_{(1,k_1)i}$ and 5$_{(1,k_1)i+1}$, ending on the D3 segment from above or from below, as in Figure \ref{T4branchIV}-b for $v^+_{IV,2}$. 
In general the monopole generators are associated with brane setups with a semi 1$_{(1,k_1)}$ string stretched between 5$_{(1,k_1)i}$ and 5$_{(1,k_1)j}$, with $i < j$, ending on the D3 segment from above or from below, and we denote them $V^\pm_{[i,j]}$. In particular we have $v^\pm_{IV,i} = V^\pm_{[i,i+1]}$. However, as in the study of the mesons for the Higgs branch, the brane setups for $V^\pm_{[i,j]}$, when $j > i +1$, come with extra insertions and are related to chiral ring relations, rather than inserting monopole operators alone.

The gauge theory definitions of the monopoles $V^\pm_{[i,j]}$ are read from the brane setups where they are inserted. For instance, for the monopole operators realized in Figure \ref{T4branchIV}-b,c, assuming $0 < k_1 < k_2$,
\be\ba
& V^+_{[2,3]} = \ol V_{0000++0} (q_5)^{k_2-k_1} \,, \cr
& V^+_{[1,4]} = \ol V_{0++++++}  (\ti q_2)^{k_1}(q_3)^{k_2-k_1}(q_5)^{k_2-k_1} \,.
\ea\ee
It can be checked that the monopole operators $V^\pm_{[i,j]}$ read from the brane setups match the gauge invariant monopole operators of smallest dimensions of the $T[4]$ theory that can take vevs on the maximal branch $\scM_{IV}$. We should also remark that the knowledge of the field theory definitions of the monopole operators is of no use in the method that we propose in this paper to describe the space of vacua. It simply allows us to make contact with the gauge theory language.

A subset of the chiral ring relations, which we call again pre-relations, are obtained from brane setups of two types. One type of setup has a semi 1$_{(1,k_1)}$ string stretched between the 5$_{(1,k_1)i}$ and 5$_{(1,k_1)j}$ branes with $j>i+1$, ending on the D3 segments from above or from below. One example is shown in Figure \ref{T4branchIV}-c, from which we read the chiral ring relation $V^+_{[1,2]}V^+_{[2,3]}V^+_{[3,4]}=V^+_{[1,4]}(\varphi_4-\varphi_5)(\varphi_6-\varphi_7)$. The insertion $V^+_{[1,2]}V^+_{[2,3]}V^+_{[3,4]}$ corresponds to interpreting the brane setup as three semi 1$_{(1,k_1)}$ strings, each inserting a monopole operator. The insertion $V^+_{[1,4]}(\varphi_4-\varphi_5)(\varphi_6-\varphi_7)$ corresponds to interpreting the setup as a single semi 1$_{(1,k_1)}$ string inserting the monopole $V^+_{[1,4]}$ and the contributions $(\varphi_4-\varphi_5)$ and $(\varphi_6-\varphi_7)$ from the semi 1$_{(1,k_1)}$ string crossing the 5$_{(1,k_1)2}$-D3 and 5$_{(1,k_1)3}$-D3 intersections. These last insertions are derived from the property that the insertion from a 5$_{(1,k_1)i}$-D3 intersection with a 1$_{(1,k_1)}$ crossing the 5$_{(1,k_1)2}$ and ending on the D3 segment from above (resp. from below) is the same as when the 1$_{(1,k_1)}$ crosses the D3 and ends on the  5$_{(1,k_1)2}$ from the right (resp. from the left), and is therefore equal to $\varphi_{i-1} - \varphi_i$ (resp. $\varphi_{i} - \varphi_{i-1}$). This property relies on  $SL(2,\bZ)$ string dualities relating this setup to the one with an F1 string and a D5 brane (see \cite{Assel:2017hck}).
We obtain $2.\binom{K}{2} - 2(K-1)=(K-1)(K-2)$ pre-relations, where $K$ is the number of 5$_{(1,k_1)}$ branes. Here $K=4$.

The second kind of brane setups to consider has a full 1$_{(1,k_1)}$ string stretched between consecutive 5$_{(1,k_1)i}$-5$_{(1,k_1)i+1}$ branes, as in Figure \ref{T4branchIV}-d for $i=2$. They lead to pre-relations of the form $V^+_{[i,i+1]}V^-_{[i,i+1]}=\cdots$. The setup of Figure \ref{T4branchIV}-d yields the relation $V^+_{[2,3]}V^-_{[2,3]}=(\varphi_4-\varphi_5)(X_5)^{|k_1-k_2|}\varphi_6(\varphi_7-\varphi_6)$. The insertion of $V^+_{[2,3]}V^-_{[2,3]}$ comes from the interpretation with two semi 1$_{(1,k_1)}$ strings. The insertion of $(\varphi_4-\varphi_5)(X_5)^{|k_1-k_2|}\varphi_6(\varphi_7-\varphi_6)$ comes from the interpretation as a single 1$_{(1,k_1)}$ string with the various factors coming from the 1$_{(1,k_1)}$-D3$_{4}$, 1$_{(1,k_1)}$-5$_{(1,k_2)}$, 1$_{(1,k_1)}$-D5 and 1$_{(1,k_1)}$-D3$_{7}$ modes in order. 
This leads to $K-1$ more pre-relations.

In total we obtain $(K-1)^2$ pre-relations, with here $K=4$. Explicitly we have 
\be\ba
&  V^\pm_{[1,2]}V^\pm_{[2,3]}=V^\pm_{[1,3]}(\varphi_4-\varphi_5) \,, \cr
& V^\pm_{[2,3]}V^\pm_{[3,4]}=V^\pm_{[2,4]}(\varphi_6-\varphi_7) \,, \cr
& V^\pm_{[1,2]}V^\pm_{[2,3]}V^\pm_{[3,4]}=V^\pm_{[1,4]}(\varphi_4-\varphi_5)(\varphi_6-\varphi_7) \,, \cr
& V^+_{[1,2]}V^-_{[1,2]}=(\varphi_1-\varphi_2)(X_2)^{|k_1|}(X_3)^{|k_1-k_2|}\varphi_4(\varphi_5-\varphi_4) \,, \cr
& V^+_{[2,3]}V^-_{[2,3]}=(\varphi_4-\varphi_5)(X_5)^{|k_1-k_2|}\varphi_6(\varphi_7-\varphi_6) \,, \cr
& V^+_{[3,4]}V^-_{[3,4]}=(\varphi_6-\varphi_7)\varphi_7 \,.
\ea\ee
It can be checked that the dimension of the monopole operators, extracted from \eqref{MonopDim}, are consistent with these relations.
 From these pre-relations one can obtain the full set of chiral ring relations by manipulating the pre-relations and allowing for division by $\varphi_i-\varphi_{i+1}$ operators, similarly to what we did for the Higgs branch. This generates many relations that we do not list here. All the information about the algebraic variety is already contained in the pre-relations.
 
From these chiral ring (pre-)relations one obtains the ring (pre-)relations on $\scM_{IV}$ by plugging in the constraints \eqref{MIV}. This yields (absorbing factors into operator redefinitions)
\be\ba
& \underline{\scM_{IV} \ \text{(pre-relations)} }:  \cr
& V^\pm_{[1,2]}V^\pm_{[2,3]}=V^\pm_{[1,3]}(X_{IV,1}-X_{IV,2}) \,, \cr
& V^\pm_{[2,3]}V^\pm_{[3,4]}=V^\pm_{[2,4]}(X_{IV,2}-X_{IV,3}) \,, \cr
& V^\pm_{[1,2]}V^\pm_{[2,3]}V^\pm_{[3,4]}=V^\pm_{[1,4]}(X_{IV,1}-X_{IV,2})(X_{IV,2}-X_{IV,3}) \,, \cr
& V^+_{[1,2]}V^-_{[1,2]}=(X_{IV,1})^{|k_1|+|k_1-k_2|+2}(X_{IV,2}-X_{IV,1}) \,, \cr
& V^+_{[2,3]}V^-_{[2,3]}=(X_{IV,2})^{|k_1-k_2|+1}(X_{IV,1}-X_{IV,2})(X_{IV,3}-X_{IV,2}) \,, \cr
& V^+_{[3,4]}V^-_{[3,4]}=(X_{IV,2}-X_{IV,3})X_{IV,3} \,.
\ea\ee
The (quaternionic) dimension of $\scM_{IV}$ is equal to the number of displaced D3 segments at a generic point, here dim$_{\bH}\scM_{IV}=3$. In general it is equal to $K-1$, where $K$ is the number of 5 branes of the type corresponding to the maximal branch.

The subspaces where components of other branches may open are associated with certain D3 segments being placed at the origin and correspondingly some vevs set to zero. The subspaces of codimension one and two are
 \be\ba
 & \scM_{IV,-i} \equiv   \scM_{IV} \cap \{ X_{IV,i} = 0 \,, \, V^\pm_{[j,k]} = 0 \, | \, j \le  i < k  \,  \}  \ , \, i=1,2,3 \,, \quad  \text{dim}_\bH =2 \,, \cr
 & \scM_{IV,-ij} \equiv   \scM_{IV,-i} \cap \scM_{IV,-j} \ , \, i\neq j \,, \quad  \text{dim}_\bH =1 \,.
 \label{MIVsubspaces}
\ea\ee
This completes the study of maximal branches. 
\medskip

We can now return to the claims in point \ref{Claims}. From the above analysis we find that the number of generators for a maximal branch of type $\ell$, $\scM_{\ell}$, is $K_\ell -1 + 2\binom{K_\ell}{2} =(K_\ell)^2 -1$, corresponding to the $X_{\ell,i}$ and the $V^\pm_{\ell,[i,j]}$, and $K_\ell$ is the number of 5 branes of type $\ell$. The total number of pre-relations is $K_\ell -1 + 2\binom{K_\ell}{2} - 2(K_\ell-1) = (K_\ell -1)^2$, where we added the pre-relations from full strings stretched between consecutive 5 branes and pre-relations from semi strings stretched between distant 5 branes. Moreover the dimension of $\scM_{\ell}$ is given by the number of displaced D3 segments, which, in abelian linear quiver theories, is simply $K_\ell -1$.
The complex dimension $2(K_\ell -1)$ matches the difference between the number of generators and the number of pre-relations: 
\be
\text{dim}_{\bC} \scM_\ell = 2( K_\ell -1) = (K_\ell)^2 -1 - (K_\ell -1)^2 = \#(\text{generators}) - \#(\text{pre-relations}) \,.
\ee

\medskip

From the knowledge of the maximal branches we are able to build all the branches $\scB_n$ of the moduli space. Each branch corresponds to a selection of factors $\scM^{(\ell)}_n$ associated with the motion of D3 segments along the 5 branes, leaving only unmovable D3 segments at the origin (stretched between 5 branes of different types). For instance, in the $T[4]$ theory we can move a D3 segment stretched between the two NS5s, a D3 segment stretched between the two 5$_{(1,k_2)}$s and a D3 segment stretched between the D5$_2$ and the D5$_3$ altogether, corresponding to a branch of vacua $\scM_{I} \times \scM_{II} \times \scM_{III,-1}$.
We find, by examination of the possible D3 motions, the following seven branches of the $T[4]$ theory, in no particular order,
\be\ba
& \scB_1 = \scM_{I} \times \scM_{II} \times \scM_{III,-1}  \,,  \cr
& \scB_2 = \scM_{I} \times \scM_{II} \times \scM_{IV,-12}  \,,  \cr
& \scB_3 = \scM_{I} \times \scM_{III}  \,,  \cr
& \scB_4 = \scM_{I} \times \scM_{III,-2} \times \scM_{IV,-12}  \,,  \cr
& \scB_5 = \scM_{I} \times \scM_{IV,-1}  \,,  \cr
& \scB_6 = \scM_{III,-1} \times \scM_{IV,-23}  \,,  \cr
& \scB_7 = \scM_{IV}  \,,  \cr
\ea\ee
where we suppressed factors which are single points (the origins of maximal branches).

The above analysis can be applied to any abelian linear quiver theory, leading to the algebraic description of its maximal branches and full moduli space of vacua.

\bigskip

We pause here to summarize what we have achieved so far. Using type IIB brane constructions we have provided an algorithm to describe the moduli space of vacua of any abelian linear quiver theory, where each factor is given by a set of generators and ring relations.  
The algorithm starts by identifying the maximal branches from the brane realization of the theory and studying them. For each maximal branch we have a set of generators related to a set of brane setups and we extract the ring pre-relations from another set of brane setups. The full set of ring relations can be deduced from the pre-relations. Subspaces of the maximal branches where transverse branches open are easily found. The description of the complete branches of vacua is then achieved by observing which products of factors are allowed by the brane configuration of the theory.
The splitting into various branches can be derived in the purely field theoretic approach from the F-term relations \ref{PotentialConstraints} and the constraints \ref{MonopConstraint2} on monopole operators\footnote{We leave as an exercise to the enthusiastic reader to check that the F-term constraints \eqref{PotentialConstraints} reproduce the splitting of the moduli space of vacua into the four branches studied for the $T[4]$ theory and to check that the monopoles operators entering into the branch descriptions are gauge invariant and of dimensions consistent with the ring relations that we derived.}, although it is much more quickly found from the brane approach.
The important information that one extracts from this brane approach is the knowledge of a minimal set of generators and of the quantum ring relations involving monopole operators, which are not given by the field theory analysis. Even though we spent some time explaining how to proceed, it must be noticed that, once the algorithm is known, the above recipe is very efficient in explicit cases, giving the algebraic spaces without actual computations.

\bigskip

To conclude this section we would like to contrast the results we have obtained for the general structure of $\N=3$ linear quivers with that of $\N=4$ linear quivers. Although we studied only abelian theories, we believe that the branch structure of the moduli space is the same for non-abelian linear quivers. This can be guessed from studying the motions of D3 segments in brane configurations realizing non-abelian linear quivers.
As explained in Section \ref{sec:Branes}, the restriction to $\N=4$ theories comes from considering only two types of 5-branes. This leads to having only two maximal branches $\scM_{\ell_1},\scM_{\ell_2}$. The moduli space is then of the form 
\be
\underline{\N=4 \ \text{theories}:} \quad 
\scM = \bigcup_{n=1}^B \, \scM^{(\ell_1)}_n \times \scM^{(\ell_2)}_n \,.
\label{NEqual4ModSp}
\ee
This is familiar for $\N=4$ Yang-Mills theories where the two maximal branches are the Coulomb branch and the Higgs branch. For $\N=4$ Chern-Simons theories the two branches are Coulomb-like, in the sense that the gauge group is only partially broken at a generic point, and they are both parametrized (in the UV Yang-Mills-Chern-Simons theory) by monopole operators satisfying quantum chiral ring relations.


\section{Abelian circular quivers}
\label{sec:CircQuiv}

In this section we extend our analysis to abelian circular quivers. When the sum of the Chern-Simons levels vanishes, we find that there is an additional branch, which we call {\it geometric branch}, following \cite{Benini:2009qs}, with different features than the maximal branches of linear quivers.
We first study the abelian ABJM theory, which is a special case with enhanced $\N=6$ supersymmetry and where this geometric branch is the full moduli space. We then study a second, more generic, example and we give our conclusions on the structure of the vacuum space of $\N=3$ circular quiver theories.

\subsection{Abelian ABJM}
\label{ssec:ABJM}

\begin{figure}[h!]
\centering
\includegraphics[scale=0.8]{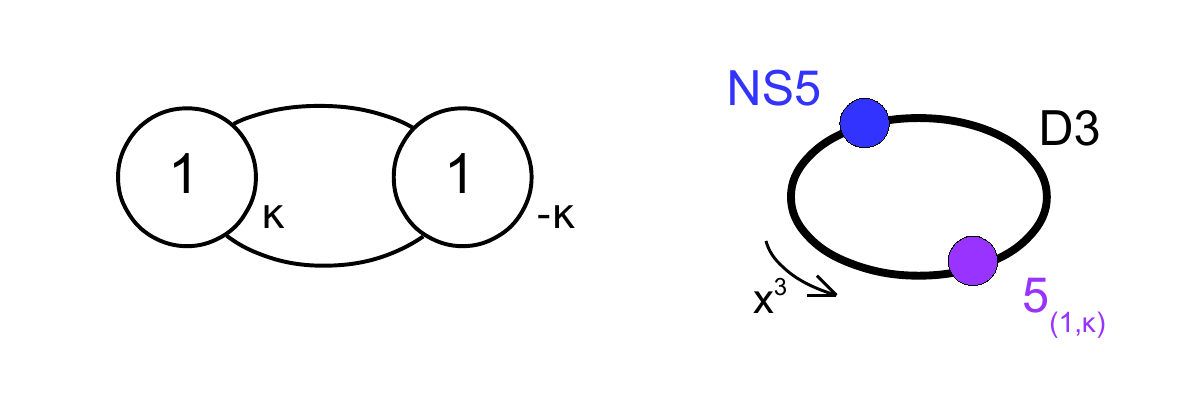} 
\vskip -0.5cm
\caption{\footnotesize Quiver and brane realization of the abelian ABJM theory at CS level $\kappa$.}
\label{ABJM}
\end{figure}

The abelian ABJM theory is a $U(1)_{\kappa}\times U(1)_{-\kappa}$ circular Chern-Simons quiver, with a specific superpotential engineered to preserve $\N=6$ supersymmetry \cite{Aharony:2008ug}. It arises in the infrared limit of the $U(1)_{\kappa}\times U(1)_{-\kappa}$ circular quiver theory with the standard $\N=3$ superpotential which is realized by the brane configuration of Figure \ref{ABJM} where the D3 brane wraps the compact $x^3$ direction. We assume $\kappa > 0$.

The abelian ABJM theory describes the low energy theory of a single M2 brane probing a $\bC^4/\bZ_\kappa$ orbifold singularity in M-theory. The moduli space of vacua is well known to be exactly $\bC^4/\bZ_\kappa$ parametrizing motions of the M2 brane in the transverse geometry. Let us see how this moduli space emerges from the brane analysis.

\begin{figure}[h!]
\centering
\includegraphics[scale=0.75]{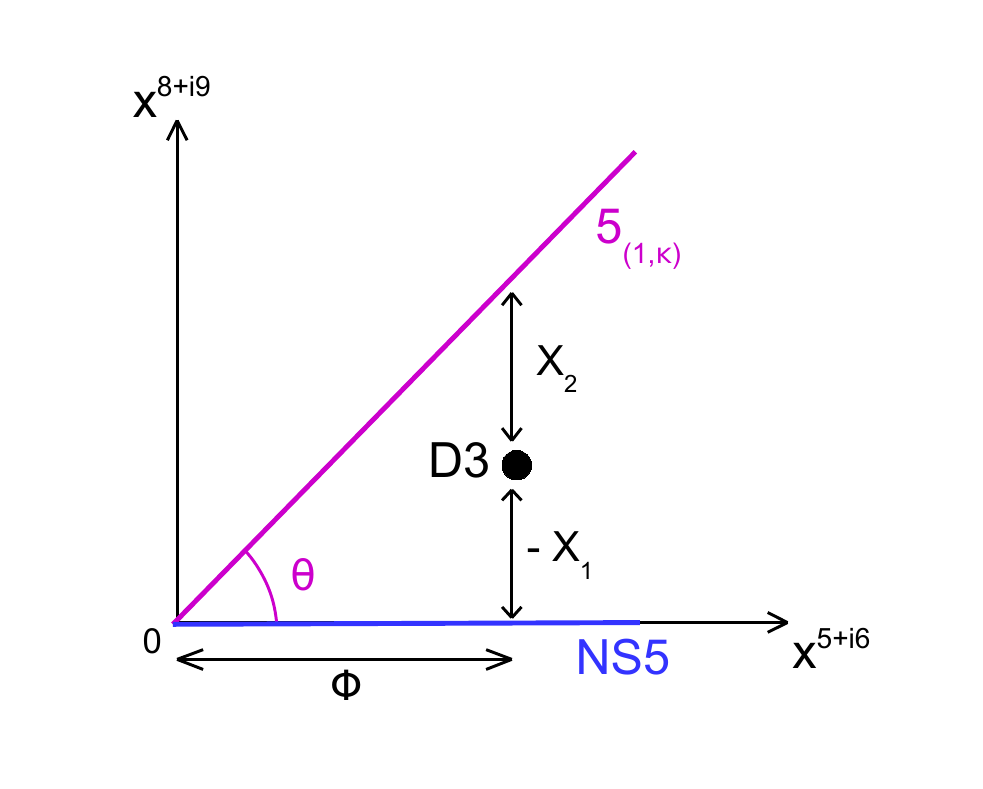} 
\vskip -0.5cm
\caption{\footnotesize In the (complexified) plane $x^{5+i6}-x^{8+i9}$ the reconnected D3 brane is a point. We read the constraint $X_2 - X_1 = \tan\theta \, \Phi = \kappa \Phi$ on the geometric branch.}
\label{ABJMXandVarphi}
\end{figure}

We denote $\varphi_{i=1,2}$ the complex vector multiplet scalars, $(q_j,\ti q_j)_{j=1,2}$ the two bifundamental hypermultiplets, and $X_j = q_j \ti q_j$, $Y=q_1 q_2$, $\ti Y=\ti q_1 \ti q_2$, the four chiral mesons. We also denote $\ol V_{n_1 n_2}$ the bare monopoles of magnetic charges $(n_1,n_2)$.
There is a single NS5 brane and a single 5$_{(1,\kappa)}$ brane, so, according to the discussion of the previous section, there is no branch of vacua associated to the motion of D3 segments along 5 branes. Instead the two D3 segment can reconnect across the $x^3$ circle and move freely in the transverse $x^{456789}$ directions, away from both 5-branes. This is the sign of an extra branch (compared to the case of linear quivers): the geometric branch $\scB_{\rm geom}$. 
The reconnection of the D3 segments translates into the constraints (see Figure \ref{ABJMXandVarphi})
\be
\underline{\scB_{\rm geom}:} \quad \varphi_1 = \varphi_2 \equiv \Phi \,, \quad X_2-X_1 = \kappa \Phi \,,
\ee
matching the F-term constraints in the field theory analysis.
The second relation can be used to solved for $\Phi$, leaving unconstrained $X_1,X_2$ meson operators.

\begin{figure}[h!]
\centering
\includegraphics[scale=0.75]{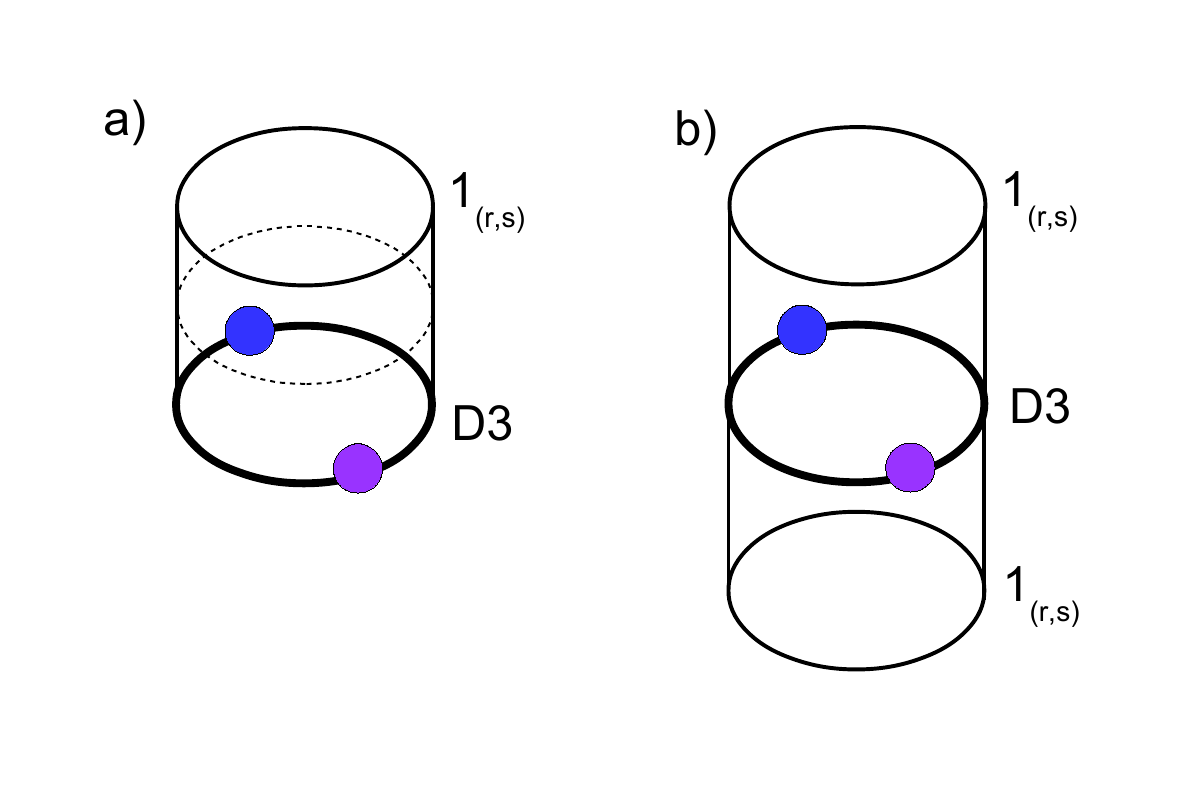} 
\vskip -1cm
\caption{\footnotesize a) Brane setup with a semi 1$_{r,s}$ string ending on the D3 from above, inserting the monopole operator $V^+_{(r,s)} $. b) Setup with a full 1$_{(r,s)}$ string leading to the relation $V^+_{(r,s)}V^-_{(r,s)}= (X_1)^{|s|}(X_2)^{|s-\kappa r|}$. The vertical direction is that spanned by the 1$_{(r,s)}$ string in the $x^{47}$ plane.}
\label{ABJMgeomBr}
\end{figure}

In addition there are monopole operators $V_{(r,s)}$ which are associated with new brane setups where a semi 1$_{(r,s)}$ string wraps the $x^3$ circle and ends on the reconnected D3 segments 
as in Figure \ref{ABJMgeomBr}-a. Importantly the half-string need not end on any 5 brane and therefore can be at any angle in the $x^{58}$ and $x^{69}$ planes compatible with supersymmetry. This means that $(r,s)$ can be any couple of coprime integers (see Section \ref{sec:Branes}).
Such a string carries $r$ units of D1 charge and therefore inserts the bare monopole $\ol V_{rr}$ which has magnetic charge $r$ for both gauge nodes. The matter dressing of the monopole arises from the intersections of the 1$_{(r,s)}$ string and the 5 branes. There are several cases to deal with. According to the rules found in previous sections, the 1$_{(r,s)}$-NS5 intersection yields the insertion of $(\ti q_1)^{s}$ if $s>0$ and $(q_1)^{-s}$ if $s<0$. The 1$_{(r,s)}$-5$_{(1,\kappa)}$ intersection yields the insertion of a factor $(\ti q_2)^{s-\kappa r}$ if $s>\kappa r$ and $(q_2)^{\kappa r -s}$ if $s<\kappa r$. This leaves us with two special cases when $(r,s)=(1,0)$ (D1 string) or $(r,s)=(1,\kappa)$, corresponding to the string being aligned with the NS5 or
the 5$_{(1,\kappa)}$ brane in the $x^{47}$ plane.
The rules of previous sections dictate that there is an extra insertion $\varphi_1-\varphi_2$ from the semi D1-NS5 or semi 1$_{(1,\kappa)}$-5$_{(1,\kappa)}$ intersection, which evaluates to zero on this geometric branch. However we also concluded that this extra factor is not part of the monopole operator inserted but rather multiplies it in chiral ring relations associated with the brane setup. Therefore the monopole operators $V^+_{(1,0)}$ and $V^+_{(1,\kappa)}$ related to these two brane setups are simply dressed with bifundamental scalars coming from one intersection with one 5 brane, while the other intersection yields no dressing factor.  This leads to the dictionary
\be
V_{(r,s)} = \left\lbrace
\begin{array}{cc}
\ol V_{rr}(\ti q_1)^{s}(\ti q_2)^{s-\kappa r} \,, & \quad    0 \le s \ \text{and} \ \kappa r \le s  \,, \\
\ol V_{rr}(\ti q_1)^{s}(q_2)^{\kappa r -s} \,, & \quad   0 \le s \le \kappa r \,, \\
\ol V_{rr}(q_1)^{-s}(\ti q_2)^{s-\kappa r} \,, & \quad   \kappa r \le s \le  0 \,, \\
\ol V_{rr}(q_1)^{-s}(q_2)^{\kappa r - s} \,, &  \quad   s \le 0 \ \text{and} \  s \le \kappa r  \,.
\end{array} \right.
\label{MonopRS}
\ee
The insertions of $V_{(r,s)}$ and $V_{(-r,-s)}$ are associated with half strings at opposite angles $\alpha$ and $\alpha+\pi$ (with $\tan\alpha=\frac{s}{r}$), which can recombine into a single 1$_{(r,s)}$ string. For $r \ge 0$, we denote $V^+_{(r,s)}=V_{(r,s)}$ and $V^-_{(r,s)}=V_{(-r,-s)}$. The set of monopoles is then spanned by the $V^\pm_{(r,s)}$ with $r \ge 0$. The monopole operators of zero magnetic charge inserted by semi F1 strings are nothing but the two chiral mesons $V^+_{(0,1)} = \ti Y$, $V^-_{(0,1)} = Y$.

\begin{figure}[h!]
\centering
\includegraphics[scale=0.75]{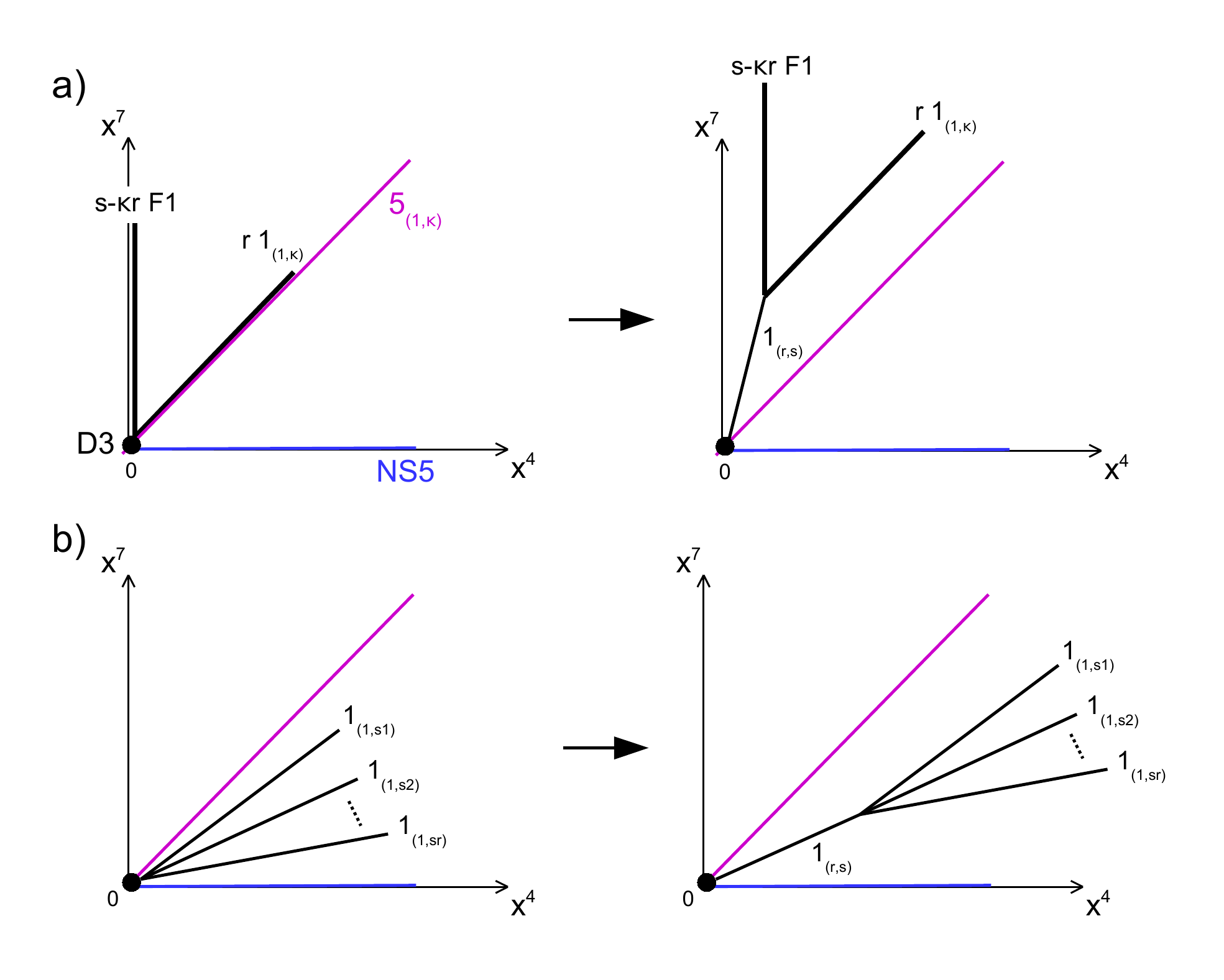} 
\vskip -1cm
\caption{\footnotesize a) When $\kappa r \le s$, a setup with $s-\kappa r$ F1 and $r$ 1$_{(1,\kappa)}$ strings can be deformed into a junction with a 1$_{(r,s)}$ string ending on the D3. b) When $0 \le s \le \kappa r$, a setup with 1$_{(1,s_i)}$ strings, $i=1,\cdots, r$, $0\le s_i \le \kappa$, can be deformed into a junction with a 1$_{(r,s)}$ string ending on the D3, with $s=\sum_i s_i$.}
\label{Junctions}
\end{figure}

There is however a large redundancy in these monopole operators $V^\pm_{(r,s)}$. From \eqref{MonopRS} we observe, for instance in the case $s \ge \kappa r $, the relation $V^+_{(r,s)} = \ol V_{rr} (\ti q_1)^{s}(\ti q_2)^{s-\kappa r} = \ol V_{rr} (\ti q_1)^{\kappa r + n} (\ti q_2)^n  \ti Y^{s-\kappa r -n}=V^+_{(r,n)} \ti Y^{s-\kappa r -n}$, for any $n$ such that $s-\kappa r -n >0$ and $(r,n)$ are coprime integers. This shows that the monopole operator inserted by a 1$_{(r,s)}$ string is the equal to the product of the operators inserted by a 1$_{(r,n)}$ string and $s-\kappa r -n$ F1 strings. This phenomenon has a simple interpretation in the brane picture: a semi 1$_{(r,n)}$ and  $s-\kappa r -n$ semi F1 strings ending on the D3-segment can be deformed to make a string junction with a single 1$_{(r,s)}$ string stretched between the junction and the D3 segment. The equality between the operator insertions indicates that the infrared theory is unaffected by the deformation of to the string junction. This readily generalizes to more complicated string junctions involving several types of strings. For $s \ge \kappa r$ we can consider the setup of Figure \ref{Junctions}-a  with $r$ 1$_{(1,\kappa)}$ strings and $s-\kappa r$ F1 strings ending on the D3s with different angles in the $x^{47}$ plane. They get deformed into a string junction with a single 1$_{(r,s)}$ string ending on the D3s, implying  the relation
\be
V^+_{(r,s)} = (V^+_{(1,\kappa)})^r \ti Y^{s-\kappa r} \,, \quad  \kappa r \le s  \,.
\ee
This can be traced back to the relations between bare monopoles $\ol V_{rr} = (\ol V_{++})^r$ (with $r>0$).
For $s \le 0$ the brane setup involves a junction between the 1$_{(r,s)}$ string, $r$ semi D1 strings and $-s$ semi F1 strings coming from below (along the $x^7$ direction) and leads to the relation
 \be
V^+_{(r,s)} = (V^+_{(1,0)})^r  Y^{-s} \,, \quad  s \le 0 \,.
\ee
For  $0 \le s \le \kappa r$ one must be careful to select brane setups where the various strings coming out of a junction do not intersect the 5 branes (so that the operator insertion comes only from the 1$_{(r,s)}$ string ending on the D3s). The best one can do is to consider a junction as in Figure \ref{Junctions}-b where the 1$_{(r,s)}$ string splits into a collection of $r$ 1$_{(1,s_i)}$ strings with $\sum_i s_i =s$ and $0 \le s_i \le \kappa$, leading to the relations
 \be
V^+_{(r,s)} = \prod_{i=1}^r V^+_{(1,s_i)} \,, \quad  0 \le s_i \le \kappa \,, \ \sum_{i=1}^r s_i = s \,,  \quad    0 \le s \le \kappa r  \,.
\ee
All these relations actually follow from simple relations between bare monopole operators $\ol V_{rr} = (\ol V_{++})^r$. Similar relations hold for the $V^-_{(r,s)}$ operators. 
The conclusion is that the operators $Y,\ti Y$ and $V^{\pm}_{(1,s)}$ , with $0 \le s \le \kappa$, form a basis of the monopole operators. In total the geometric branch is parametrized by the vevs of $X_1,X_2,Y,\ti Y$  and $V^{\pm}_{(1,s)}$, with $s=0, \cdots, \kappa$.

\medskip

In addition there are ring relations. The brane setups corresponding to the chiral ring relations are illustrated in Figure \ref{ABJMgeomBr}-b, with a full 1$_{(r,s)}$ string crossing the wrapped D3 brane. One interpretation is the insertion of the product $V^+_{(r,s)}V^-_{(r,s)}$. The second interpretation comes from integrating out the light 1$_{(r,s)}$-NS5 and 1$_{(r,s)}$-5$_{(1,\kappa)}$ modes. This yields the factors $(X_1)^{|s|}$ and $(X_2)^{|s-\kappa r|}$ respectively. We obtain the chiral ring relations
\be
V^+_{(r,s)}V^-_{(r,s)} = (X_1)^{|s|}(X_2)^{|s-\kappa r|} \,.
\ee
For $(r,s)=(0,1)$, this is nothing but the trivial relation $\ti Y Y = X_1X_2$ following from the definition of the mesons in terms of elementary fields.
We only need to consider relations between the vevs of operators in the basis $X_1,X_2,Y,\ti Y$  and $V^{\pm}_{(1,s)}$, $s=0, \cdots, \kappa$:
\be\ba
& \underline{\scM_{\rm geom}:} 
\quad  Y \ti Y = X_1 X_2 \,, \quad    V^+_{(1,s)}V^-_{(1,s)} = (X_1)^{s}(X_2)^{\kappa - s} \,, \quad s=0, \cdots , \kappa \,.
\ea\ee
These relations were already found in \cite{Cremonesi:2016nbo} from the study of Hilbert series. Here we recovered the relations from the brane approach.
They can be solved in terms of four complex variables $z_{i=1,2,3,4}$, with $Y=z_1 z_4$, $\ti Y = z_2 z_3$, $X_1 =z_1 z_3$, $X_2 = z_2 z_4$, $V^+_{(1,s)}= (z_1)^{s} (z_2)^{\kappa-s}$, $V^-_{(1,s)}= (z_3)^{s} (z_4)^{\kappa-s}$, subject to the identification $(z_1, z_2, \bar z_3, \bar z_4) \sim e^{\frac{2\pi i}{\kappa}} (z_1, z_2, \bar z_3, \bar z_4)$. This is nothing but the space $\bC^4/\bZ_{\kappa}$ as advertised.

\subsection{$T_{\rm circ}$ theory}
\label{ssec:Tcirc}

\begin{figure}[h!]
\centering
\includegraphics[scale=0.75]{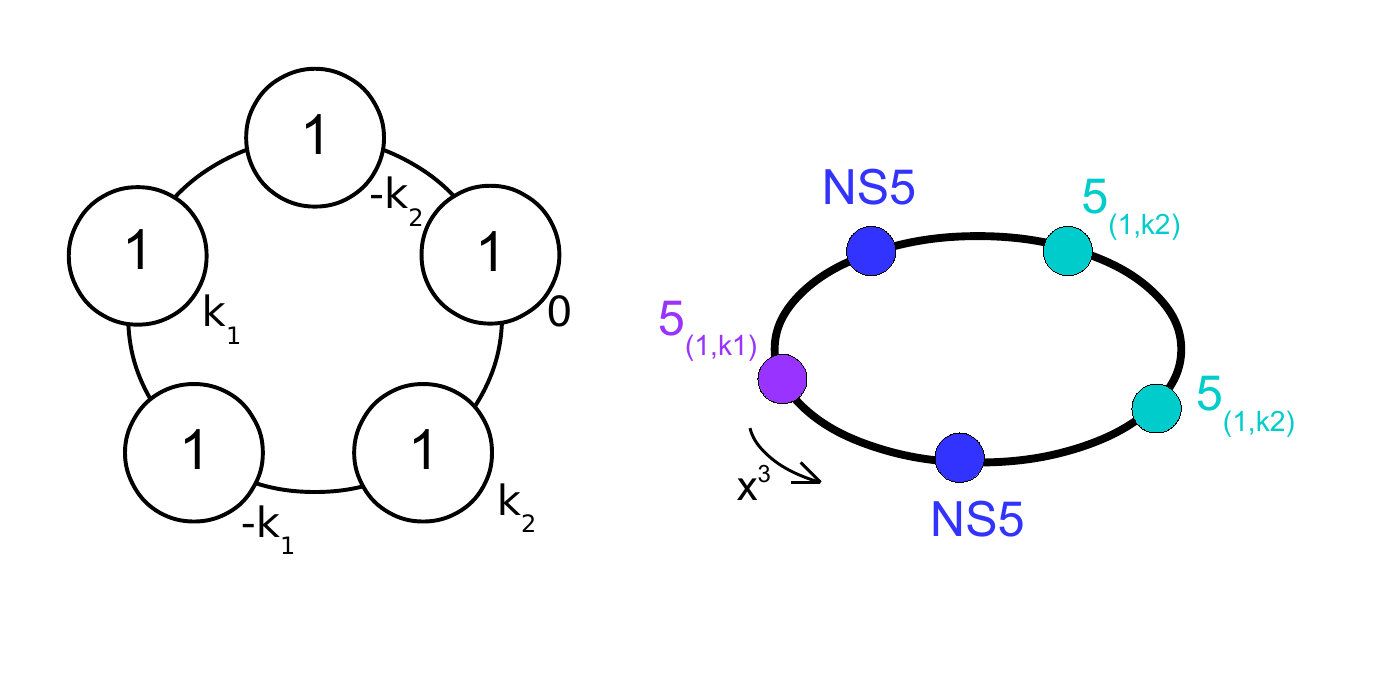} 
\vskip -1cm
\caption{\footnotesize Quiver and brane realization of the $T_{\rm circ}$ theory.}
\label{Tcirc}
\end{figure}

To explore further the properties of the space of vacua in circular quivers, we consider now a more generic theory, with only $\N=3$ supersymmetry and several branches of vacua. To keep the analysis simple enough we choose a theory whose brane realization has at most two 5 branes of a given type. We consider the abelian circular quiver theory $T_{\rm circ}$ with $U(1)_{k_1}\times U(1)_{-k_1}\times U(1)_{k_2} \times U(1)_0 \times U(1)_{-k_2}$ gauge group and no fundamental hypermultiplet. The quiver diagram and the brane realization are shown in Figure \ref{Tcirc}. We assume $k_2>k_1>0$. 

From the brane realization of $T_{\rm circ}$ we find that there are three maximal branches of vacua: a Coulomb-like branch corresponding to the motion of D3 segments along the NS5 branes, that we will call branch I or $\scM_I$, a Coulomb-like branch corresponding to the motion of D3 segments along the 5$_{(1,k_2)}$ branes, that we will call branch II or $\scM_{II}$, and the ``geometric" branch corresponding to the reconnection of all D3 segments with motions along the $x^{456789}$ directions, that we will call $\scM_{\rm geom}$. These maximal branches are (isomorphic to) actual branches of the $T_{\rm circ}$ theory. In addition there is a mixed branch $\scB_4$ corresponding the motion of one D3 segment between the NS5s and one D3 segment between the 5$_{(1,\kappa_2)}$.

We denote $\varphi_{i=1,\cdots,5}$ the vector multiplet scalars and  $(q^-_j,q^+_j)_{j=1,\cdots,5}$ the bifundamental scalars, with the mesons $X_j = q^+_j q^-_j$, ordered clockwise along the quiver diagram (or along the $x^3$ direction in the brane configuration) starting with $\varphi_1$ for the $U(1)_{k_1}$ node and $(q^-_1,q^+_1)$ for the $U(1)_{k_1}\times U(1)_{-k_1}$ bifundamental scalars. In addition to the $X_j$ there are two``long" meson operators $Y^\pm = \prod_{j=1}^5 q^\pm_j$.
The constraints on (non-monopole) chiral operators on each maximal branch is read from the constraints on D3 segment motions as in previous sections, giving
\be\ba
& \underline{\scM_I:} \cr
& \varphi_1 = \varphi_2 = \frac{X_1}{k_1} \equiv X_{I,1} \,, \quad \varphi_3 = \varphi_4 = \varphi_5 = \frac{X_3}{k_2} = \frac{X_4}{k_2} \equiv X_{I,2} \,, \quad X_2 = X_5 =0 \,. \cr
& \underline{\scM_{II}:} \cr
& \varphi_1 = \varphi_2 = \varphi_3 = \varphi_5 = \frac{X_1}{k_1-k_2} = -\frac{X_2}{k_2} = -\frac{X_5}{k_2} \equiv X_{II,1} \,, \quad \varphi_4 \equiv X_{II,2} \,, \quad X_3 = X_4 =0 \,. \cr
& \underline{\scM_{\rm geom}:} \cr
& \varphi_1 = \varphi_2 = \varphi_3 = \varphi_4 = \varphi_5 = \Phi \,, \cr
& X_2 = X_5 \equiv X_{III} \,, \quad X_1 = X_{III} + k_1 \Phi \,, \quad X_3 = X_4 = X_{III} + k_2 \Phi \,. 
\label{TcircBr0}
\ea\ee
We now analyse each branch in turn, providing less details than in previous sections.
\medskip

On the branch I there are six monopole operators
\be\ba
& v_{I,1}^{\pm} = \ol V_{\pm\pm 000} (q^\pm_1)^{k_1} \,, \quad  v_{I,2}^{\pm} = \ol V_{00\pm\pm\pm} (q^\pm_3)^{k_2}(q^\pm_4)^{k_2} \,, \cr 
& V^\pm_{(1,0)} = \ol V_{\pm\pm\pm\pm\pm}  (q^\pm_1)^{k_1} (q^\pm_3)^{k_2}(q^\pm_4)^{k_2} \,.
\ea\ee
\begin{figure}[h!]
\centering
\includegraphics[scale=0.75]{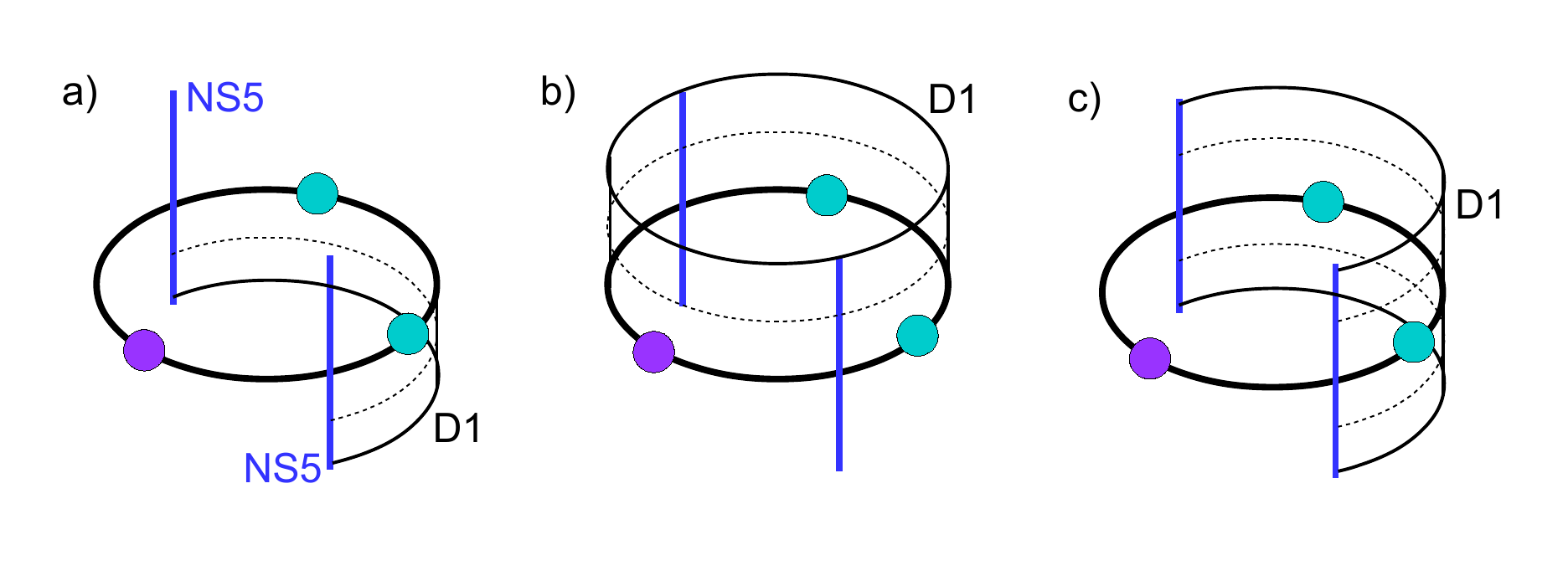} 
\vskip -1cm
\caption{\footnotesize a) Brane setup with a semi-infinite D1, inserting $v_{I,2}^-$. b) Brane setup with a semi-infinite D1 wrapping $x^3$, leading to the relation $v_{I,1}^{+}v_{I,2}^{+} = V^+_{(1,0)}(\varphi_2-\varphi_3)(\varphi_5-\varphi_1)$. c) Brane setup leading to the relation $v_{I,2}^{+}v_{I,2}^{-} = (\varphi_1-\varphi_5)(\varphi_2-\varphi_3)X_3^{k_2}X_4^{k_2}$. The vertical direction is $x^4$.}
\label{TcircBrI}
\end{figure}
The brane setups realizing the monopoles $v_{I,i}^{\pm}$ have a semi D1 string stretched from one NS5 to the other, ending on the D3 segments from above or from below. The setup for $v^-_{I,2}$ is shown in Figure \ref{TcircBrI}-a. 
The monopoles $V^\pm_{(1,0)}$ are related to brane setups with a semi D1 wrapping the whole $x^3$ circle, as in the example of Figure \ref{TcircBrI}-b. Such brane setups produce chiral ring relations involving $V^\pm_{(1,0)}$. Additional chiral ring relations are read from the two brane setups with a full D1 stretched from one NS5 to the other, as in the example of Figure \ref{TcircBrI}-c. We obtain the four chiral ring pre-relations
\be\ba
& v_{I,1}^{\pm}v_{I,2}^{\pm} = V^\pm_{(1,0)}(\varphi_2-\varphi_3)(\varphi_5-\varphi_1) \,, \cr
& v_{I,1}^{+}v_{I,1}^{-} = (\varphi_5-\varphi_1)(\varphi_3-\varphi_2)X_1^{k_1} \,, \cr
& v_{I,2}^{+}v_{I,2}^{-} = (\varphi_1-\varphi_5)(\varphi_2-\varphi_3)X_3^{k_2}X_4^{k_2} \,,
\ea\ee
leading to the algebraic ring pre-relations on $\scM_I$, with coefficients absorbed into operator redefinitions,
\be\ba
 \underline{\scM_I:} \quad  & v_{I,1}^{\pm}v_{I,2}^{\pm} = V^\pm_{(1,0)}(X_{I,1}-X_{I,2})^2 \,, \cr
& v_{I,1}^{+}v_{I,1}^{-} = (X_{I,1}-X_{I,2})^2 (X_{I,1})^{k_1} \,, \cr
& v_{I,2}^{+}v_{I,2}^{-} = (X_{I,1}-X_{I,2})^2 (X_{I,2})^{2 k_2} \,.
\ea\ee
\medskip

\begin{figure}[h!]
\centering
\includegraphics[scale=0.75]{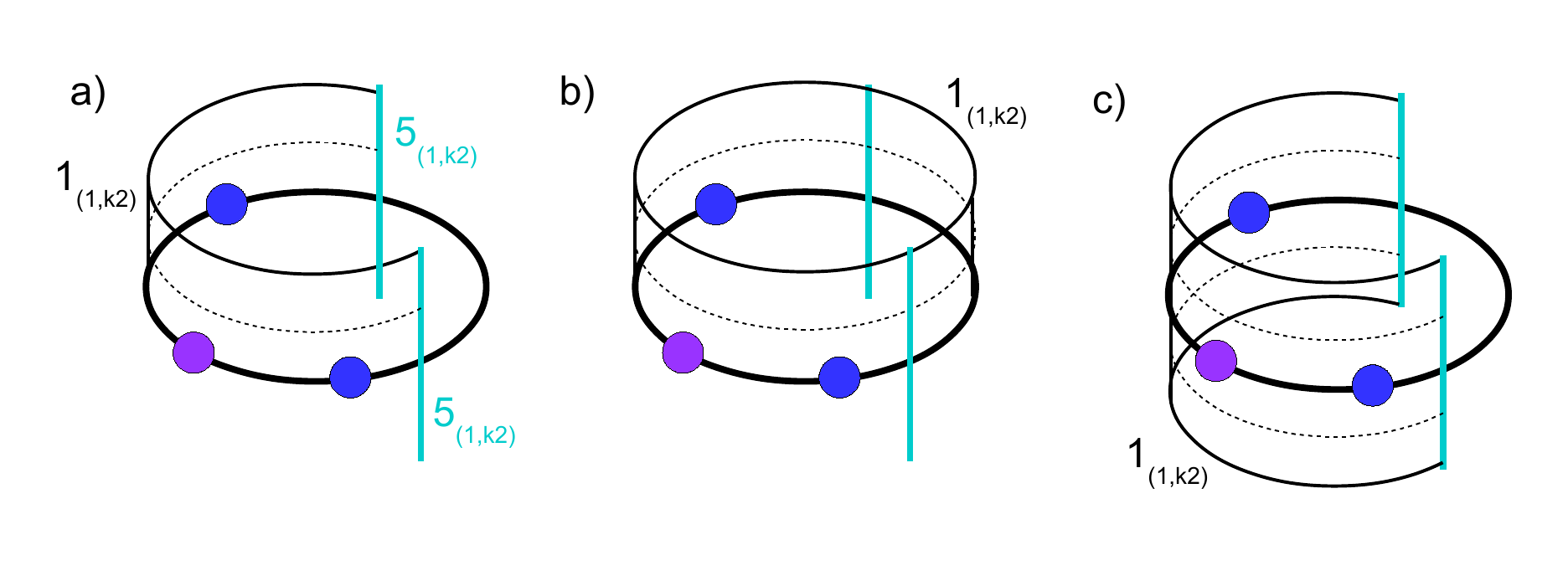} 
\vskip -1cm
\caption{\footnotesize a) Brane setup with a semi-infinite 1$_{(1,k_2)}$, inserting $v_{II,1}^+$. b) Brane setup leading to the relation $v_{II,1}^{+}v_{II,2}^{+} = V^+_{(1,k_2)}(\varphi_3-\varphi_4)(\varphi_4-\varphi_5)$. c) Brane setup leading to the relation $v_{II,1}^{+}v_{II,1}^{-} = (\varphi_4-\varphi_3)(\varphi_4-\varphi_5) X_1^{k_1-k_2}X_2^{k_2}X_5^{k_2}$. The vertical direction is that spanned by the 5$_{(1,k_2)}$ in $x^{47}$.}
\label{TcircBrII}
\end{figure}

On the branch II there are also six monopole operators
\be\ba
& v_{II,1}^{\pm} = \ol V_{\pm\pm\pm 0\pm} (q^\pm_1)^{k_1-k_2}(q^\mp_2)^{k_2}(q^\mp_5)^{k_2} \,, \quad  v_{II,2}^{\pm} = \ol V_{000\pm 0} \,, \cr 
& V^\pm_{(1,k_2)} = \ol V_{\pm\pm\pm\pm\pm} (q^\pm_1)^{k_1-k_2}(q^\mp_2)^{k_2}(q^\mp_5)^{k_2} \,.
\ea\ee
The brane setups realizing the monopoles $v_{II,i}^{\pm}$ have a semi 1$_{(1,k_2)}$ string stretched from one 5$_{(1,k_2)}$ to the other, ending on the D3 segments from above or from below. The setup for $v^+_{II,1}$ is shown in Figure \ref{TcircBrII}-a. 
The monopoles $V^\pm_{(1,k_2)}$ are related to brane setups with a semi 1$_{(1,k_2)}$ wrapping the whole $x^3$ circle, as in Figure \ref{TcircBrII}-b. Such brane setups produce chiral ring relations involving $V^\pm_{(1,k_2)}$. Additional chiral ring relations are read from the two brane setups with a full 1$_{(1,k_2)}$ stretched from one 5$_{(1,k_2)}$ to the other, as in Figure \ref{TcircBrII}-c. We obtain the four chiral ring pre-relations
\be\ba
& v_{II,1}^{\pm}v_{II,2}^{\pm} = V^\pm_{(1,k_2)}(\varphi_3-\varphi_4)(\varphi_4-\varphi_5) \,, \cr
& v_{II,1}^{+}v_{II,1}^{-} = (\varphi_4-\varphi_3)(\varphi_4-\varphi_5) X_1^{k_1-k_2}X_2^{k_2}X_5^{k_2} \,, \cr
& v_{II,2}^{+}v_{II,2}^{-} = (\varphi_3-\varphi_4)(\varphi_5-\varphi_4) \,,
\ea\ee
leading to the algebraic pre-relations on $\scM_{II}$, with coefficients absorbed into operator redefinitions,
\be\ba
 \underline{\scM_{II}:} \quad  
 & v_{II,1}^{\pm}v_{II,2}^{\pm} = V^\pm_{(1,k_2)}(X_{II,1}-X_{II,2})^2 \,, \cr
& v_{II,1}^{+}v_{II,1}^{-} = (X_{II,1}-X_{II,2})^2 (X_{II,1})^{k_1+k_2} \,, \cr
& v_{II,2}^{+}v_{II,2}^{-} = (X_{II,1}-X_{II,2})^2 \,.
\ea\ee
\medskip

\begin{figure}[h!]
\centering
\includegraphics[scale=0.75]{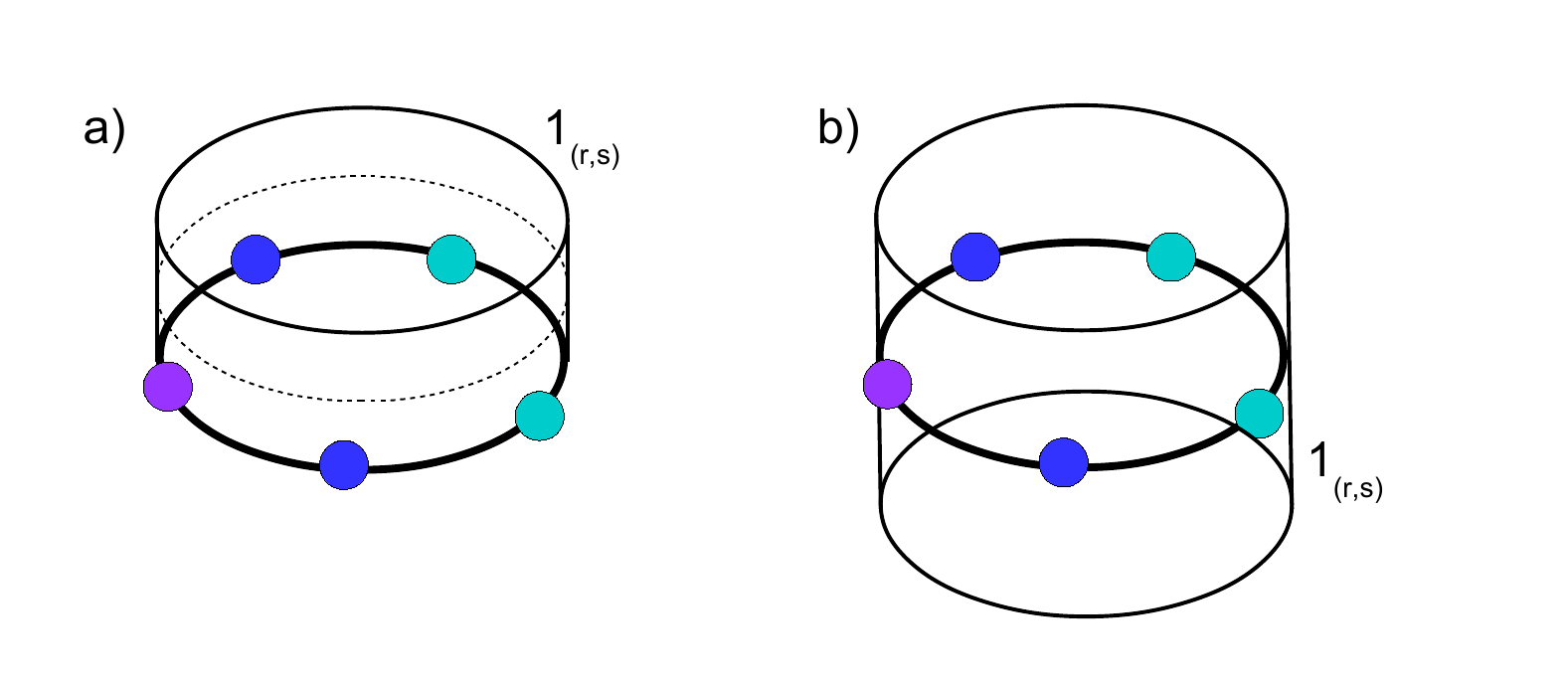} 
\vskip -0.5cm
\caption{\footnotesize a) Brane setup with a semi-infinite 1$_{(r,s)}$, inserting $V^+_{(r,s)}$. b) Brane setup leading to the relation $V^+_{(r,s)}V^-_{(r,s)} = (X_1)^{|s-k_1 r|} (X_2)^{|s|} (X_3)^{|s-k_2 r|} (X_4)^{|s-k_2 r|} (X_5)^{|s|}$. The vertical direction is that spanned by the 1$_{(r,s)}$ string in the plane $x^{47}$.}
\label{TcircBrGeom}
\end{figure}

On the geometric branch there are monopole operators $V^\pm_{(r,s)}$ as in the ABJM theory studied before. They are labeled by two coprime integers $(r,s)$ with $r>0$. For $s\ge k_1 r \, (> k_2 r)$, we have
\be
V^\pm_{(r,s)} =  \ol V_{\pm r \pm r \pm r \pm r \pm r} (q^\pm_1)^{s-k_1 r} (q^\pm_2)^{s} (q^\pm_3)^{s-k_2 r} (q^\pm_4)^{s-k_2 r} (q^\pm_5)^{s}   \,.
\ee
If  $s < k_1 r$, the factor $(q^\pm_1)^{s-k_1 r}$ must be replaced by $(q^\mp_1)^{k_1 r-s}$, and similar replacements for the other factors when their exponent becomes negative. These operators are realized by brane setups with a semi infinite 1$_{(r,s)}$ string wrapping the $x^3$ circle and ending on the D3 segments, as in Figure \ref{TcircBrGeom}-a, and are subject to the chiral ring relations
\be
V^+_{(r,s)}V^-_{(r,s)} = (X_1)^{|s-k_1 r|} (X_2)^{|s|} (X_3)^{|s-k_2 r|} (X_4)^{|s-k_2 r|} (X_5)^{|s|} \,,
\ee
which follow from the brane setups with a full 1$_{(r,s)}$ string wrapping the $x^3$ circle as shown in Figure \ref{TcircBrGeom}-b.
The zero-charge monopoles inserted by semi F1 strings wrapping the $x^3$ circle are nothing but the long mesons $Y^{\pm} = V^\pm_{(0,1)}$. The above chiral ring relation $Y^+ Y^- = X_1X_2X_3X_4X_5$ follows from the definition of the mesons in terms of elementary fields.
Notice that the operators $V^\pm_{(1,0)}$ and $V^\pm_{(1,k_2)}$ are also part of the branch I and II monopoles respectively, meaning that they can take vevs on two branches. 

\begin{figure}[h!]
\centering
\includegraphics[scale=0.75]{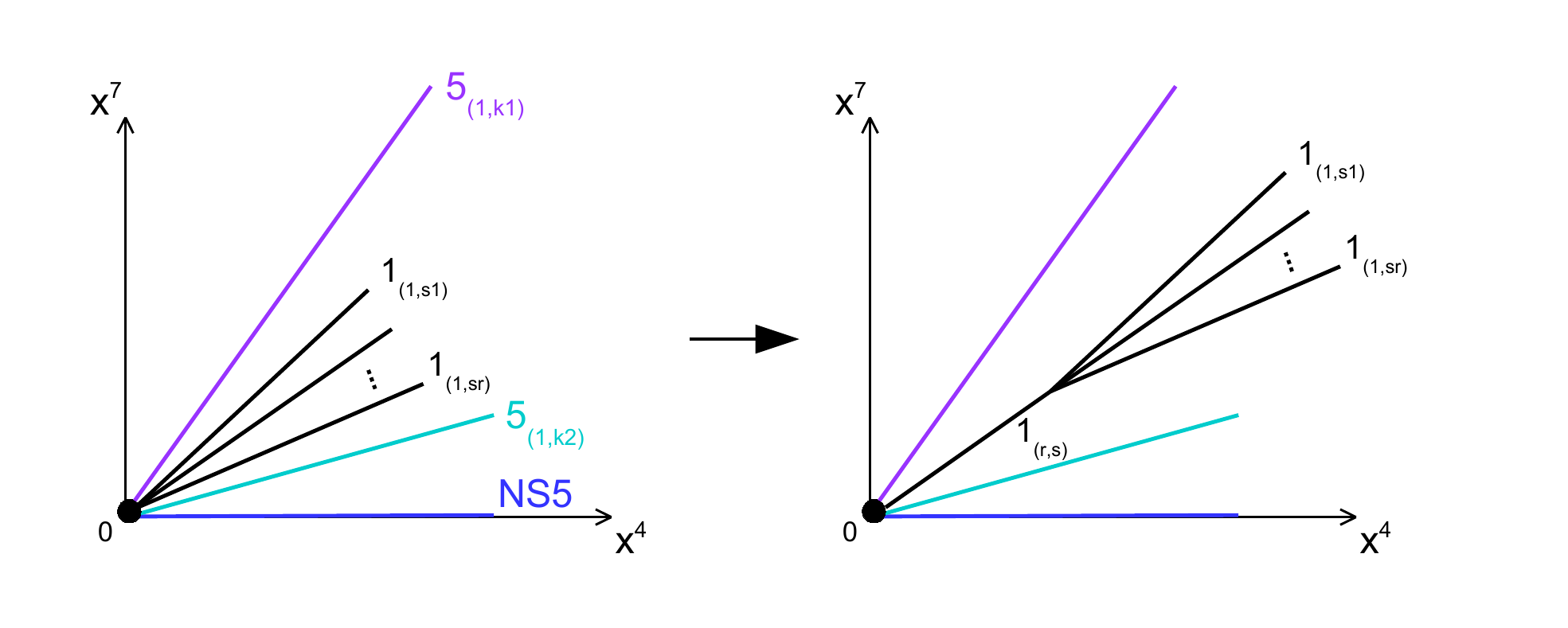} 
\vskip -0.5cm
\caption{\footnotesize  When $k_2 r \le s \le k_1 r$,  a setup with 1$_{(1,s_i)}$ strings, $i=1,\cdots, r$, $k_1\le s_i \le k_2$, can be deformed into a junction with a 1$_{(r,s)}$ string ending on the D3, with $s=\sum_i s_i$.}
\label{Junctions2}
\end{figure}

As in the ABJM theory there are more relations allowing to solve for many $V^\pm_{(r,s)}$ in terms of a smaller basis of monopoles, however the analysis is slightly complicated by the presence of more 5 branes of different types.
There are four cases to deal with: $ k_1 r \le s$, $k_2 r \le s \le k_1 r$, $0 \le s \le  k_2 r$ and $s \le 0$, corresponding to the four positions of the 1$_{(r,s)}$ string relative to the 5 branes in the $x^{47}$ planes. 
When $k_1 r \le s$ the 1$_{(r,s)}$ string can arise from a junction with $r$ 1$_{1,k_1}$ strings and $s-k_1 r$ F1 strings, as in Figure \ref{Junctions}-a with $\kappa=k_1$, leading to the relations
\be
V^\pm_{(r,s)} = (V^\pm_{(1,k_1)})^r (Y^{\pm})^{s-k_1 r} \,, \quad  k_1 r \le s  \,.
\ee
For $k_2 r \le s \le k_1 r$ the 1$_{(r,s)}$ string can arise from a junction with a collection of $r$ 1$_{(1,s_i)}$ strings with $k_2 \le s_i \le k_1$ and $\sum_{i=1}^r s_i =s$, as shown in Figure \ref{Junctions2}, leading to the relations
\be
V^\pm_{(r,s)} = \prod_{i=1}^r V^\pm_{(1,s_i)} \,, \quad k_2 \le s_i \le k_1\,, \ \sum_{i=1}^r s_i =s \,, \quad  k_2 r \le s \le k_1 r \,.
\ee
For $0 \le s \le  k_2 r$ the 1$_{(r,s)}$ string can arise from a junction with a collection of $r$ 1$_{(1,s_i)}$ strings with $0 \le s_i \le k_2$ and $\sum_{i=1}^r s_i =s$, as in Figure \ref{Junctions}-a with $\kappa=k_2$, leading to the relations
\be
V^\pm_{(r,s)} = \prod_{i=1}^r V^\pm_{(1,s_i)} \,, \quad 0 \le s_i \le k_2 \,, \ \sum_{i=1}^r s_i =s \,, \quad  0 \le s \le k_2 r \,.
\ee
For $s \le 0$ the 1$_{(r,s)}$ string can arise from a junction with $r$ D1 strings and $s$ F1 strings leading to the relations
\be
V^\pm_{(r,s)} = (V^\pm_{(1,0)})^r (Y^{\pm})^{s} \,, \quad  s \le 0  \,.
\ee
This leaves us with a basis of monopole operators $V^\pm_{(1,s)}$ with $s=0, \cdots, k_1$, together with the mesons $Y^{\pm}$, $\Phi$, $X_{III}$, to parametrize the geometric branch. Using \eqref{TcircBr0} the algebraic ring relations for this basis read
\be\ba
& \underline{\scM_{\rm geom}:} \cr
& Y^+Y^- =  (X_{III}+k_1\Phi) X_{III}^2 (X_{III}+k_2\Phi)^2 \,, \cr
& V^+_{(1,s)}V^-_{(1,s)} = (X_{III}+k_1\Phi)^{k_1-s} (X_{III})^{2s} (X_{III}+k_2\Phi)^{2|s-k_2|} \,, \quad s=0,\cdots, k_2, \cdots, k_1\,.
\ea\ee

We have described the three maximal branches (which are actual branches of the theory). The ``mixed" branch $\scB_4$ is the direct product of subspaces of the branch I and II:
\be\ba
& \scB_4 = \scM_{I,-2} \times \scM_{II,-1} \,, \cr
& \scM_{I,-2} = \{ X_{I,2}= v^\pm_{I,2} = V^\pm_{(1,0)} = 0 \} \cap \scM_{I} \,, \cr
& \scM_{II,-1} = \{ X_{II,1}= v^\pm_{II,1} = V^\pm_{(1,k_2)} = 0 \} \cap \scM_{II} \,.
\ea\ee
This completes the study of the moduli space of vacua of the $T_{\rm circ}$ theory.
\bigskip

The study of the $T_{circ}$ theory generalizes easily to any abelian circular quiver theory and allows us to draw some conclusions about the structure of their vacuum space. In essence they are much like the vacuum space of linear quiver theories, with branches $\scB_n=\prod_{\ell=1}^L \scM_{n}^{(\ell)}$, where $\scM_{n}^{(\ell)}$ are hyperk\"ahler cones labeled by a type $\ell$ of 5 brane. The factors $\scM_{n}^{(\ell)}$ at fixed $\ell$ obey an inclusion relation with a larger space $\scM_{\ell}$, which is the maximal branch associated to the 5 brane of type $\ell$. These maximal branches are Coulomb like branches, parametrized by the vevs of dressed chiral monopoles, and possibly a Higgs branch parametrized by the vevs of chiral mesons. The only difference with the linear quivers arises in the special situation when the sum of all Chern-Simons levels is vanishing.  Then there is an extra maximal branch $\scM_{\rm geom}$ -- the geometric branch -- parametrized by monopole operators with equal magnetic charges in all $U(1)$ gauge nodes and some mesonic operators. In abelian theories the geometric branch does not mix with other maximal branches, but we expect this to occur in non-abelian theories. The geometric branch is supposed to match the transverse geometry probed by M2 branes, realizing the $\N=3$ CS theory on their worlvolume.

Notice that the brane configuration for circlar quivers, with D3 branes wrapping the $x^3$ circle and crossing a sequence of 5 branes, automatically encodes the necessary condition on the CS levels $\kappa_i$ for the existence of a geometric branch $\sum_{i} \kappa_i =0$. If this condition is not realized in the gauge theory the bare monopoles with equal magnetic charges in all nodes have a non-vanishing (electric) gauge charge which cannot be canceled by a matter dressing, namely they do not satisfy \eqref{MonopConstraint1}. Therefore there is no such monopole in the theory and consequently no geometric branch.\footnote{We are not aware of the existence of any holographic dual for such theories.} Such quivers can still be realized in type IIB string theory by adding an extra twist by a spacial rotation in the three planes $x^{47},x^{58},x^{69}$ as we compactify the $x^3$ direction. In that case the configurations with 1$_{(r,s)}$ strings wrapping the $x^3$ circle cease to exist and the geometric branch is lifted, in agreement with the field theory analysis. The other branches are unaffected by this twist.
\medskip

From the brane construction we expect these features to hold in non-abelian theories as well (see also \cite{Gaiotto:2009tk,Benini:2009qs,Jafferis:2009th}), however the derivation of the chiral ring basis and relations is a harder task that we postpone to the future.


\section{Theories realized with 5$_{(p,q)}$ branes}
\label{sec:5pq}

To conclude this discussion we would like to present a few results on an extension of this work to quiver theories realized with general 5$_{(p,q)}$ branes. From the point of view of the brane analysis, there is no essential difference compared to theories realized with 5$_{(1,k)}$ branes. The difficulty comes from the gauge theory interpretation, since, at least in the non-abelian case, the corresponding theories do not have known Lagrangian formulation. In \cite{Gaiotto:2008ak} a description of these theories was proposed in terms of quivers involving gluings with the $T[U(N)]$ SCFT by gauging its $U(N)\times U(N)$ global symmetry with $U(N)$ nodes of the quiver chain.\footnote{See also \cite{Assel:2014awa} for tests of this construction with exact 3-sphere partition functions.}  
The brane approach may prove useful to studying the vacuum space of such non-Lagrangian non-abelian theories.
Here we will focus on abelian theories, which do admit Lagrangian realizations. In that case the $T[U(1)]$ theory is simply made of an $\N=4$ BF coupling between two background $U(1)$ gauge multiplets \cite{Kapustin:1999ha}, which can be gauged with $U(1)$ nodes of the quiver. We will analyse the vacuum space of such theories from branes in two examples.

\begin{figure}[h!]
\centering
\includegraphics[scale=0.75]{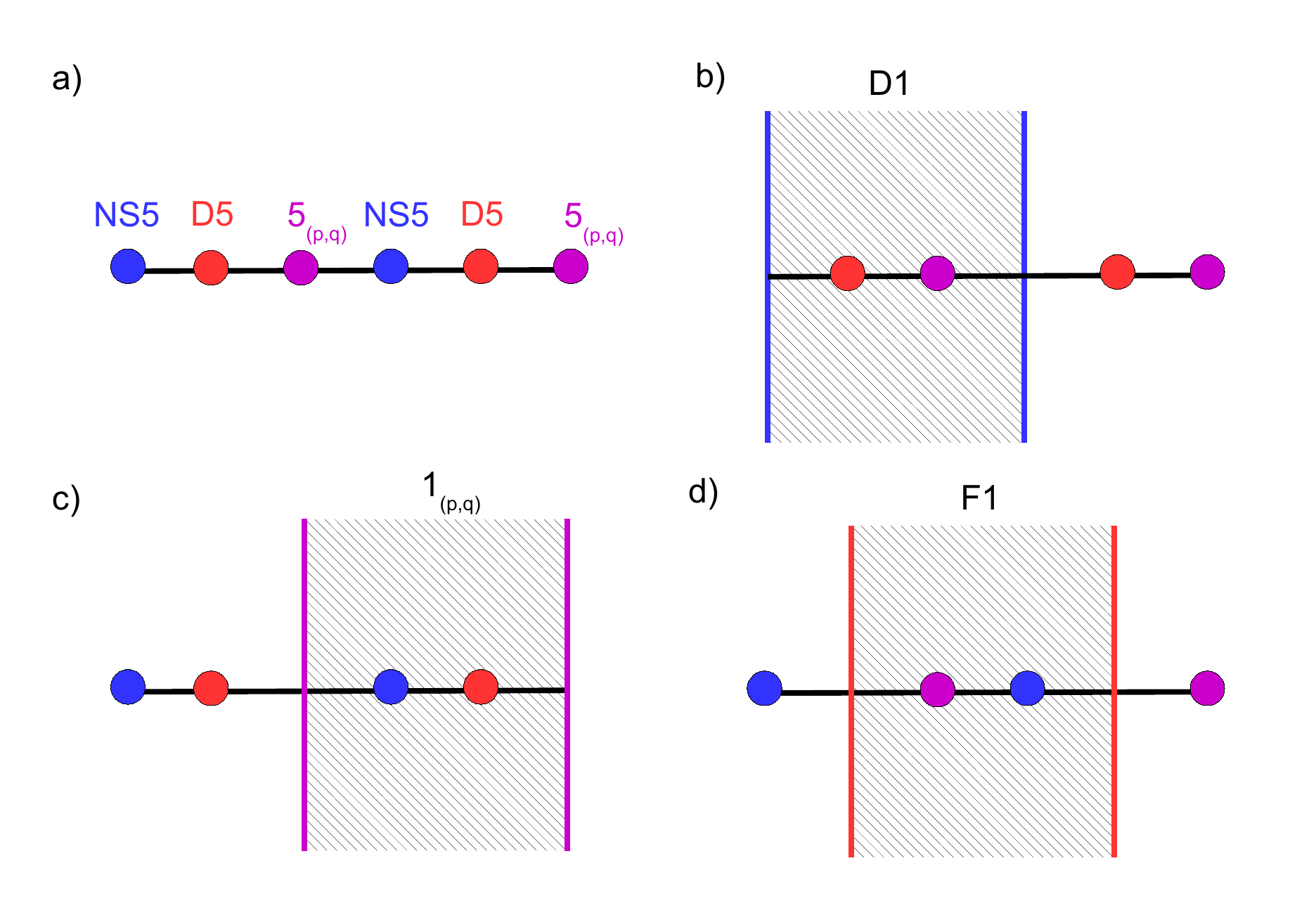} 
\vskip -0.5cm
\caption{\footnotesize  Brane realization of the $T_{(p,q)}[3]$ (a) and brane setups for the ring relations on the branch I (b), branch II (c) and branch III (d).}
\label{T3pq}
\end{figure}

We first consider the linear quiver theory realized by the brane setup of Figure \ref{T3pq}-a, with a sequence of 5 branes: NS5-D5-5$_{(p,q)}$-NS5-D5-5$_{(p,q)}$, where $p>0$, $q \neq 0$ and $p,q$ are coprime integers. This is a generalization of the $T[3]$ theory studied in Section \ref{ssec:T3} obtained by replacing 5$_{(1,\kappa)}$ branes by 5$_{(p,q)}$ branes and we call this theory $T_{(p,q)}[3]$. We refer to \cite{Gaiotto:2008ak} for the explicit (rather complicated) description of the linear quiver theory corresponding to this brane setup. The knowledge of the explicit quiver theory is inessential to the derivation of the moduli space of vacua from the brane analysis. Let us simply mention that the part of the quiver associated to the 5$_{(p,q)}$ brane is a sequence of $U(1)$ nodes with Chern-Simons couplings linked either by bifundamental matter or by $\N=4$ BF terms (mixed CS coupling).

The analysis of the space of vacua from the brane picture is essentially identical to that of the $T[3]$ theory, so we will not repeat it.
There are three maximal branches, corresponding to the three branches of vacua. The branches I, II and III are associated to the motion of D3 segments along the NS5s, 5$_{(p,q)}$s and D5s respectively. Each branch has a basis of three chiral operators, $v^\pm_I, X_I$ for the branch I, $v^\pm_{II}, X_{II}$ for the branch II, and $\pm Z^\pm \cong v^\pm_{III}, X_{III}$ for the branch III, where we have renamed $\pm Z^\pm$ of the into $v^\pm_{III}$ to emphasize the fact that in the $T_{(p,q)}[3]$ theory these operators have monopole charges generically. The brane setups inserting these operators are as in the $T[3]$ theory with 1$_{(1,\kappa)}$ strings replaced by 1$_{(p,q)}$ strings. The difference with respect to the $T[3]$ theory arises in the ring relations, which are read from the setups presented in Figure \ref{T3pq}-b,c,d. The D1 string intersects the 5$_{(p,q)}$ brane $|q|$ times, the 1$_{(p,q)}$ string intersects the NS5 brane $|q|$ times and the D5 brane $p$ times, and the F1 string intersects the 5$_{(p,q)}$ brane $p$ times, leading to the modified relations (with respect to \eqref{T3BranchesFinal})
\be\ba
& \text{branch I:} \quad v^+_I v^-_I = (X_I)^{|q|+2} \,, \cr
& \text{branch II:} \quad v^+_{II} v^-_{II} = (X_{II})^{p+|q|+1} \,, \cr
& \text{branch III:} \quad v^+_{III} v^-_{III}= (X_{III})^{p+3} \,. 
\label{T3pqBranches}
\ea\ee
The vacuum space of the $T_{(p,q)}[3]$ theory is therefore $(\bC^2/\bZ_{|q|+2})\cup (\bC^2/\bZ_{p+|q|+1})\cup(\bC^2/\bZ_{p+3})$, with the three branches meeting at their origins.

\medskip

\begin{figure}[h!]
\centering
\includegraphics[scale=0.75]{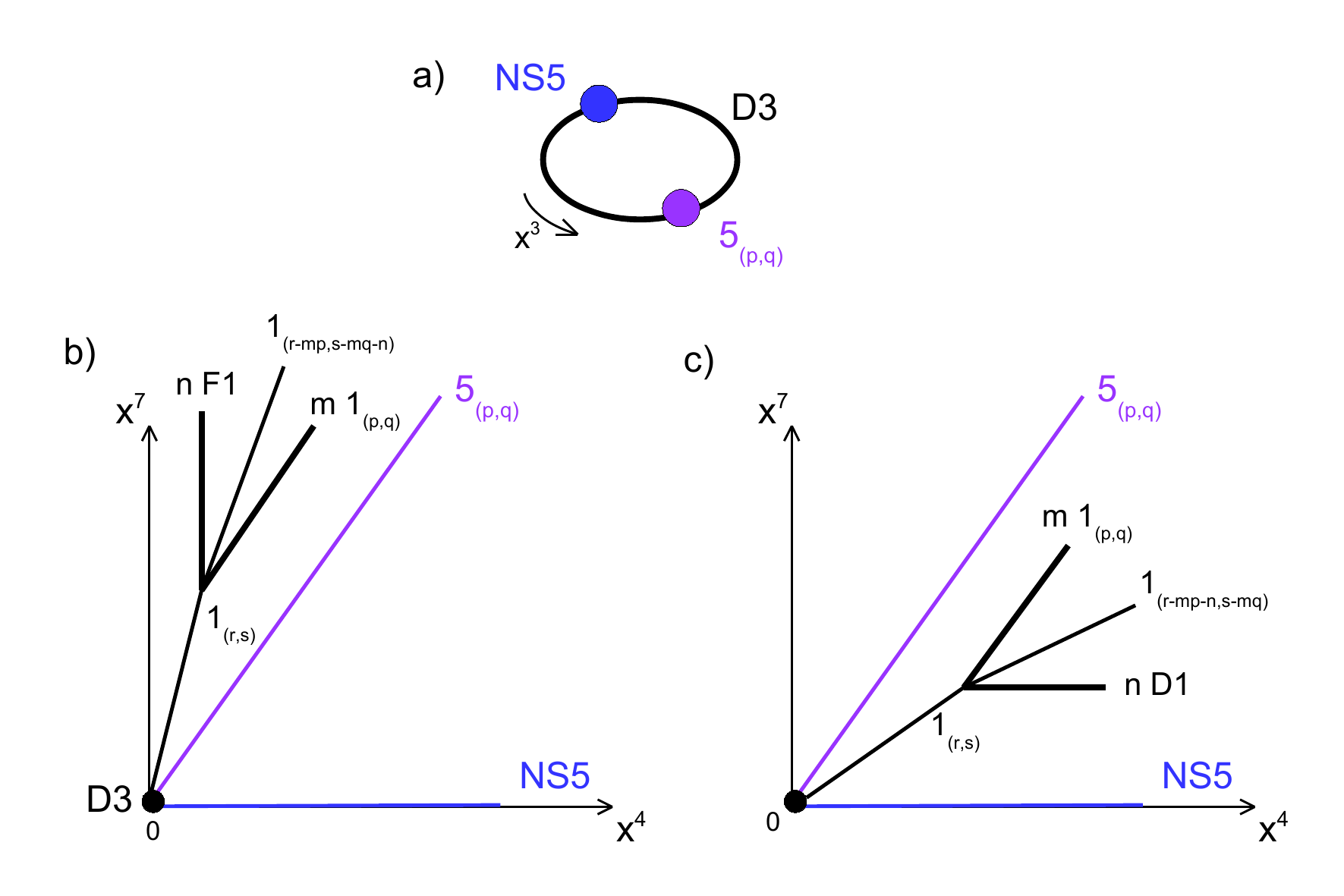} 
\vskip -0.5cm
\caption{\footnotesize  a) Brane realization of the $(p,q)$-generalized abelian ABJM theory. b) String junction for $\frac{q}{p} \le \frac{s}{r}$. c) String junction for $0 \le \frac{s}{r} \le \frac{q}{p}$.}
\label{ABJMpq}
\end{figure}

Finally we study an example of a circular quiver and we choose a simple non-trivial theory which can be thought of as a generalization of the abelian ABJM theory. It is realized from the ABJM brane setup with the 5$_{(1,\kappa)}$ replaced by a 5$_{(p,q)}$, with $p,q$ coprime integers (with $p>0$ and $q\neq 0$), as in Figure \ref{ABJMpq}-a. Since it has only two kinds of 5 branes, we know that the low-energy theory on the D3 branes has $\N=4$ supersymmetry. This theory is mentioned in \cite{Jafferis:2009th} where it is explained that the moduli space of vacua can be described as the quotient $\bC^4/\Gamma_{(p,q)}$ where $\Gamma_{(p,q)}$ acts on the complex coordinates $z_{i=1,2,3,4}$ by
\be
\underline{\Gamma_{(p,q)}:} \quad (z_1,z_2,z_3,z_4) \sim (e^{\frac{2\pi ip}{q}} z_1, e^{\frac{2\pi i}{q}} z_2,e^{-\frac{2\pi ip}{q}} z_3,e^{-\frac{2\pi i}{q}} z_4) \,. 
\label{Gammapq}
\ee
The M-theory holographic dual background should then correspond to the near horizon geometry of an M2 brane placed at the singularity of $\bC^4/\Gamma_{(p,q)}$.

The brane analysis is identical to that of the abelian ABJM theory presented in Section \ref{ssec:ABJM}. We find a single geometric branch parametrized by the vevs of the $X_1$, $X_2$ operators, which are related to the relative positions of the D3, NS5 and 5$_{(p,q)}$ branes along $x^{8+i9}$ as before, and of the monopole operators $V^{\pm}_{(r,s)}$, with $r>0$, $(r,s)$ coprime integers, realized by adding semi-infinte 1$_{(r,s)}$ strings wrapping the $x^3$ circle and ending on the D3. From the brane setup with a full 1$_{(r,s)}$ string crossing the D3, one finds the modified ring relations
\be
V^+_{(r,s)}V^-_{(r,s)} = (X_1)^{|s|}(X_2)^{|ps-qr|} \,,
\ee
where the exponent $|ps-qr|$ corresponds to the intersection number between the 1$_{(r,s)}$ string and the 5$_{(p,q)}$ brane.
The determination of redundant monopole operators is modified in the presence of the 5$_{(p,q)}$ brane. Let us assume $q>0$. 

For $\frac{q}{p} \le \frac{s}{r}$ the 1$_{(r,s)}$ semi infinite string inserting $V^+_{(r,s)}$ can end on a junction with $n$ F1s, $m$ 1$_{(p,q)}$ and a 1$_{(r-mp,s-mq-n)}$ string,  as shown in Figure \ref{ABJMpq}-b, provided the positive integers $m,n$ satisfy $\frac{q}{p} \le \frac{s-mq-n}{r-mp}$. This leads to
\be
V^\pm_{(r,s)} = (V^\pm_{(0,1)})^n (V^\pm_{(p,q)})^m V^\pm_{(r-mp,s-mq-n)} \,, \quad  \frac{q}{p} \le \frac{s-mq-n}{r-mp} \,.
\ee
A basis of operators in the range $\frac{q}{p} \le \frac{s}{r}$ is then given by the $V^\pm_{(r,\lceil \frac{qr}{p} \rceil)} \equiv U^\pm_r$, $r=1,\cdots, p$, together with $V^\pm_{(0,1)}\equiv U^\pm_0$.\footnote{If $r,\lceil \frac{qr}{p} \rceil$ are not coprime integers, the corresponding monopole operator can be expressed as a product of other monopoles and thus must be removed from the basis.}

For $0 \le \frac{s}{r} \le \frac{q}{p}$, the 1$_{(r,s)}$ string can end on a junction with $n$ D1s, $m$ 1$_{(p,q)}$s and a 1$_{(r-mp-n,s-mq)}$ string, as shown in Figure \ref{ABJMpq}-c, provided the positive integers $m,n$ satisfy $0 \le \frac{s-mq}{r-mp-n} \le \frac{q}{p}$. This leads to
\be
V^\pm_{(r,s)} = (V^\pm_{(1,0)})^n (V^\pm_{(p,q)})^m V^\pm_{(r-mp-n,s-mq)} \,, \quad  0 \le \frac{s-mq}{r-mp-n} \le \frac{q}{p} \,.
\ee
A basis of operators in the range $0 \le \frac{s}{r} \le \frac{q}{p}$ is then given by the $V^\pm_{(\lceil \frac{ps}{q} \rceil,s)}\equiv W^\pm_s$, $s=1,\cdots, q-1$, and $V^\pm_{(1,0)}\equiv W^\pm_0$.\footnote{Same remark if $s,\lceil \frac{ps}{q} \rceil$ are not coprime integers.}

For $\frac{s}{r} \le 0$, the 1$_{(r,s)}$ string can end on a junction with $r$ D1s and $-s$ F1s coming from ``below". This does not lead to any new monopole operator in the basis. 

The geometric branch of this generalized ABJM theory is therefore
\be\ba
 \underline{\scM_{\rm geom}:} & \cr
& U^+_0 U^-_0  =  X_1(X_2)^{p} \,, \cr
& U^+_r U^-_r =  (X_1)^{\lceil \frac{qr}{p} \rceil}(X_2)^{p\lceil \frac{qr}{p} \rceil-qr} \,, \quad r=1, \cdots, p \,, \cr
& W^+_0 W^-_0 =  (X_2)^{q} \,, \cr
& W^+_s W^-_s =   (X_1)^{s}(X_2)^{q\lceil \frac{ps}{q} \rceil -ps}\,, \quad s=1, \cdots, q-1 \,.
\ea\ee
To match the description of the geometric branch as the quotient $\bC^4/\Gamma_{(p,q)}$, we can solve these equations in terms of four complex coordinates $z_i$ with 
\be\ba
& X_1 = z_1 z_3 \,, \quad X_2 = z_2 z_4 \,, \cr
& U_0^+ = z_3 (z_2)^p \,, \quad U_0^-= z_1 (z_4)^p \,, \cr
& U_r^+ = (z_3)^{\lceil \frac{qr}{p} \rceil} (z_2)^{p\lceil \frac{qr}{p} \rceil - qr} \,, \quad  U_r^- = (z_1)^{\lceil \frac{qr}{p} \rceil} (z_4)^{p\lceil \frac{qr}{p} \rceil - qr} \,, \cr
& W_0^+ = (z_2)^q \,, \quad  W_0^- = (z_4)^q \,, \cr
& W_s^+ = (z_1)^s (z_2)^{q\lceil \frac{ps}{q} \rceil -ps} \,, \quad W_s^- = (z_3)^s (z_4)^{q\lceil \frac{ps}{q} \rceil -ps} \,, 
\ea\ee
with the $z_i$ subject to the identification $\Gamma_{(p,q)}$ \eqref{Gammapq}.

\medskip

We have obtained the algebraic description of the moduli space of vacua of two simple theories realized with 5$_{(p,q)}$ branes, a linear quiver and a circular quiver. The method generalizes without difficulty to any abelian linear or circular quiver. The moduli space of such theories has the same structure than that of theories realized with 5$_{(1,\kappa)}$ branes: there are maximal branches associated to 5 branes of different type present in the brane realization and an extra geometric branch for circular quivers.


\section{Future directions}
\label{sec:Future}

There is a number of extensions of the present work that can be envisioned. We give here a non-exhaustive list.

\begin{itemize}
\item It would be nice to confirm the results presented in this paper with Hilbert series computations \cite{Cremonesi:2016nbo}.
\item From the results for linear and circular quivers, it should be possible to study the space of vacua of abelian quivers with arbitrary shapes, and possibly the space of vacua of any abelian quiver. We believe that a general formula for chiral ring relations for monopole operators in abelian $\N=3$ theory can be found. Such a general formula was found for $\N=4$ abelian theories in \cite{Bullimore:2015lsa}.
\item An important direction to explore in the future is the extension to non-abelian theories. It was already found in \cite{Assel:2017hck} that the brane setups can reproduce the abelianized Coulomb branch relations of \cite{Bullimore:2015lsa} in $\N=4$ theories, from which the non-abelian ring relations can be extracted. The method still has to be developed. It is not clear at the moment whether or not the non-abelian relations can be directly extracted from brane setups.
\item Another important direction would be to explore $\N=2$ theories which admit brane realizations. These theories are subject to many infrared dualities and sometimes involve superpotentials made out of monopole operators \cite{Aharony:1997gp,Benini:2017dud,Amariti:2017gsm}. It would be interesting to test the dualities and explore the space of vacua of this very large class of theories.
\end{itemize}


\section*{Acknowledgments}

We wish to thank Stefano Cremonesi for providing many useful insights on various technical and conceptual issues. We also thank Cyril Closset for interesting discussions related to the topic.



\bibliography{Benbib}
\bibliographystyle{JHEP}

\end{document}